\documentclass[article,12pt]{article}

\usepackage[titletoc]{appendix}
\usepackage[utf8x]{inputenc}
\usepackage[T1]{fontenc}
\usepackage[english,spanish]{babel}
\usepackage{amsmath, amssymb, amsthm, amsfonts}
\usepackage{color}
\usepackage{calc}
\usepackage{lscape}
\usepackage{supertabular}
\usepackage{gensymb}
\usepackage{MnSymbol} 
\usepackage{wasysym}
\usepackage{hyperref}
\usepackage{tabulary}
\usepackage{array}
\usepackage{hhline}

\usepackage{lineno}
\usepackage{setspace}
\usepackage{cancel}

\DeclareMathOperator{\sech}{sech}
\usepackage{tikz}
\usepackage{lipsum}
\usepackage{scrextend}

\usepackage[a4paper, total={6.5in, 8.7in}]{geometry}

\definecolor{micolor}{rgb}{0.8, 0.12,0}
\deffootnotemark{\textbf{\textsuperscript{\color{micolor}(\thefootnotemark)}}}  
\deffootnote{2em}{1.5em}{\textbf{\color{micolor}\scriptsize(\thefootnotemark)\enskip}}   %
\newtheorem{condition}{Condition}

\newtheorem{definition}{Definition}
\newtheorem{lemma}{Lemma}

\definecolor{hypecolor}{RGB}{0,255, 0}
\definecolor{linkcolor}{RGB}{64,0, 64}
\hypersetup{colorlinks=true, linkcolor=linkcolor, filecolor=linkcolor, urlcolor=linkcolor}

\definecolor{LOgreen}{RGB}{35.445, 165.2, 0}

\newcommand{\abs}[1]{\left| {#1} \right| }

\newcommand{\comillas}[1]{\textquotedblleft{}{#1}\textquotedblright{}}

\newcommand{\EqF}{Eqs. \textlbrackdbl(\ref{E:CartesianPrediction}) $ \equiv $ (\ref{E:SimplisticPrediction})\textrbrackdbl{ }}
\newcommand{\e}{\mathrel{e\mspace{-7.5mu}e}}
\newcommand{\G}[1]{{\text{\textsf{N}}}{[#1]}}

\newcommand{\given}{\;\mid\;}
\newcommand{\GNU}{\includegraphics[height=0.4cm]{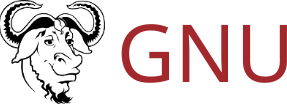}{ }}
\newcommand{\GogSch}{\includegraphics[height=0.33cm]{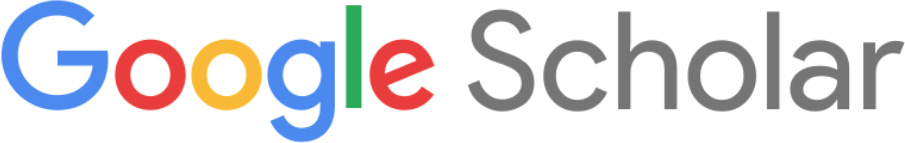}{ }}

\newcommand{\lindep}{\perp \!\!\! \perp }
\newcommand{\lqqd}{\boldsymbol{\blacksquare}}

\newcommand{\LOv}{{\color{LOgreen}\text{\textbf{Libre}}}Office (v 7.1.4.2-2)}
\newcommand{\mbb}[1]{\mathbb{#1}}
\newcommand{\mf}[1]{\mathfrak{#1}}
\newcommand{\nlindep}{\not{\lindep}}

\newcommand{\R}{\mbb{R}}

\newcommand{\sign}[1]{\text{sgn}{[#1]}}
\newcommand{\sindep}{\mathrel{\perp \mspace{-10mu}\perp \mspace{-10mu}\perp}}

\newcommand{\then}{\text{\wasytherefore{\quad }}}
\newcommand{\TXs}{\includegraphics[height=0.35cm]{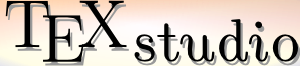}{ }v. 4.0.0 (git 4.0.0alpha1-2-g74caae43) }

\newcommand{\vect}[1]{\overrightarrow{#1}}
\newcommand{\vmag}[1]{\abs{\abs{{#1}}}}
\newcommand{\vrdif}{\neq\mspace{-8mu}\neq}



\begin{document}
	\selectlanguage{english}
	\bibliographystyle{plain}

		\title{First derivatives at the optimum analysis (\textit{fdao}): An approach to estimate the uncertainty in nonlinear regression involving stochastically independent variables.}
			
		\author{
			C. Sevcik\thanks{
				C. Sevcik , SciMeDAn, Av. Paral\textperiodcentered{}lel 124, Ent. 2B, Postal Code 08015, Barcelona, Spain. Phone: +34 697 66 84 0. eMail: \url{carlos.sevcik.s@gmail.gmail}. ORCID: 0000-0003-3783-6541.}
			// MD, PhD, Professor Emeritus, //
			Centro de Biof\'{\i}sica y Bioqu\'{\i}mica,//
			Instituto Venezolano de Investigaciones Cient\'{\i}ficas (IVIC),//
			Caracas, Venezuela.
		}
	
		\maketitle
		
		\begin{abstract}
			An important problem of optimization analysis surges when parameters such as $ \vect{\{\theta\}_j}_{j=1,\, \dots ,k }$, determining a function $ y=f(x \given \vect{ \{\theta\}_j}_{j=1,\, \dots ,k }) $, must be estimated from a set of observables $ \{ x_i,y_i\}_{i=1,\, \dots \,,m} $. Where $ \{x_i\} $ are independent variables assumed to be uncertainty-free. It is known that analytical solutions are possible if $ y=f(x\given\theta_j) $ is a linear combination of $\{ \vect{\theta}_{j=1,\, \dots \,,k} \}.$ Here it is proposed that determining the uncertainty of parameters that are not \textit{linearly independent} may be achieved from derivatives $ \tfrac{\partial f(x \given \{ \vect{\theta}_j\})}{\partial \theta_j} $ at an optimum, if the parameters are \textit{stochastically independent}.
		\end{abstract}

	
	%For numbers numerados por orden de citación
	% end For numbers
	%\bibliographystyle{model2-names}
	%\bibliographystyle{model4-names}
	%\bibliographystyle{model5-names}
	%\bibliographystyle{elsarticle-num}
	%\bibliographystyle{model3d-num-names}
	
\newpage
\tableofcontents
%\listoffigures
%\listoftables
%\newpage

\section{Introduction.}\label{S:Intro}

\subsection{Formulation of the problem.}

Data from the physical world always contains uncertainty which does not result from measurement error. The sources are many: Limitations of the observation instrument make the measurements hazy; optical instruments are a very intuitive example of fuzziness; actually, measuring instrument introduces fuzziness, which is dependent on the instrument’s limit of resolution; impossibility of accurately measuring something, a classical example deals withe the velocity and position of a particle \cite{Heisenberg1927}; observing reality with a scope (aim or purpose) modifies the object observed \cite{Sassoli2013a}, this is specially relevant to quantum physics, but applies to any measurement (draining current, compressing with a caliper, heating, etc.) to, hopefully, a minor extent; uncertainty is essential to life, otherwise any noxious factor would affect equally a whole species population making its extinction likelier, thus any parameter measured on living beings is significantly variable, uncertain, hazy \cite{Conrad1977}; fuzziness appears also when the object measured changes more or less cyclically in time, the height of the Mont Blanc peak (like most other mountains) is a well known case \cite{Gilluly1949, Evans2015}; since temperature ($ T $) is $ T > 0 \text{\textdegree  K} $, molecules vibrate and rapidly change between conformations and molecular properties are hazy too \cite{Sherwood1972, Morters2009, Morters2010}; in high energy physics the existence of a particle was evidenced by an energy peak which had to be differentiated from background noise \cite{CERN2015}; all processes of chemical or electrical intercellular communication are stochastic in nature \cite{delCastillo1954a, Armstrong1974, Fishman1975, Neher1976}. Regression analysis is a fundamental tool tool to calibrate the dependency between variates and to calibrate the uncertainty involved in the dependency, if it exists. 

An important problem of optimization analysis surges when it is desired to guess the parameters $ \vect{\{\theta\}}_{i=1,\, \dots ,k} $, determining a function $ y=f(x \given \{ \theta_{i=1,\, \dots \,,k}\}) $ (also called the \textit{objective function}, \textit{obf}) must be determined from a set of observables $ \{ x_j,y_j\}_{j=1,\, \dots \,,m} $. Where $ \{x\} $ are independent variates assumed to be uncertainty-free (called explanatory variables),  and their associated observed dependent variates (response variables), $\{ y\} $. Analytical solutions are possible, and many are well known (see for example \cite{Montgomery2012}), when $ f(x \given \{ \theta\}) $ involves a linear combination of $ \vect{\{\theta\}}$. In the case of non linearly independent \cite{Seber1989, Schittkowski2002} $ \{x_i.y_i\} $ pairs:

\begin{quotation}
	\textquotedblleft{}In general, there is no closed-form expression for the best-fitting parameters, as there is in linear regression. Usually numerical optimization algorithms are applied to determine the best-fitting parameters. Again in contrast to linear regression, there may be many local minima of the function to be optimized and even the global minimum may produce a biased estimate. In practice, estimated values of the parameters are used, in conjunction with the optimization algorithm, to attempt to find the global minimum of a sum of squares \textquotedblright{}.
\end{quotation}

The problem of local minima is inherent to the function fitted and cannot be avoided. Yet, efficient minimization algorithms, which start searching from a set of user provided $ \{ \theta_{i} \}_{init} $ parameters in case of many \textquotedblleft{}well behaved functions\textquotedblright{} converge towards the global optimum if $ \{ \theta_{i} \}_{init} $ is within a certain boundary of of the global optimum$ \{ \theta \}_{opt} $. There is, however, no analytical solution to the problem of knowing the width the boundary of guaranteed convergence to the global optimum (minimum or maximum) for a given \textit{obf}.

If $ \{  \vect{\theta\}} $ is a set of orthogonal Cartesian variable set with Euclidean metric, then the gradient is the vectorial sum 
\begin{equation}\label{E:CartesianSum}
	\vect{\nabla f\left( x \given \{\theta \}\right) }_{opt}=\sum_{i=0}^{k} \left( \frac{\partial f\left( x \given \{ \theta_i\}\right)_{opt} }{\partial \theta_i} \boldsymbol{u_i}\right),
\end{equation}
where $ \boldsymbol{{u}_i} $ are the standard unit vectors in the directions of each of the coordinate; the vector\textquoteright{}s magnitudes is then
\begin{equation}\label{E:CartesianMod}
	\left\|  \nabla f_{opt}\right\| =\sqrt{\sum_{i=0}^{k} \left[\frac{\partial f\left( x \given \{ \theta_i\}\right)_{opt} }{\partial \theta_i} \right]^2}.
\end{equation}
When the Hessian matrix (see Appendix \ref{S:Hessian})  of $ f\left( x \given \{ \vect{ \theta} \} \right) $ is diagonal this equation is a linear function.

As a concrete example of practical importance we will consider the Hill equation \cite{Bancroft1910, Hill1910b, Hill1913, Langmuir1918, Segel1975, Weiss1997, Gesztelyi2012, Reeve2013} in its original form is [Also see Eqs. (\ref{E:MassActionLaw}) through (\ref{E:HillMod})]:
\begin{equation}\label{E:Hill}
	y\left( [D]\given\{y_m,K_m,n\} \right) =\frac{y_m}{1+\left( \tfrac{K_m}{[\mathrm{D}]}\right)^n}=\mf{H}.
\end{equation}

The Hill equation is used in enzyme kinetics and in pharmacology to represent the interaction of one or more molecules of substrate with the catalytic site of an enzyme, or of a drug molecule with its receptor site \cite{Ariens1964, Segel1975}. Under these conditions $n$ is called the \textit{molecularity }of the reaction. Yet, $ n $, is also used in situations where properties of the enzyme or drug receptor are modified during the interaction, the, so called, cooperative schemes, where $ n \in \R$ is plainly named \textit{Hill coefficient} without further molecularity implications  \cite{Monod1965, Segel1975, Abeliovich2005}. There are also situations in which when drug effects are studied in cells, tissues or cell fragments, there may exists several kind of drug \cite{Sevcik1976b, Sevcik1982} or enzyme receptors, following kinetics such as the one described by Eq. (|ref{E:ChemEquil})  could result in $ n \in \R  $ if the diversity is not recognized. Combining the notion of gradient of a function and the Hill equation we get the set of all its first partial derivatives, it may be represented as a vector 
\begin{equation}\label{E:GradVect}
	\begin{matrix}
		\vect{\nabla f\left( x \given  \vect{\{\theta\}}\right)}  &= \left[ 
		\begin{matrix}
			\frac{\partial f\left( x \given \vect{\{\theta\}}_{opt}\right) }{\partial \theta_1}\\
			\frac{\partial f\left( x \given \vect{\{\theta\}}_{opt}\right) }{\partial \theta_2}\
			\vdots\\
			\frac{\partial f\left( x \given \vect{\{\theta\}}_{opt}\right) }{\partial \theta_k}\\
		\end{matrix}
		\right]  \implies \vect{\nabla \mf{H}}= \zeta 
		\left[ 
		\begin{matrix}
			&1\\
			&-\frac{n y_m \mf{D}^n}{K_m} \zeta\\
			&-y_m \mf{D}^n \log \left(\mf{D}\right)\zeta
		\end{matrix}
		\right]  ,
	\end{matrix}
\end{equation}
where $ \mf{H} $ replaces the Hill equation \cite{Hill1909, Langmuir1918, Segel1975} as in Ec. (\ref{E:Hill}), and $ \zeta =\tfrac{1}{1+\left(\tfrac{K_m}{[D]} \right)^n} $, where $ \mf{D} =\tfrac{K_m}{[D]}$ [Deduced in Eqs. (\ref{E:dy0}) through (\ref{E:dn})] may be called \textit{reciprocal normalize concentration}. An interesting particular case is
\begin{equation}\label{E:GradVect_DigKm}
	\begin{matrix}
		\vect{\nabla \mf{H}}_{\mf{D}=1}=
		\left[ 
		\begin{matrix}
			&\frac{1}{4}\\
			&-\frac{n \cdot y_m }{4 \cdot K_m}\\
			&0
		\end{matrix}
		\right].
	\end{matrix}
\end{equation}

The Hill equation is an example where equation parameters represent physical entities that are indecent from one another,  that in, spite of any exception, there are situations where 
\begin{equation}\label{E:NotEntangled}
	(\theta_i \sindep \theta_{j}) \; \forall (i \neq \; j),
\end{equation}
where $\sindep$ indicate stochastic independence, in which the $ \vect{\theta} $ get \textit{entangled}\footnote{Plain English, nothing to do with quantum entanglements.}, \textit{interweaved} or \textit{intertwined}  when considered in connection with $ f\left( x \given \{\vect{\theta} \}\right) $ and where $ (\theta_i \nlindep \theta_{j}) $, where, again, $ \nlindep $ means \textit{not linearly independent}. This produces the duality in in equations such a E. (\ref{E:Hill}), it does not have a diagonal Hessian matrix [see Eq. (\ref{E:HillHessian})], and $ y_m $, $ K_m $ and $ n $,  
\begin{equation} \label{E:StocIndTriangle}
	\\ 
	\includegraphics[width=1.5cm]{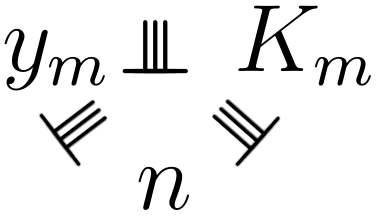}  
\end{equation}
in many empirical situations, no not determine each others values, in words: \textit{they are stochastically. independent from each other}, but they are entangled in Ec. (\ref{E:Hill}), where\
\begin{equation} \label{E:NotLinDepTriang}
	\\ 
	\includegraphics[width=1.5cm]{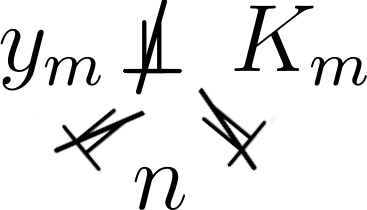}
\end{equation}
not linearly independent in Ec. (\ref{E:MassActionLaw}), they are entangled in this equation. A similar situation is found in the Boltzmann equation \cite{Sevcik2017a}. Another example may be foun for a Gaussian variable such as
\begin{equation}\label{E:GaussVariab}
	\text{N}[u,s^2]=\dfrac{1}{\sqrt{2 \pi s^2}} \e^{-\tfrac{1}{2} \left( \tfrac{x-u}{s}\right) ^2}
\end{equation}
which has a non diagonal Hessian matrix which indicates that $ (u \nlindep s)  $ and yet $ (u \sindep s) $, changes in $ u $ do ot modify $ s $, and vise versa. A similar situation happens with the Cauchy \cite{Cramer1991, Pitman1993, Wolfram2003} distribution changes: in its median ($ \hat{\mu} $) have no effect on its width factor ($ \lambda $), and yet the Cauchy distribution has a non diagonal Hessian matrix, indicating that its median  and its width factor are  ($ \hat{\mu} \nlindep \lambda  $). And if you keep dining, you will uncover many more cases. The inter-molecular reaction parameters of the Hill equation are scholastically independent: the maximum effect (or speed of catalysis) ($ y_{m} $) does not depend on the affinity constant of the reactants ($ K_m $), and none of them depends on the molecularity of the reaction (number of molecules of one kind reacting with a molecule of another kind, $n$) \cite{Ariens1964, Segel1975, Erdi1989}. A similar reasoning can be used in connection with the Boltzmann equation [Eq. (\ref{E:BoltzElect})] where $ V_{\text{\textonehalf}} $ and $ {v} $ are mechanistically independent.

The $ \vect{\theta}_i$  in  $y= f(x \given  \vect{\{\theta \}}) $ in all these circumstances are stochastically and causally in dependent, but entangled or intertwined when measured. How does this \textit{entanglement} or \textit{intertwining} occurs is unclear, but Eqs. (\ref{E:Hill}), (\ref{E:GaussVariab}) (\ref{E:DensCauchy}) and (\ref{E:BoltzElect}), have one thing in common, \textit{the stochastically and causally independent variables are are each entangled or intertwined with another parameter which is plotted as abscissa: also called the \textbf{independent determining} variable when the function is plotted \cite{DependentIndependent2021},  $ [D] $, $ x $ or $ V $ (depending on which of these equations discussed here, we consider)}. 

\begin{figure}[h!]
	\centering
	\includegraphics[width=12cm]{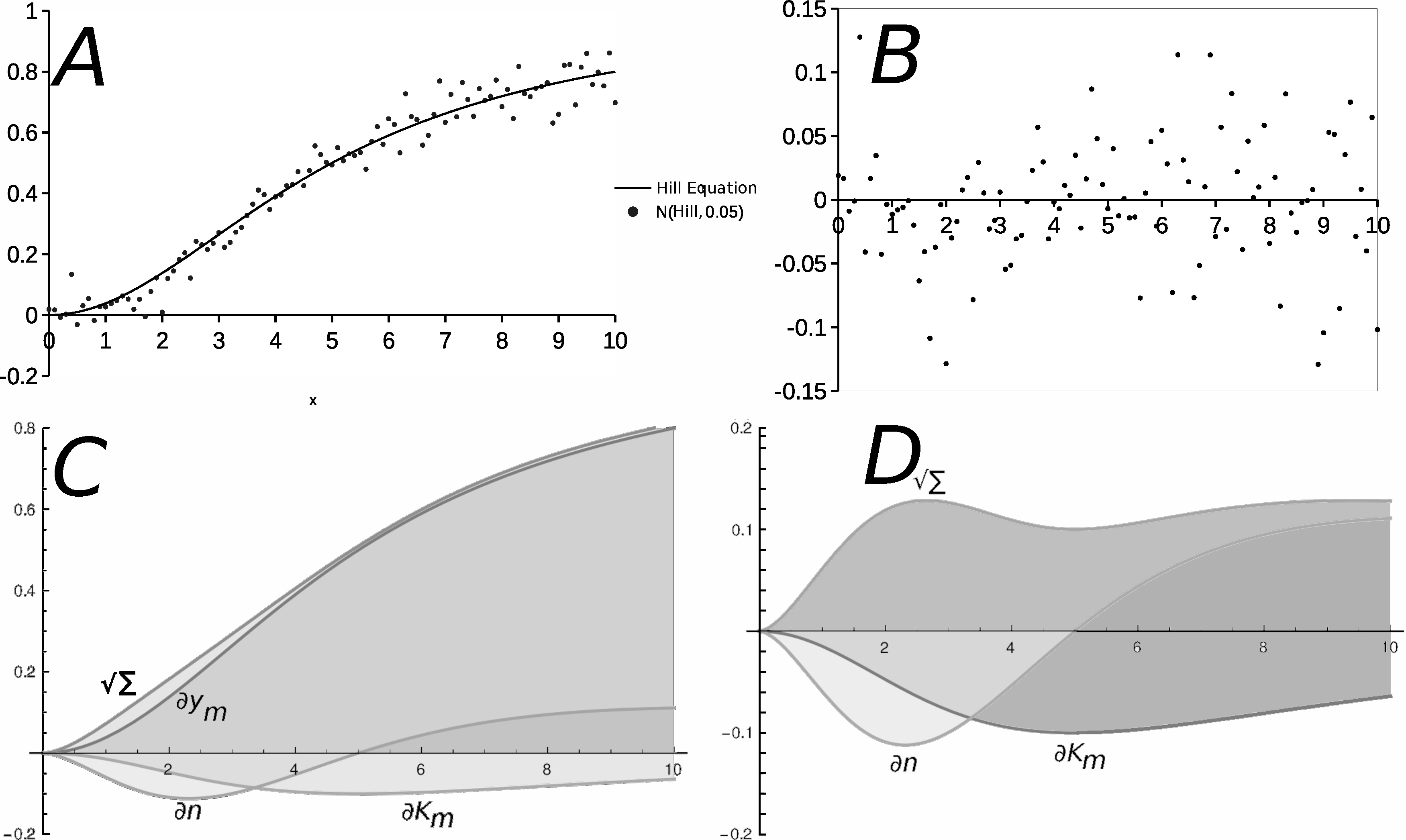}
	\caption{\textbf{Example of first derivatives as vectors. } The figure is a simulated fit to a Hill equation,. \textbf{Panel A}: Black line $ \mf{H}=f(x \given \{y_m,K_m,n\}) $, a Hill equation wit $ y_m=1 $, $ K_m=5 $ and $ n=2  $. Dots ($ \bullet $) are a of 100  Gasussian, $ \G{\mu,\sigma^}= \G{\mf{H},,0.05} $ points where $ \mu $ is mean and $ \sigma^2 $ is variance. \textbf{Panel B}: Residuals, i.e. dots ($ \bullet $) are the same in panel A, after subtracting the black line in the same panel from each dot. \textbf{Panel C}: Present the partial derivatives of $ \mf{H} $, to keep the figure uncluttered, each derivative, and th corresponding vector are represented by the partial $ \partial $ denominator of the derivative ($ \equiv $, indicates equivalence): $ \partial y_m \equiv \tfrac{\partial \mf{H}}{\partial y_m} $; $ \partial K_m \equiv \tfrac{\partial \mf{H}}{\partial K_m} $; $ \partial n \equiv \tfrac{\partial \mf{H}}{\partial n}$ (Also in Definition \ref{D:DefinitionsAndLabels}). $ \Sigma \sqrt{}\equiv  \vmag{\nabla\mf{H}}= \sqrt{(\partial y_m)^2+( \partial K_m)^2+(\partial n)^2}  $ is the magnitude \cite{Magnitude2021} of a Cartesian vector [See also Eq. (\ref{E:ScaCartGrad})]. \textbf{Panel D:} Like Panel C but with  $ y_m =1 $ set constant. Data for this figure were generated with Monte Carlo simulation \cite{Dahlquist1974, Metropolis1987} using the Box and Muller algorithm \cite{Box1958} and using the  	\LOv Calc RAND.NV() function to generate uniform random variates of type $ U(0,1) $; this function was used only in this figure.}
	\label{F:Vectors} 
	\end{figure}

Equations such as Ec, (\ref{E:GradVect}) indicate than the gradient o $ f \left(x \given \{ \theta \} \right) $  is a continuous vector function of $ x $, the entangling or intertwining variable should be taken into account when expression the \textit{obj} function gradient, which should be properly expressed as
\begin{equation}\label{E:EntangGradVect}
	\vect{\nabla_{x,\{\vect{\theta}\}}} f\left( x \given \vect{\{\theta\}}\right) 
	\underset{\textit{opt}}{\longrightarrow} 
	\left[ 
	\begin{matrix}
		\frac{\partial f\left( x \given \vect{\{\theta\}}\right) }{\partial \theta_1}\\
		\frac{\partial f\left( x \given \vect{\{\theta\}}\right) }{\partial \theta_2}\\
		\vdots\\
		\frac{\partial f\left( x \given \vect{\{\theta\}}\right) }{\partial \theta_k}
	\end{matrix}
	\right]  = \vect{\nabla f}\left( x \given \vect{\{\theta\}}\right)_{opt} .
\end{equation}
This means that a optimization procedure will modify the elements of  $\vect{\{\theta\}}  $ until $  \vect{\nabla_{x,\vect{\{\theta\}}}} f\left( x \given \{\vect{\theta}\}\right)$,   cannot be further reduced, you are reduced to the system\textquoteright{}s inherent uncertainty, the system entropy in a Boltzmann sense \cite{Boltzmann1964} or to the system information in the Shannon sense \cite{Shannon1948}. In statistical terms the system\textquoteright{}s residual variably which is seen as unpredictable, random, variation of some sort, this is seen as $ \bullet  $ in \textbf{Figures \ref{F:Vectors}A}, \textbf{\ref{F:Vectors}B} and \textbf{\ref{F:GradientAndVectors}B}. 

\textbf{Figure \ref{F:GradientAndVectors}A} is a Cartesian scheme illustrating the relationship between the gradient $  \vect{\nabla \mf{H}} $ [labeled $ \boldsymbol{\Sigma \sqrt{ }} $  in the figure and and in Eq. (\ref{E:CartGrad})], with other labels as in  Definition \ref{D:DefinitionsAndLabels}  with a graphical representation of the $ \boldsymbol{\alpha, \; \beta} \text{ and } \boldsymbol{\gamma} $ angle labels used in Eq. (\ref{E:EmpGradToCoord}).

To facilitate understanding of the analysis proposed here we introduced \textbf{Figures  \ref{F:Vectors}} and \textbf{\ref{F:GradientAndVectors}}. \textbf{Figure \ref{F:Vectors}A} presents a black line representing Eq. (\ref{E:Hill}), aka $ \mf{H} $, surrounded by 100 Gaussian dots ($ \bullet $) all with a Gaussian probability density function (\textit{fdp}) $ \\G{\mu,\sigma)} = \\G{\mf{\mf{H}}, 0.05} $, the black line is meant to represent an objective function surrounded by \comillas{data} (simulating residual deviations or plainly \textit{residuals}) after an \comillas{optimum fit} is found, the quote marks are used here to stress that Figure \ref{F:Vectors} is realty a simulation. It is important to notice that variance was set to be same along the whole span of the curve \comillas{fitted}, a condition called \textit{homoscedasticity}.

 \textbf{Figure \ref{F:Vectors}B} presents the residuals after subtracting $ \mf{H} $ value to make visually evident the homoscedasticity. The residuals deviation fro $ \mf{H} $ may stem from, at least, two sources. The empirical situation described by the objective function parameters may be fuzzy, may vary due to physical factors such as temperature, with random thermally induced vibrations of the object studied,  the parameters \comillas{vibrate}, The other source, may be random errors (inaccuracies)  when $ \mf{H} $ is determined some times called \comillas{experimental errors}. Both factors make fuzzy or inaccurate the estimates os $ \{\vect{\theta} \} $, the unknowns we want to determine. Whatever the origin of the residuals observed in Figures \ref{F:Vectors}A, \ref{F:Vectors}B or \ref{F:Vectors}F, they will determine that optimal estimates of $ \{y_m, K_m, n\} $ will be fuzzy, uncertain. The central purpose of this paper is to delimit this uncertainty. If the origin of the uncertainty is a random changes in $ \mf{H} $ value must occur in the \textit{surface of solutions}, all the possible values of $ \mf{H} $ given the possible $ \{ y_m, K_m, n \} $, very large or infinite. 

\subsection{Analysis of first derivatives at the optimum in a Cartesian system of coordinates.}

\begin{figure}[h!]
	\centering
	\includegraphics[width=12cm]{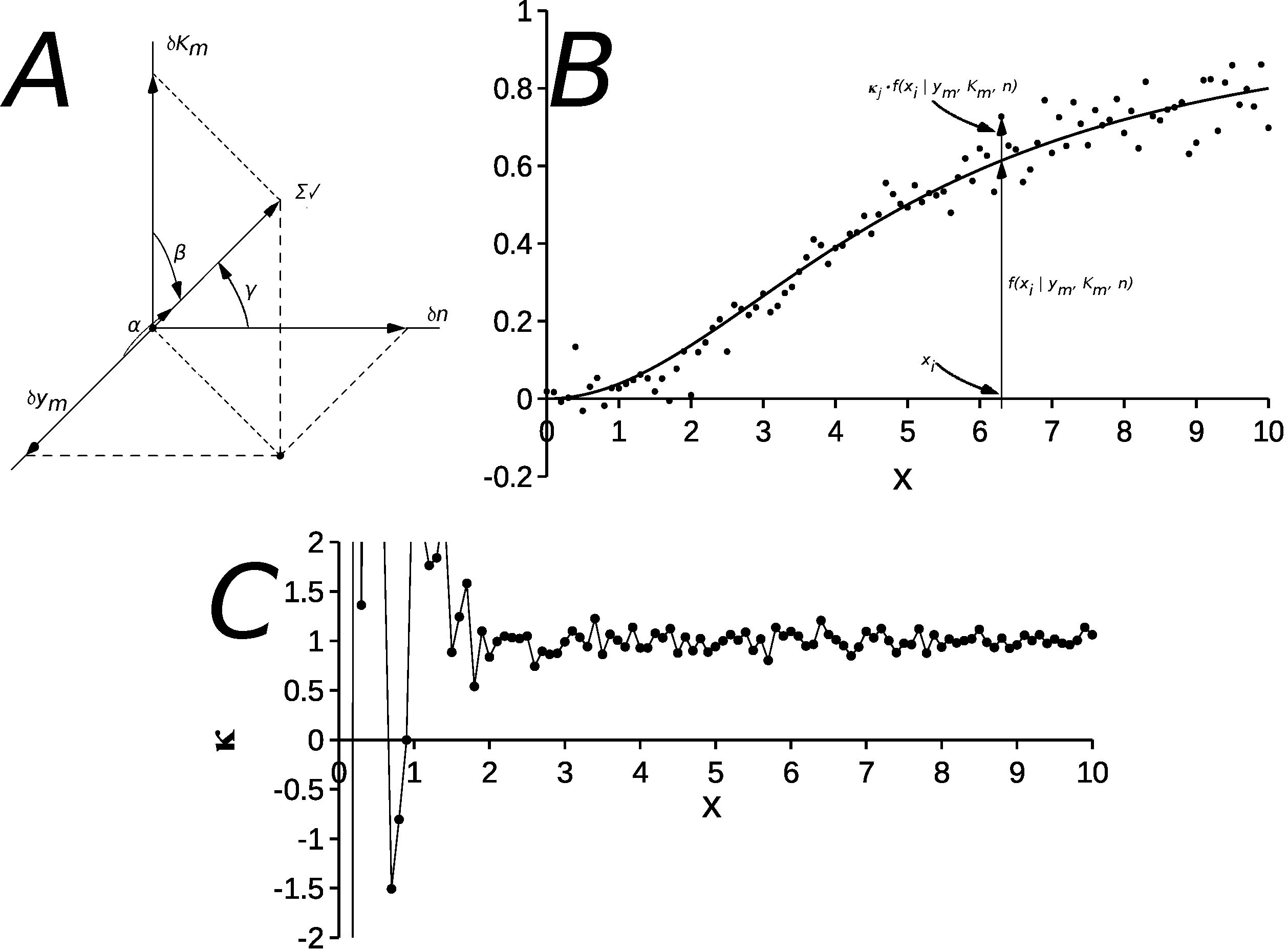}
	\caption{\small{\textbf{A schematic view to understand the relationship between $ \mf{H}=f(x\given \{y_m,K_m,n\} ) $,  $ \mf{H}_i=\mf{f}(x_i \given \{y_m,K_m,n\} ) $ and gradient, $ \nabla \mf{H} $, in a tridimensional Cartesian space.} Symbols : $ \partial y_m $, $ \partial K_m $,  $\partial n$ and $ \sqrt{} \Sigma $ , have same meaning as in  Figure \ref{F:Vectors}. \textbf{Panel A}:Greek letters $ \alpha $, $ \beta $ and $ \gamma $, aside of curved arrows, indicate angles between the gradient vector labelled $ \Sigma\sqrt{} $ and each coordinate. \textbf{Panel B:} Is a reproduction of Figure \ref{F:Vectors}A, but with two arrows added at an arbitrarily choose point $ x_i $. A \textbf{black vertical arrow} spans the distance between  0 and $ \mf{H}_i $}, and a \textbf{gray arrow}, underlying the black one, spans from 0 to one of the \comillas{empirical} data point having coordinates $ \left[ x_{i},( \kappa_j \cdot  \mf{H_i} ) \right]  = \left[ x_i, \mf{H}_{i,j}\right] $, where, obviously, $ \kappa_{i,j} = \frac{\mf{H}_{i,j}}{\mf{H}_i}$ is a scaling parameter for $ x_i $ and the data point $ j $ observed at $ x_i $. \textbf{Panel C:} Is a plot of $ \kappa_{j,i} $ values. At values of $ x_i <2 $,  $\kappa_{j,i} $ is ill behaved [range: ($-170.3  $, $ 27.1 $)].  See the text for the \textit{meaning of the steel ball placed in Panel D} and other details.  }\label{F:GradientAndVectors} 
\end{figure}

 Separating the sources of variation may be impossible, at least while the experimental conditions are kept constant. Thus the sources of variation will be lumped. Yet, since if you determine $ \mf{H} $ keeping $ [D] $ constant, it seems obvious that the variability is independent from the entangling or determining, independent, variable.
\begin{definition}\label{D:DefinitionsAndLabels}
	The following symbols for derivatives are used in the figures and sometimes refereed in the text.
	\begin{equation}
		\begin{matrix}
			\frac{\partial \mf{H}}{\partial y_m} &\equiv& \overbrace{\partial y_m \ }^{\text{\tiny{symbols in figures}}} \\
			\frac{\partial \mf{H}}{\partial K_m} &\equiv& \partial K_m  \\
			\frac{\partial \mf{H}}{\partial  n}&\equiv& \partial n  \;\; 
		\end{matrix}
	\end{equation}
\end{definition}
where equivalences at right (as well as the symbols between parentheses) are presented to connect link with symbols used in \textbf{Figure \ref{F:Vectors}C}. Data in \textbf{Figure \ref{F:Vectors}C} are particularly interesting since derivatives and gradients determine the spontaneous evolution of random systems. A system changes spontaneously only if there is a gradient, and changes in the direction minimizing system  free energy and maximizing its entropy, and these are not just thermodynamic concepts \cite{Szilard1929, Shannon1948, Renyi1959, Boltzmann1964, Szilard1964, Wang2008}. Since $ \tfrac{\partial \mf{H}}{\partial K_m} $ and $ \tfrac{\partial \mf{H}}{\partial n} $, specially at low $ [D] $,  with low $ [D] $ will make changes more likely.

\textbf{Figure \ref{F:Vectors}C} ialso contains a plot of the Cartesian gradient vector magnitude
\begin{equation}\label{E:CartGrad}
	\vmag{\nabla \mf{H}} = \sqrt{\left( \tfrac{ \partial \mf{H}}{\partial y_m}\right) ^2+	\left( \tfrac{\partial \mf{H}}{\partial K_m}\right) ^2 + \left( \tfrac{\partial \mf{H}}{\partial n}\right) ^2} \quad  \equiv \overbrace{  \sqrt{}\Sigma}^{\text{\tiny{symbols in figures}}}, 
\end{equation}
and using the Greek letters in \textbf{Figures  \ref{F:Vectors}},   and \textbf{\ref{F:GradientAndVectors}}, where the equivalence also presented to connect link with symbol used un \textbf{Figures \ref{F:Vectors}C}, to represent graphically a Cartesian gradient ($ \sqrt{}\Sigma $  in the figure) and its component vectors ($  \partial y_m $, $  \partial K_m $ and $  \partial n $ in the figure).

From either Eq. (\ref{E:EntangGradVect}) or Eq. (\ref{E:CartGrad}) It is trivial to show that if we scale the gradient predict from optimized $ \{y_m, K_m, n\} $ all its vector components will be scaled bu the same factor:
\begin{equation}\label{E:ScaCartGrad}
	\begin{small}
		\boldsymbol{\then \kappa  \vmag{\nabla( \mf{H})} = \sign{\kappa} \sqrt{\kappa^2\left[ \left(  \tfrac{ \partial \mf{H}}{\partial y_m}\right) ^2+	\left( \tfrac{\partial \mf{H}}{\partial K_m}\right) ^2 + \left(  \tfrac{\partial \mf{H}}{\partial n}\right) ^2\right] } \given (\kappa \neq 0) \in \R   \;  \lqqd}
	\end{small}
\end{equation}
where the \textit{sign}, $ \sign{x} $,  function is defined
\begin{equation}\label{E:sfn}
	\sign{x}=
	\begin{cases}
		-1 &\impliedby x<0 \\
		0  &\impliedby xa=0 \\
		1 &\impliedby x>0 \\
	\end{cases} \quad .
\end{equation}
Equation (\ref{E:ScaCartGrad}) is not only reasonably obvious, but agrees with a fundamental property of metric spaces: \textit{the topology of a metric space does not change under linear transformation} \cite{Barnsley1993}. Eq. (\ref{E:ScaCartGrad}) has an additional interesting meaning if considered in connection with the arrows in Figure \ref{F:GradientAndVectors}B, if there is an optimum $ \mf{H}_{opt,i}$ for a given $x_i= [D]_i $, and $ 1 \leqslant j \leqslant m $ \comillas{empirical} points $ \mf{H}_{i,j}$  we define 
\begin{eqnarray}
	\kappa_{i,j} =& \dfrac{\mf{H}_{i,j}}{\mf{H}_{opt,i}} \equiv \dfrac{\mf{f}_{j,i}}{\mf{f}_{opt,i}}\label{E:kappa_ji} \\
	\delta_{j,i}=& \mf{H}_{i,j} -\mf{H}_{opt,i}\label{E:delta_ji} \equiv \mf{f}_{i,j} -\mf{f}_{opt,i}\label{E:delta_ji}
\end{eqnarray}
where $ \kappa_{j,i} $ is the ratio, and $ \delta_{j,i} $ the difference, between the overlapping straight vertical arrows, drawn at an arbitrary $ x_i $ in \textbf{Figure \ref{F:GradientAndVectors}B}. Where $ \mf{f} $ refers to any function, equal or different from the Hill equation.

Where the \textit{opt} sub indexes indicate the parameter values defining the black line in \textbf{Figures \ref{F:Vectors}A and \ref{F:GradientAndVectors}B}; the $ \mf{f} $ notation refers to an function in a more general manner, beyond the Hill equation. Letter $ \boldsymbol{j=1,\ldots, d_i } $ in de sub index indicates the $ \boldsymbol{j} $\textquoteright{}s pair at $ \boldsymbol{x_i} $. $ \boldsymbol{\kappa} $ will generally be $ \kappa > 0 $ in the region not excessively variable ($ x \geqslant2  $) in \textbf{Figure \ref{F:GradientAndVectors}C}, but may be $ \kappa \gtreqqless 0 $ at lower $ x $ values.

As suggested by the black and gray arrows in \textbf{Figure \ref{F:GradientAndVectors}B},it is possible to scale the optimized value of $ \mf{H} $ at abscissa value $ x_i $, which we will cal $ (x_i, \mf{H}_i) $, to match the $ j $st observation of a set of size $ m $ also occurring at $ x_i $, say $ \mf{H}_{i,j} $,  by a $ kappa_{j,i} $ factor to obtain the parameters of $ \mf{H}_{i,j} $ scaling by $ \kappa_{j,i} $ the parameters of the optimized $ \mf{H}_i $. Figure \ref{F:GradientAndVectors}C is a plot of $ \kappa_{j,i} $ values for all the $ \bullet $ in \textbf{Figures \ref{F:Vectors}A}, \textbf{\ref{F:Vectors}B} and \textbf{\ref{F:GradientAndVectors}B}.  Probably due to subtractive cancellation \cite{LossOfSignificance2021}, there is considerable uncertainty regarding $ \kappa_{j,i} $ at $ x_i \leqslant 2 $, but the variability of $ \kappa_{j,i} $  is narrowly bound within a ($ 0.744 $, $ 1.226 $) range at $ x_i \geqslant 2 $. Variability at low $ x_i $ is to be expected, because we divide the random variates by decreasing values of $ \mf{H_i} $, and because optimization at low $ x_i $ if difficult in this region where gradient due to $ \vect{\{y_m,K_m,n\}} $ as expressed by Eqs. (\ref{E:CartGrad})  and (\ref{E:ScaCartGrad}) components (see Figure \ref{F:Vectors}C)  become very large, and even under homeostatic conditions variations (\textbf{Figures \ref{F:Vectors}A, \ref{F:Vectors}B  and \ref{F:GradientAndVectors}B}) in $ x_i $ may have huge impact on $ \vect{\nabla \mf{H}}_{i.opt}  $ in this range.

It is interesting to scale factor $ \kappa $ properties. When tested for Gaussianity The Jarque-Bera \cite{Jarque1980, Bera1981, Bera1981a, Giles2014},  the robustified Jarque-Bera test of Gel-Gastwirth \cite{Gel2008} and the Shapirp-Wilk  test \cite{Shapiro1965}, residuals in Figure \ref{F:Vectors}B  were found to be Gaussian $ (P>0.05) $. Yet the residuals had a kurtosis, $ ku = 2.759 $ and skewedness, $ sk = 0.113$ which the robustified Jarque-Bera by Gel and Gastwirth test, considered that te set of residuals may be not Gaussian $ (P  = 4 \cdot 10^{-4}) $, which was  below $ P=0.005 $, considered today a safer threshold of significance \cite{Benjamin2018, Ioannidis2018}. Values of $ \kappa_{j,i} $ for $ x_i \geqslant 2 $ were only weakly non Gaussian when tested with the robustified Jarque-Bera by Gel and Gastwirth test $ (P =  1.7 \cdot  10^{-3}) $, in this range: $ ku = 4.292$ and $ sk = 0.311  $. The situation was dramatically different when $ 0 \leqslant x_i \leqslant 10 $ were considered, all Gaussianity test rejected Gaussianity,  $ P \leqslant 10^{-4}  $.  in this range: $ ku = 50.743$ and $ sk = 5.181  $. All these statistical analyses, together with Eqs. (\ref{E:EntangGradVect}) through (\ref{E:kappa_ji}), indicate that it is possible to deduce $ \vect{\{y_m, K_m, n\}} $  for $ \mf{H}_{i,j} $ from the parameters of $ \mf{H}_i $ scaling its  parameters by $ \kappa_{j.i} $, if $  \vect{\nabla \mf{H}}_i $ is not too steep, with $ x_i   \geqslant 2  $, in the present case.

\textbf{Figure \ref{F:Vectors}C} suggests that the gradient of the Hill equation si largely dependent on $ \tfrac{\partial \mf{H}}{\partial y_m} $, if $ y_m $ is a constant known without error, as set in \textbf{Figure \ref{F:Vectors}D}, the gradient not only became smaller, but even had a broad maximum at $ x_i=2 $.

When $ \vect{ \{ \theta_{i={1,\ldots,k}} \}}   \equiv \vect{\{ y_m, K_m, n  \}} $ are stochastically and causally independent the condition in Eq. (\ref{E:GradVect}) holds, and using the Greek letters identifying angles in \textbf{Figure \ref{F:GradientAndVectors}} we may see that
\begin{equation}\label{E:EmpGradToCoord}
	\begin{matrix}
		\sin(\alpha)&=& \tfrac{\left( \tfrac{\partial \mf{H}}{ \partial y_m} \right)  }{ \vmag{\nabla \mf{H}(x_i)}} &=& \mf{s}_1 &= & \tfrac{\mf{H}_{\theta_1}'}{\vmag{\mf{\nabla H}(x_i)} } &\implies&  \mf{H}_{\theta_1}' =   \mf{s}_1 \cdot    \vmag{\nabla \mf{H}(x_i)} \\ 
		\sin(\beta)&=& \tfrac{\left( \tfrac{\partial \mf{H}}{ \partial K_m} \right)  }{ \vmag{\nabla \mf{H}(x_i)}} &=& \mf{s}_2 &= & \tfrac{\mf{H}_{\theta_2}'}{\vmag{\mf{\nabla H}(x_i)} } &\implies & \mf{H}_{\theta_2}' =  \mf{s}_2 \cdot    \vmag{\nabla \mf{H}(x_i)} \\ 
		\sin(\gamma)&=& \tfrac{\left( \tfrac{\partial \mf{H}}{ \partial n} \right)  }{ \vmag{\nabla \mf{H}(x_i)}} &=& \mf{s}_3 &= & \tfrac{\mf{H}_{\theta_3}'}{\vmag{\nabla \mf{H}(x_i)} } &\implies & \mf{H}_{\theta_3}'  =   \mf{s}_3 \cdot    \vmag{\nabla \mf{H}(x_i)} \\
		&{ }& { } & { }& \vdots & && &&   \\
		& & \tfrac{\left( \tfrac{\partial \mf{f}}{ \partial \theta_i} \right)  }{ \vmag{\nabla \mf{f}(x_i)}} & = & \mf{s}_i &=&  \frac{\mf{f}_{\theta_i}'}{\vmag{\nabla\mf{f}(x_i)} } &\implies & \boldsymbol{\mf{f}_{\theta_i}' =  \mf{s}_i \cdot    \vmag{\nabla \mf{f}(x_i)}} \\
		&{ }& { } & { }& \vdots & && &&   \\
		& & \tfrac{\left( \tfrac{\partial \mf{f}}{ \partial \theta_k} \right)  }{ \vmag{\nabla \mf{f}(x_i)}} & = & \mf{s}_k &=&  \tfrac{\mf{f}_{\theta_k}'}{\vmag{\nabla \mf{f}(x_i)} } &\implies & \mf{f}_{\theta_k}' =  \mf{s}_k \cdot  \vmag{\nabla \mf{f}(x_i)} ,
	\end{matrix}
\end{equation}
where $ \mf{f}_k $ represents some  angle dependent  function with $ k $ $ \theta_i $ parameters. In Eq. (\ref{E:EmpGradToCoord}), and onward, the nation stresses that gradient magnitude is a function of $ x_i $. In 2- and 3-dimensional  Cartesian spaces $ \mf{s} $ are sine functions.  Thus, from Eq. (\ref{E:EmpGradToCoord}) obviously
\begin{eqnarray}
		\dfrac{\partial \mf{f}}{\partial \theta_i} &=& \mf{s}_i \cdot    \vmag{\nabla \mf{f}(x_i)} \label{E:GradDeriv} \\ 
		\dfrac{\partial \mf{f}}{\partial \theta_i } &=&  \mf{s}_i \cdot    \vmag{\nabla \mf{f}(x_i)} \approxeq \dfrac{\Delta \mf{f}} {\Delta \theta_i} \implies \boldsymbol{\Delta \theta_i \approxeq \dfrac{\Delta \mf{f}}{ \mf{s}_i ^{\circ}\cdot    \vmag{\nabla \mf{f}(x_i)}}\label{E:Aproximacion}}\quad \cdot
\end{eqnarray}
The $ \mf{s}_k^{\circ} $  notation was introduced for cases (and spaces) of $ k $ dimensions, and the $^\circ $ indicate that it is approximated using the parameters obtained from the optimized objective function
\begin{equation}\label{E:SinCor}
\mf{s}_k^{\circ}= \tfrac{ \mf{f}_{\theta_i}'( x_i \given \vect{\{\theta\}})_{opt}}{\vmag{\nabla f(x _i \given \vect{\{\theta\}})_{opt}}}\,  \cdot
\end{equation}
Eq. (\ref{E:SinCor}) implies that Ec. (\ref{E:Aproximacion}) is really equivalent to:
\begin{equation}\label{E:CartesianAProximation}
\Delta \theta_i \approxeq \dfrac{\Delta \mf{f}}{ \mf{s}_i ^{\circ}\cdot    \vmag{\nabla \mf{f}(x_i)}} \approxeq \dfrac{\Delta \mf{f}}{\cancel{\vmag{\nabla \mf{f}(x_i)}}} \cdot \tfrac{\cancel{\vmag{\nabla f(x _i \given \vect{\{\theta\}})_{opt}}}}{\mf{f}_{\theta_i}'( x_i \given \vect{\{\theta\}})_{opt}}
\end{equation}
and thus
\begin{equation}\label{E:CartesianPrediction}
	\then \boldsymbol{\theta_i \left( x_i\right)  \approxeq\; \theta_{i_{opt}}\left(x_i \right) + \frac{\Delta \mf{f}(x_i)}{\mf{f}_{\theta_i}' (x_i)}} \cdot \qquad \lqqd
\end{equation}

This result demonstrates that calculating the first derivatives $ \mf{f}_{\theta_i}'\left( x \given \{ \vect{\theta} \}\right)  $ of the optimized objective function $ \mf{f}\left( x \given \{ \vect{\theta} \}\right)  $, is the only requirement to estimate the parameters describing the outliers at each $ \boldsymbol{x_i } $, with the only condition that $ \boldsymbol{\kappa_{i,j}} $ and $ \delta_{}j,i $ [Figure \ref{F:GradientAndVectors}B and Eqs. (\ref{E:ScaCartGrad}) and (\ref{E:kappa_ji})] are not too large. Here $ \mf{f} $ stands for the optimized function.

\subsection{A simpler finite differences approximation.}

The Cartesian analysis described so far, is formal and based on vector algebra, yet another approximation using finite differences \cite{Hogg1978, Ghorbal2002, LeVeque2007} which seems simpler:
\begin{equation}\label{E:Simpler}
	\dfrac{\partial \mf{f}(x_i)}{\partial \theta_i} \approxeq \dfrac{\Delta \mf{f}(x_i)}{\Delta \theta_i } \implies \Delta \theta_i  \approxeq \frac{\Delta \mf{f}(x_i)}{\mf{f}_{\theta_i}'(x_i) }\;,
\end{equation}
which this predicts
\begin{equation}\label{E:SimplisticPrediction}
	\then \boldsymbol{\theta_i \left( x_i\right)  \approxeq\; \theta_{i_{opt}}\left(x_i \right) + \frac{\Delta \mf{f}(x_i)}{\mf{f}_{\theta_i}' (x_i)}} \cdot \qquad \lqqd
\end{equation}

This rather long introduction, tries to do a graphical description, easy to grasp intuitively on the approach this paper follows to tackle the problem of establishing limits to the uncertainty of $ \{\theta_{j,i}\} $ of a function $ \mf{f} \left( x \given \{\theta\} \right) $ after reaching an optimum. Several well known functions will be studied using their derivatives, and Monte Carlo simulation \cite{Dahlquist1974, Metropolis1987}, as well as empirical experimental data from a study of marine bioactive compounds \cite{Quintana2017}.

\section{Methods.}\label{S:Methods}

\subsection{Monte Carlo variate simulation.}\label{S:MonteCarlo}

To test the goodness of fitting curves to data, random data with known statistical properties were generated using Monte Carlo simulation \cite{Dahlquist1974, Metropolis1987}. For this purpose sets of pairs $[x_i, f(x_i)]$ were generated as 
\begin{equation}\label{E:RanNor}
	f_{rnd}(x_j) = f(x_j) + \epsilon_j =  f(x_j) + \psi \left( 0,\sigma \lor \gamma \right) 
\end{equation}
where, as said, $ \sigma^2 $ is the variance and $ \gamma $ is the Cauchy pdf scale factor. Thus for population having defined mean and variance:
\begin{equation}
	\psi[E(x),\sigma]=f_{rnd}(x) =\psi(\mu,\sigma)
\end{equation}

When needed, Gaussian pseudo-variates were generated using the Box and Muller \cite{Box1958} algorithm as modified by Press et al. \cite{Press2007}. Fundamental to all Monte Carlo simulations \cite{Dahlquist1974} is a good uniform (pseudo) random (PRNG) number generator. Data for all numerical simulations carried out in this work were produced using  random numbers ($r$) with continuous rectangular (uniform) distribution in the closed interval [0,1] or $U[0,1]$. Except for illustrative purposes such as shown in Figures \ref{F:Vectors} or \ref{F:GradientAndVectors}, all $U[0,1]$  were generated using the 2002/2/10 initialization-improved  623-dimensionally equidistributed uniform pseudo random number generator MT19937 algorithm \cite{Matsumoto1998, Panneton2006, MersenneTwister2021}. The generator has passed the stringent DIEHARD statistical tests \cite{Marsaglia2003, Bellamy2013}. It uses 624 words of state per generator and is comparable in  speed to other generators.  It has a Mersenne prime period of $2^{19937} -1$ ($\approxeq10^{6000}$).  
The MT19937 requires an initial starting value called \textit{seed}. The seed used was a 64-bit unsigned integer obtained using the \textit{/dev/random} Ubuntu Linux PRNG , which saves environmental noise from device drivers and other sources into an entropy pool. Device \textit{/dev/random} gets temporarily blocked, and stops producing random bytes, when the entropy of the device gets low, and commences producing output again when it recovers to safe levels. No such delays were perceived during this work. Using \textit{/dev/random} seed makes exceedingly unlikely ($P = 2^{-64} \approxeq 5.4 \cdot 10^{-20}$) that the same sequence, $\{r_i\}$, of $U[0,1]$ is used twice. Calculations were programmed in C++14, using  amd64 GNU C++ compiler with C++14 standards, under \GNU Linux{} Mint 20.1 Ulyssa running on a \textbf{LENOVO}\textsubscript{\texttrademark}{} product: F0D0001USP: . Linux{} Kernel: 5.4.0-72-generic x86\_64 bits. GNU g++ 64 compiler: gcc v: 9.3.0 Desktop: Cinnamon 4.8.6. Topology: Quad Core model: Intel\textsuperscript{\textregistered}{} Core\textsuperscript{\texttrademark}{} i7-7700T $ \times $ 8 CPU 800/3800 MHz Core speeds, type: MT MCP arch: Kaby Lake  rev: 9 L2 cache: 8192 KiB nad a 7 TB disk. 

\subsection{Minimizing programming rounding errors.}\label{S:TruncationErr}

There is a caveat in regard with the evaluation of Eqs.  (\ref{E:CartGrad}) for $ \mf{f}_{\theta_i}' $ as expressed in the equation set (\ref{E:GradVect}), Eqs (\ref{E:dym}) through (\ref{E:dn}), and similar equations. When solving this equations it may require calculating quotients of very large or very small terms, leading to large truncation errors. All calculations involving in this communication were programmed in C++17 and all $ \mf{f}_{\theta_i}' $ were evaluated using  using  \textbf{long double} C++ variables. In a modern personal computer uses 12 byte (i.e. 96 bit) precision to minimize severe rounding errors in computer memory. C++ l\textbf{ong double}  variables allow the representation of numbers in the range $ 3.36210\cdot 10^{-4932} \text{ -- } 1.18973 \cdot 10^{4932} $ and $ \approx 25 $ decimal digits accuracy, with g++ Linux compilers in 64 bit computers \cite{GCCVariables2021a, GCCVariables2021}. Evaluating critical parts of Eqs. (\ref{E:CartGrad}), (\ref{E:EmpGradToCoord}) and \EqF in logarithmic form was employed too. \textit{Gross errors may occur if these precautions are not taken}, this is specially true for $ \mf{H}_{K_m}' $ and $ \mf{H}_{n}' $ in equation set (\ref{E:GradVect}).

\subsection{Statistical procedures.}\label{S:StatProc}

\subsubsection{Fitting functions to data.}\label{S:CurveFit}

Functions were adjusted to data using a simplex minimization \cite{Nelder1965}. The simplex procedure was designed to minimize differences between empirical data assumed to obey a function such as $ g(\{x_{i=1,2, \, \dots \,, m}\} \given \{ \vect{\theta}_{j=1, \, \dots \,, k}\}) $, where $ \{x_{i=1,2, \, \dots \,, m}\} $ is a set of observables, and a model function $ f(\{x_{i=1, \, \dots \,, m}\} \given \{ \vect{\theta}_{i=1, \, \dots \,, k}\}) $. In this work the simplex was designed to minimize
\begin{equation}\label{E:LNorm}
	\varepsilon_a = \sum_{i=1}^m \abs{y_i - f(x_i \given \{ \vect{\theta}_{j=1,2, \, \dots \, , k}\})}.
\end{equation} 
Differences $ y_i - f(x_i \given \{ \vect{\theta}_{j=1,2, \, \dots \, ,k}\}) $ are commonly called \textit{residuals}.  A common alternative to Eq. (\ref{E:LNorm}) is 
\begin{equation}\label{E:SqNorm}
	\varepsilon_{s} = \sum_{i=1}^m \left[y_i - f(x_i \given  \{ \vect{\theta}_{j=1,2, \, \dots \, ,k}\})\right] ^2  
\end{equation} 
used, for example, in the so called \textit{least squares} minimization \cite{Montgomery2012}. Minimizing least squares has a bias to give unduly high weights to outliers, which may be merely an extreme manifestation of the random variability inherent in the data, but could also stem from gross deviation from a prescribed experimental procedure or to error in calculating or recording the numerical value \cite{Grubbs1969}. Giving unduly weight to outliers is avoided by using the absolute values of the deviations as done in Eq. (\ref{E:LNorm}). 

In this work the optimization continued looping while the following condition was true:
\begin{condition}\label{Cnd:StopOpt}
	Keep looping while
	\begin{equation}\label{E:EqStopOpt}
		\varepsilon_{stop} =  \left|   \frac{\varepsilon_{a_{l+1}}-\varepsilon_{a_{l}}}{\varepsilon_{a_{l}}}\right|  \substack{ \geqq \\ \circlearrowright } 10^{-8},
	\end{equation}
	with $ \varepsilon_{a_l} $ calculated as indicated in Eq. (\ref{E:LNorm}).
\end{condition}
\noindent{} or else until
\begin{condition}\label{Cnd:TooLong} 
	Kept looping while $l \leqq 1024000$ in Eq. (\ref{E:EqStopOpt})
\end{condition}
\noindent{}was used to stop optimization, in order to prevent the algorithm from running forever.
The simplex was implemented also to provide a set $ \left\lbrace  \delta_{j\cdot}\right\rbrace$, used to calculate the uncertainties of $  \{\vect{\theta}\}_{j=1,\cdots,k}  $ estimated as described in Section \ref{S:Results}.

The simplex algorithm requires not only a set of initiation parameters, $ \{\vect{\theta}_j\} $, but an initial increment value, $ \Delta_{init}, $ to start modifying the initial parameters.  $\Delta_{init}  $ is the initial fraction to increment the parameters which is subsequently modified by the algorithm as tle optimization continues \cite{Nelder1965}.

\subsubsection{On statistical procedures utilized.}\label{S:StatProc}

Gaussianity of data was tested with the Jarque-Bera test, which also provides data on skewedness and kurtosis of data \cite{Jarque1980, Bera1981, Bera1981a, Giles2014}, with the socalled robustified Jarque-Bera test by Gell and Gastwirth  \cite{Gel2008} and with the Shapiro and Wilk \cite{Shapiro1965} test. Unless otherwise is indicated, data are presented as medians and their 95\% confidence intervals (95\% CI) calculated using nonparametric Moses \cite{Hollander1973} statistics. Other data are presented as medians and their 95\% confidence interval calculated with the procedure of Hodges and Lehmann \cite{Hollander1973}. Statistical significance of differences was decided with Mann–Whitney (Wilcoxon) test. Multiple comparisons were done with the nonparametric Kruskall-Wallis analysis of variance. See Hollander and Wolde \cite{Hollander1973} for all not specified details of nonparametric methods used. Statistical differences between samples were considered significant when the probability that they stem from chance was $ \leq 5\% $ ($ P  \leq  0.05 $).

\section{Results and discussion.}\label{S:Results}
\begin{figure}[h!]
	\begin{small}
		\centering
		\includegraphics[width=12cm]{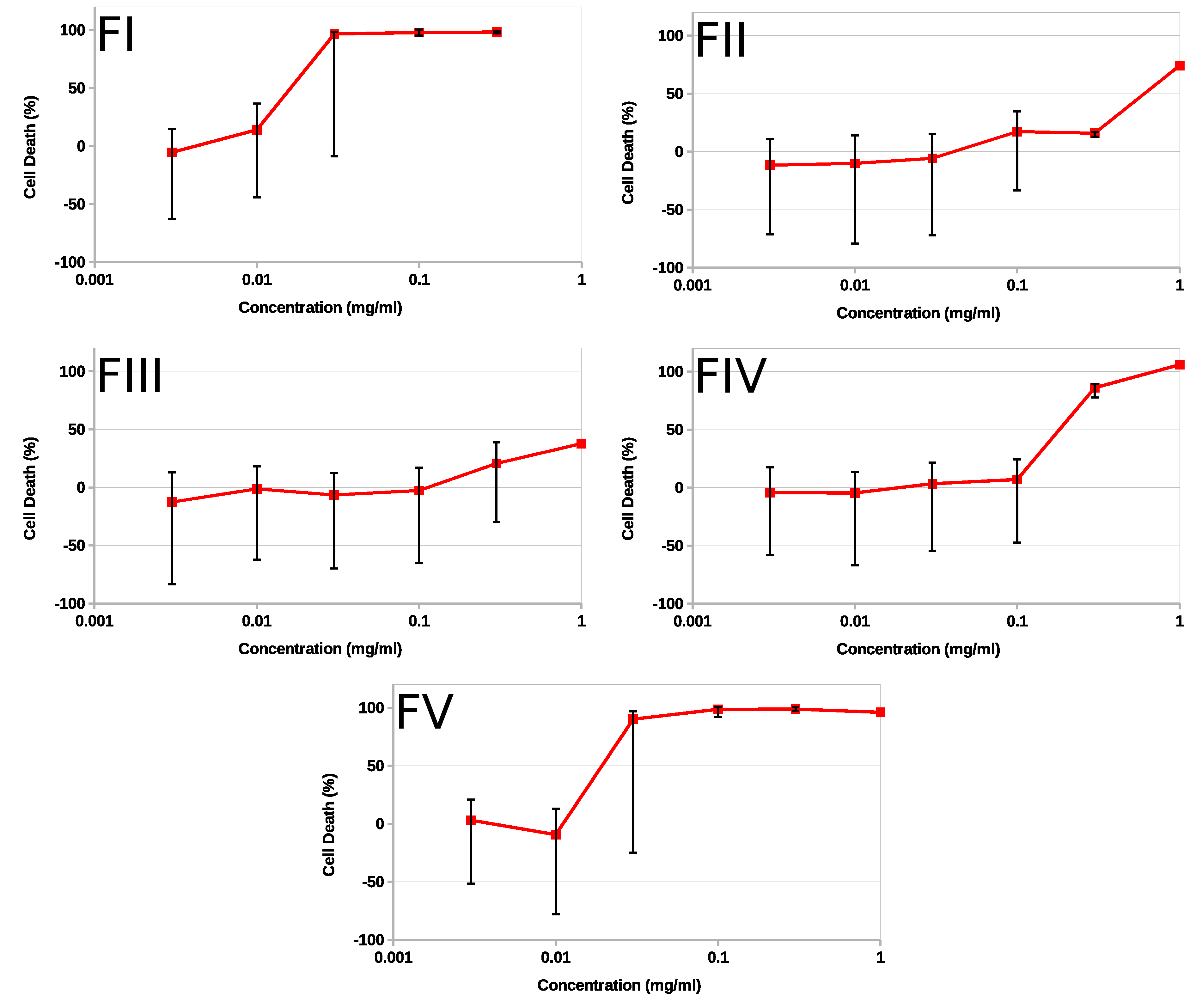}
		\caption{\textbf{Cell death induced by fractions isolated from \textit{P. constellatum} ({S}avigny, 1816) in 4T1 breast cancer cell cultures  \cite{Quintana2017}.} Percent of death calculated with Eq. (\ref{E:PercEf}). Ordinate is the percentage of dead cells, abscissa is concentration ([D] in mg/mL) of fraction tested, plotted in decimal logarithmic scale. Data presented as medians{  }($\blacksquare$) and their 95\% confidence interval (bracket lines) calculated as indicated by Hodges and Lehmann \cite{Hodges1963}. Straight lines were used to connect medians to help interpretation. The number of data processed for each fraction concentration was $ nf \cdot nd \cdot nb^2=24000 $ (\textit{nb} = 10, \textit{nd} = 48, \textit{nf} = 5). For details on cytological and biochemical procedures see Appendix \ref{S:Biochem}. Other details in the text of the communication}
		\label{F:FracNoFit}
	\end{small}
\end{figure}

\begin{figure}[h!]
	\begin{small}
		\centering
		\includegraphics[width=12cm]{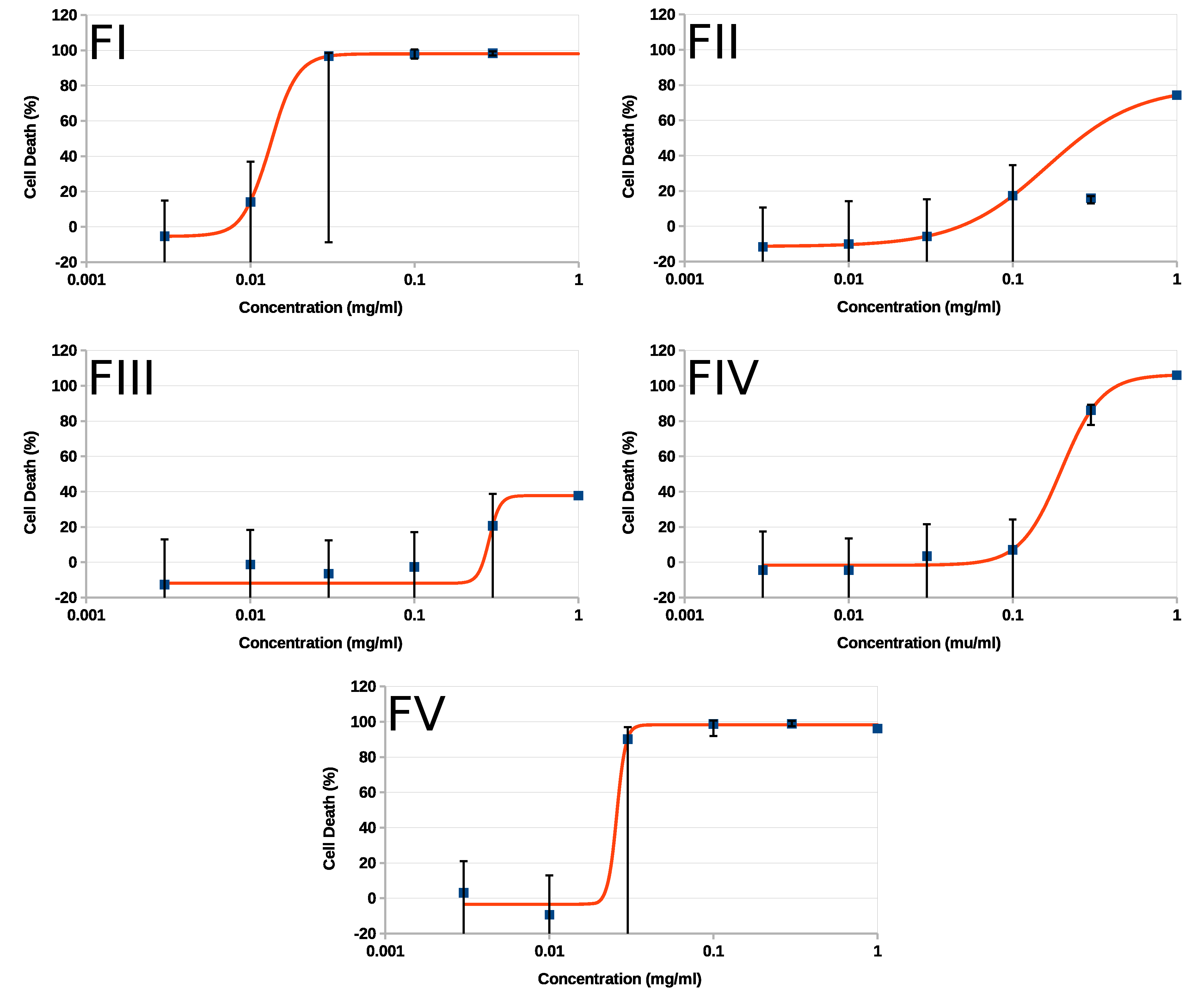}
		\caption{\textbf{\textbf{Cell death induced by fractions isolated from \textit{P. constellatum} ({S}avigny, 1816) in 4T1 breast cancer cell cultures  \cite{Quintana2017}.}} Ordinate was clipped at -20\% to increase visibility of the biologically meaningful range of effects. All other details are equal to Figure \ref{F:FracNoFit} except for the sigmoid curves drawn which were calculated fitting Eq. (\ref{E:HillMod}) to the data with a simplex optimization procedure minimizing deviations between curves and $ \{y_i\} $ data as indicated by Eq. (\ref{E:LNorm}). Values used to draw the curves are in Table \ref{T:HillSimplexPar}. The simplex algorithm was initialized with the same set of values for the five fractions: $ y_0 = -10\%, y_m = 100\%, K_m = 0.1\; \text{mg/mL},\; n=2  \text{ and } \Delta_{init}=0.1.  $}
		\label{F:FracFit}
	\end{small}
\end{figure}

\subsection{A challenging data set obtained with a procedure commonly used in cell biology.}\label{S:CitotDat}

\begin{table}
	\begin{center}
		\begin{small}
			\caption{\textbf{Parameters characterizing the regression curves in Figure \ref{F:FracFit}}. }\label{T:HillSimplexPar}
			\begin{supertabular}{m{1.5cm}*{4}{m{3.0cm}}}
				\hhline{*{5}=}
				\textbf{Fraction} & 
				\centering \textit{\textbf{y\textsubscript{0}}} &
				\centering \textit{\textbf{y\textsubscript{m}}} & 
				\centering \textit{\textbf{K\textsubscript{m}}} & 
				\centering \textit{\textbf{n}} \cr
				& \centering \textbf{(\%)} & \centering \textbf{(\%)} &\centering \textbf{(mg/mL)}&  
				\cr
				\hhline{*{5}-}\\
				\textbf{FI} & 
				\centering $ -5.4 $ \par $ (-5.7,\; -5.2) $ &
				\centering $ 103.4 $\par $ (103.1,\; 103.7) $  & 
				\centering $1.32 \cdot 10^{-2}$ \par $(1.05,\; 1.56) \cdot 10^{-2}$ &
				\centering $ 5.29 $\par $ (5,28,\; 5.29) $ 
				\cr\cr
				\textbf{FII}  &
				\centering $ -11.5 $ \par ($ -12.3,\; -10.7 $) &
				\centering $ 90.5 $ \par ($ 88.8,\; 92.0 $) &
				\centering $ 0.16  $\par $ (0.15,\; 0.17) $&
				\centering $ 1.59 $  \par $ (1.57,\; 1.60) $
				\cr\cr
				\textbf{FIII}  &
				\centering $ -11.9 $ \par $ (-13.4,\; -10.6) $&
				\centering $ 49.7 $ \par $ (46.6,\; 52.1) $ &
				\centering $ 0.29 $ \par $ (0.27,\; 0.30) $ &
				\centering $ 13.2 $ \par $ (13.2,\; 13.3) $
				\cr\cr
				\textbf{FIV}  &
				\centering $ -1.7 $ \par $ (-2.1,\; -1.3) $ &
				\centering $ 108.0 $ \par $ (107.5,\; 108.5) $ &
				\centering $ 0.20 $ \par $ (0.19,\; 0.20) $ &
				\centering $ 3.53 $ \par $ (3.52,\; 3.53) $ 
				\cr\cr
				\textbf{FV} &
				\centering $ -3.4 $ \par ($ -3.8,\; -3.1 $)&
				\centering $ 101.7 $ \par ($ 101.2,\; 102.1 $) &
				\centering $2.57 \cdot 10^{-2}$ \par $(2.12,\; 2.92) \cdot 10^{-2}$&
				\centering $ 16.2 $ \par ($ 16.2,\; 16.2 $) 
				\cr \cr
				\hhline{*{5}=} 
			\end{supertabular} 
		\end{small}
	\end{center}
	\begin{footnotesize}
		Parameters of the modified Hill Equation (\ref{E:HillMod}): $ y_0 $, offset parameter; $ y_m $, maximum effect; $ K_m $, concentration producing half maximum effect and $ n $, is called \textit{Hill coefficient} or \textit{molecularity} in some pharmacology and enzymology work \cite{Segel1975}. The simplex algorithm was initialized with the same set of values for the five fractions: $ y_0 = -10\%, y_m = 100\%, K_m = 0.1\; \text{mg/mL and } n=2 $; $ \Delta_{init}$ was set as 0.1. Please notice that the units of $ K_m $ are irrelevant as long as they are equal to the units of [D]. All data presented as medians and their 95\% CI between parentheses. Confidence intervals calculated with the Hodges and Lehman \cite{Hodges1963} procedure based on data determined as indicated in relation with \EqF. Sizes $  (m) $ of $\left\lbrace d_{j,i}\right\rbrace _{i=1,\, \dots \,,m}$ were : FI, $ m=81121 $: FII $ m=96018 $: FIII, $ m=114720 $; FIV,  $ m=115200 $ and FV, $ m=114721 $.  Differences in $ m $ were due to sample sizes and data peculiarities due to which Eq. (\ref{E:LivFract}) produced undefined values called \textit{NaN} (\textit{N}ot \textit{a N}umber, such as $ \tfrac{0}{0} $ or $ \sqrt{-x} \given x \in \mathbb{R} $) in C++ or in values such as $ \tfrac{x}{0} $ or $ \log(0) $ called \textit{inf} in C++. \textit{NaN} and \textit{inf} results were eliminated from the calculations. Similar precautions were taken when evaluating $ \mf{f}_{\theta_j}' (x_i)$ in equations such as Eqs. (\ref{E:GradVect}) as required by \EqF. Other details in the text of the communication.
	\end{footnotesize}
\end{table}

The data shown in Figure \ref{F:FracNoFit} (taken from Quintana \cite{Quintana2017}) are effects of several fractions (FI -- FV) isolated from \textit{P. constellatum} which were able to kill cells in culture of 4T1 breast cancer cells. Apparent effects (\% cell death) calculated with Eq. (\ref{E:PercEf}) are presented in the ordinate, as function of the concentration of fraction indicated in the abscissa (in $ \text{mg/mL} $). There are several oddities in the data, the effects at some concentrations are very disperse (As indicated by the brackets representing 95\% CI), and, notably at low concentrations, they indicate negative percentages of death. At the lowest concentrations even median values are slightly, but significantly, bellow zero; this could be expected if the background correction (Eqs. (\ref{E:LCorr}) and (\ref{E:FCorr})) actually over corrects absorbance data. The large variability could be due to subtractive cancellation \cite{Kahan1965, Kahan2004, Boldo2011}, combined with the quotient represented by questions such as \EqF and (\ref{E:LivFract}) which is prone to produce a large variance increase in $ \{ p_h\} $ [Eq. (\ref{E:LivFract})] and $ \{ y_i \} $ (Eq. (\ref{E:PercEf})). Also, as indicated in Section\textquoteright{}s \ref{S:CaveatsEqF} item 3, quotients between random variables will most likely have and unknown, probably pathological pdf, with undefined statistical moments and wide outliers such as the Cauchy pdf  \cite{Cramer1991, Pitman1993, Wolfram2003} for which concepts like mean, variance, skewedness and kurtosis are undefined and meaningless (See also section \ref{S:ModCauhData}). These problems are generally overlooked in cell biology literature.

There is no much reason to expect data in Figure \ref{F:FracNoFit} to be Gaussian, there are obvious asymmetries of 95\% Confidence Intervals (CI) about the medians in all plots.  Number of data for any toxin concentration in the figure are probably too small to perform credible Gaussianity test for any of these concentrations. Yet, with the common (and perhaps gratuitous) assumption that for each toxin plot, all errors are in the ordinate,  all data for each toxin concentration were pooled, an the polled $ \{y_l\} $ samples were tested for Gaussianity with the Jarque and Bera test \cite{Jarque1980, Bera1981, Giles2014} when sample sizes were huge and this test is very efficient \cite{Giles2014},  and also with the robustified Jarque-Bera test of Gell and Gastwirth \cite{Gel2008}, and the he Shapiro and Wilk \cite{Shapiro1965} Gaussianity tests. In all cases Gaussianity was rejected with a $ P \lll 10^{-9} $ confidence level.

\subsubsection{An example using a modified Hill equation.}\label{S:UsingModHill}

The three papers describing the colorimetric method \cite{Mosmann1983, Denizot1986, Swinehart1962} have been cited at least 64,568 times in the literature [June 18, 2021, source: \GogSch \textsuperscript{\texttrademark}{ } (\url{https://scholar.google.com})]. This indicates that, in spite of their management of uncertainty, the method has been found useful by a substantial number of researchers. Notwithstanding the oddities of data in Figure \ref{F:FracNoFit}, there are several features which are evidenced by the medians. All five fractions increased cell mortality as concentration raised, and in all cases there is a sigmoid aspect of the dose--effect log plots. 

When the Hill equation \cite{Hill1910b, Hill1913} is plotted as effect versus the logarithm o the concentration it is a sigmoid curve. In classical form, the Hill equation is used in enzyme kinetics and in pharmacology to represent the interaction of one or more molecules of substrate with the catalytic site of an enzyme, or of a drug molecule with its receptor site is related to the the law of mass action \cite{Ariens1964, Segel1975, Erdi1989}:
\begin{equation}\label{E:MassActionLaw}
	n[\mathrm{D}]+[\mathrm{R}] \rightleftarrows [\mathrm{D}_n\mathrm{R}] \implies K_m=\frac{[\mathrm{D}_n\mathrm{R}]}{[\mathrm{D}]^n \cdot [\mathrm{R}]},
\end{equation}
where brackets indicate concentrations. When the effect produced by drug $ \mathrm{D} $ binding receptor $ \mathrm{R} $ is 
\begin{equation}\label{E:MassToHill}
	y \propto  [\mathrm{D}_n\mathrm{R}] \implies y=\frac{y_m}{1+\left(  K_m/[\mathrm{D}]\right) ^n}\qquad n \in \mbb{Z}.
\end{equation} 
Under these conditions $n$ is called the \textit{molecularity }of the reaction. Also, $ n $ is used in situations where properties of the enzyme or drug receptor are modified during the interaction, the, so called, cooperative schemes, where $ n \in \mbb{R} $ is plainly named \textit{Hill coefficient} without further molecularity implications  \cite{Monod1965, Segel1975, Abeliovich2005}.

The Hill equation \cite{Hill1913} in its original form is
\begin{equation}\label{E:Hill}
	\mf{H}\left( [D] \given  \{y_m,K_m,n\} \right) =\frac{y_m}{1+\left( \tfrac{K_m}{[\mathrm{D}]}\right)^n}
\end{equation}
which does not include a term for \textquotedblleft{}offset,\textquotedblright{ }occurring when $ y([D]=0 \given \{y_m,K_m,n\}) \neq 0 $. 
Since the data in the figures seems to include an overcorrection for the basal absorbance, this modified Hill equation will be used. as a particular case, in our analysis:
\begin{equation}\label{E:HillMod}
	\mf{H}_o\left( [D] \given  \{\vect{y_0, y_m,K_m,n}\} \right) =y_0+\frac{y_m}{1+\left( \tfrac{K_m}{[\mathrm{D}]}\right)^n}
\end{equation}
its first derivatives on $\{ \vect{ \theta}_j\} $ are given in Section \ref{S:HillFstDer} as Eqs. (\ref{E:dy0}) -- (\ref{E:dn}).

Figure (\ref{F:FracFit}) shows the results of adjusting Eq. (\ref{E:HillMod}) to the data of Quintana \cite{Quintana2017}. In all cases the simplex optimization started from the same set of values:  $ y_0 = -10\%, y_m = 100\%, K_m = 0.1\; \text{mg/mL, } n=2 $ and $ \Delta_{init}=0.1 $. Since Eqs. (\ref{E:LCorr} -- \ref{E:LivFract}) produce 24000 points per concentration, the number of pairs in each fraction\textquoteright{}s regression analysis ranged 120000 -- 144000 in the plots shown in Fig. \ref{F:FracFit}. Interestingly, the curves in Fig. \ref{F:FracFit} follow, rather closely, the median percent of dead cells at each concentration in all the plots. This is particularly clear for FI and FV. The parameter values describing the curves are in Table \ref{T:HillSimplexPar}. The curves in Figure \ref{F:FracFit}, and the sets of data in Table \ref{T:HillSimplexPar} \textquotedblleft{}look good\textquotedblright{} but uncertainty estimator for the parameters are necessary to properly state which fraction effect differs from which fractions, specially if the outliers suggested by the 95\% CI and \textquotedblleft{}skewness\textquotedblright{} analysis are considered.

Medians and their 95\% CIs of Quintana-Hern\'{a}ndez\textquoteright{s} compounds  \cite{Quintana2017} $ \left\lbrace d_{j,i}\right\rbrace _{i=1,2, \, \dots \, ,m} $ sets are presented in Table \ref{T:HillSimplexPar}. All  $ \left\lbrace d_{j,i}\right\rbrace _{i=1,2, \, \dots \, ,m} $ sets used to guess residuals in Table \ref{T:HillSimplexPar} were tested for Gaussianity with the Shapiro-Wilk and Jarque-Bera methods, and both procedures predicted a probability $ P <10^{-6} $ that any of the sets is Gaussian. 
\begin{table}
	\begin{center}
		\begin{small}
			\caption{\textbf{Parameters characterizing the $ \boldsymbol{\{d_{j,i}\}_{i=1,2, \ldots, m}} $ sets used to calculate parameter uncertainty in Table \ref{T:HillSimplexPar}. $ \Delta_{init}$ was set as $ 0.1 $.}}\label{T:SkwKurt}
			\begin{supertabular}{
					*{3}{m{2cm}}
					*{2}{m{2.5cm}}
					*{1}{m{5cm}}
				}
				\hhline{*{6}=}
				\textbf{Fraction} &
				\textbf{Loops} &
				\textbf{Parameter} &
				\centering \textbf{\textit{Sk}} &
				\centering \textbf{\textit{Ku}} &
				\centering \textbf{Range} 
				\cr
				\hhline{*{6}-}
				\centering FI & 
				$ 2867 $ 
				& & &&\cr
				&&
				\centering  $ y_0 $&
				\centering  $ -3.95 $ &
				\centering  $ 11.5 $& \centering  $ -1.49,\; 0.21 $ 
				\cr
				&&\centering  $ y_m $&
				\centering  $ -6.30 $ &
				\centering  $ 75.5 $& 
				\centering  $-1.9 \cdot 10^{3}$, $6 \cdot 10^{2}$
				\cr
				&&\centering  $ K_m $&
				\centering  $ -2.95 $ &
				\centering  $ 11.5 $&
				\centering  $ -1.43,\; 0.28 $
				\cr
				& &\centering  $ n $&
				\centering  $ -2.95 $ &
				\centering  $ 11.5 $&
				\centering  $ 3.9,\; 5.6 $
				\cr
				\centering  FII &\centering  $ 1328 $&&&&\cr
				&&\centering $ y_0 $&
				\centering  $ -2.90 $ &
				\centering  $ 49.1 $& 
				\centering  $ -5.7,\; 0.3 $ 
				\cr
				&&\centering  $ y_m $&
				\centering  $ -5.66 $ &
				\centering  $ 66.8 $& 
				\centering  $ -492,\; 145 $
				\cr
				&&\centering  $ K_m $&
				\centering  $ -2.90 $ &
				\centering  $ 49.1 $&
				\centering  $ -5.38,\; 0.62 $
				\cr
				&&\centering  \textit{n}&
				\centering  $ -2.57 $&
				\centering  $ 40.3 $&
				\centering  $ -3.9,\; 2.0 $
				\cr
				\centering  FIII  &$ 13277 $&&&&\cr
				&&\centering  $ y_0 $&
				\centering  $ -0.87 $ &
				\centering  $ 27.42 $ & 
				\centering  $ -695,\; 697 $
				\cr
				&&\centering  $ y_m $&
				\centering  $ -4.35 $&
				\centering  $ 67.5 $&
				\centering  $ (-10^{26},\; 5 \cdot 10^{25}) $
				\cr 
				&&\centering  $ K_m $&
				\centering  $ -0.87 $&
				\centering  $ 27.4 $&
				\centering  $ -695,\; 697 $
				\cr
				&&\centering $  n $&
				\centering  $ -0.70 $&
				\centering  $ 23.1 $&
				\centering  $ -681.6,\; 710.4 $
				\cr
				\centering FIV&$ 1024007 $*&&&&\cr
				&&\centering  $ y_0 $&
				\centering  $ -2.48 $ &
				\centering  $ 15.04 $ & 
				\cr
				&&\centering $ y_m $&
				\centering  $ -6.56 $&
				\centering  $ 73.6 $&
				\centering  $ ( -2 \cdot 10^{6},\; 5 \cdot 10^{5} ) $
				\cr 
				&&\centering  $ K_m $&
				\centering  $ -2.48 $&
				\centering  $ 15.0 $&
				\centering  $ -0.68,\; 0.47 $
				\cr
				&&\centering  $ n $&
				\centering  $ -2.20 $ &
				\centering  $ 12.4 $ & 
				\centering $  2.6,\; 3.8 $
				\cr
				\centering FV &$ 21432 $&&&&\cr
				&&\centering  $ y_0 $&
				\centering  $ -3.85 $ &
				\centering  $ 18.1 $ & 
				\centering  $ -1.6,\; 0.24 $
				\cr
				&&\centering  $ y_m $&
				\centering  $ -2.79 $&
				\centering  $ 60.0 $&
				\centering  ($ -7\cdot 10^{14},\; 3 \cdot 10^{14} $)
				\cr 
				&&\centering $  K_m $&
				\centering  $ -3.85 $&
				\centering  $ 18.1 $&
				\centering  $ -1.6,\; 0.3 $
				\cr
				&&\centering  $ n $&
				\centering  $ -3.46 $ &
				\centering  $ 14.9 $&
				\centering  $ 14.6,\;16.4 $
				\cr
				\hhline{*{6}=} 
			\end{supertabular} 
		\end{small}
	\end{center}
	\begin{footnotesize}
		Parameters of the modified Hill Equation (\ref{E:HillMod}): $ y_0 $, offset parameter; $ y_m $, maximum effect; $ K_m $, concentration producing half maximum effect, $ n $, is called Hill constant; \textit{Ku}, kurtosis and \textit{Sk}, skewedness. \textit{Range}, \textit{newness} and \textit{kurtosis} have the usual statistical meanings. \textit{Loop}, indicates the number of times a parameter was changed during the simplex optimization \cite{Nelder1965}. For fractions I, II, III and V optimization stopped when Condition \ref{Cnd:StopOpt} was fulfilled. In case of FIV the optimization was topped fulfilling Condition \ref{Cnd:TooLong} after 3 h attempting unsuccessfully to fulfill Condition \ref{Cnd:StopOpt}. See the text for further discussion.
	\end{footnotesize}
\end{table}

\subsubsection{An insight on the complexity of the data used for this example.}
Table \ref{T:SkwKurt} presents data on the number of iterations required by the simplex algorithm to converge to the optimum reaching  Condition \ref{Cnd:StopOpt}. The only exception is data for FIV, which after 1024007 (number labeled with an asterisk in the table) loops, was still unable to reach Condition \ref{Cnd:StopOpt} and after $ \approxeq3 $ h of iterations (in the author\textquoteright{s} computer) the process was stopped after reaching Condition \ref{Cnd:TooLong}. The table also presents some statistical properties of the $ \{ d_{j,i} \}$ sets used to calculate the uncertainty of the parameters characterizing curves fitted to data in Figure \ref{F:FracFit}.

The data in the table indicates that in all cases presented $ \{ d_{j,i} \}$ sets are highly leptokurtic and very skewed, and in some cases (as indicated by the ranges presented at the leftmost column of the table), very wide ranges indicated that extreme values were observed. These extreme values are in all likelihood due to subtractive cancellation in Eqs. (\ref{E:LCorr}) and (\ref{E:FCorr}) combined with division by very small numbers in Eq. (\ref{E:LivFract}) and by the nature of the distribution of ratios \textit{per se} (see Section \ref{S:ModCauhData}). Interestingly,  most data points seem closely packed around the median value, since the 95\% CI of the medians are narrow.
\begin{table}  
	\begin{center}
		\begin{small}
			\caption{\textbf{Fitting Cauchy data generated with Eq. 
					(\ref{E:SimCauchyHill}) setting $ \mathbf{\gamma = 1/50} $ and simplex optimization to the modified Hill equation. In all cases the optimization started with $ \mathbf{y_0 =  -10\%} $, $\mathbf{y_m = 100\%}$, $\mathbf{K_m=0.1}$ and $\mathbf{n=2}$; $ \Delta_{init}$ was set as 0.1.}}\label{T:HillMonCau1}
			\begin{supertabular}{
					*{1}{m{0.8cm}}
					*{1}{m{0.8cm}}
					*{1}{m{1.5cm}}
					*{1}{m{3.cm}}
					*{1}{m{3.5cm}}
					*{1}{m{1.5cm}}
					*{2}{m{1.cm}}}
				\hhline{*{8}=}
				\centering$r$&
				\centering $ \theta_j $&
				\centering \textbf{Sim}&
				\centering \textbf{Pred}&
				\centering \textbf{Range} &
				\centering \textbf{Loops}&
				\centering \textbf{\textit{Sk}}&
				\centering \textbf{\textit{Kr}} \cr
				\hhline{*{8}-}
				\centering$ 3 $&
				\centering$ y_0 $&
				\centering  $ -5 $ &
				\centering  $ -5.86 $\par $ (-12.41,\; -4.10) $&
				\centering  $ (-26.71,\; 51.18) $&
				\centering $ 3172 $&
				\centering $ -2.44 $&
				\centering $ 11.0 $ 
				\cr
				&\centering$ y_m $&
				\centering  $ 100 $ &
				\centering  $ 102.2 $ \par $(59.3,\; 312.7) $& 
				\centering  ($ -1.7 \cdot 10^{4},\; 8.0 \cdot 10^{5}$)&
				&
				\centering $  3.25 $&
				\centering  $ 13.6 $
				\cr
				&\centering$ K_m $&
				\centering  $ 0.15 $ &
				\centering  $ 0.107  (0.042,\; 0.125) $&
				\centering  $ (-0.101,\; 0.677) $&
				&
				\centering $ 2.443 $&
				\centering $ 77.5 $
				\cr
				&\centering$ n $&
				\centering  $ 2 $ &
				\centering  $ 2.04 $ \par $ (1.98,\; 2.06) $&
				\centering  $ (1.84,\; 2.61) $&
				&
				\centering $ 2.44 $&
				\centering $ 11.04 $
				\cr
				\centering$ 10 $&
				\centering$ y_0 $&
				\centering  $ -5 $ &
				\centering  $ -4.76 $ \par  $(-5.57,\; -4.04) $& 
				\centering  $ (-25.61,\; 108.07) $&
				\centering  $ 6100 $&
				\centering  $ 4.91 $&
				\centering  $ 31.8 $
				\cr
				&\centering$ y_m $&
				\centering  $ 100 $ &
				\centering  $ 100.9$   \par $(94.5,\; 124.9) $& 
				\centering  ($-3.6 \cdot 10^{4},\; 2.2 \cdot 10^{5}$)&
				&
				\centering  $ 4.81 $&
				\centering  $ 26.7 $
				\cr
				&\centering  $ K_m $&
				\centering  $ 0.15 $ &
				\centering  $ 0.106$\par    $(0.098,\; 0.114) $&
				\centering  $ (-0.102,\; 1.235) $&
				&
				\centering  $ 4.919 $&
				\centering  $ 31.8 $
				\cr
				&\centering$ n $&
				\centering  $ 2 $ &
				\centering $ 2.135$ \par $ (2.127,\; 2.143) $&
				\centering  $ (1.927,\; 3.264) $&
				&
				\centering  $ 4.919 $&
				\centering  $ 31.8 $
				\cr
				\centering$ 100 $& 
				\centering$ y_0 $& 
				\centering  $ -5 $ &
				\centering $ -4.99$\par    $(-5.27, -4.71) $ 
				&\centering  $ (-807,\; 549) $
				& \centering  $ 633 $
				& \centering  $ -4.376 $
				& \centering  $ 170.9 $
				\cr
				& \centering$ y_m $&
				\centering  $ 100 $&
				\centering  $ 99.96 $ \par $(98.05,\; 102.10) $ &
				\centering  ($-1.2 \cdot 10^{5}, \;1.1 \cdot 10^{5}$)&
				&
				\centering  $ 1.774 $&	
				\centering  $ 125.7 $
				\cr
				&\centering  $ K_m $&
				\centering  $ 0.15 $&
				\centering  $ 0.100$ \par  $(0.097,\; 0.102) $&
				\centering  $ (-7.925,\; 5.635) $&
				&
				\centering  $ -4.376 $&
				\centering  $ 170 $
				\cr
				&\centering $  n $&
				\centering  $ 2 $ &
				\centering  $ 2.017$ \par   $(2.014,\; 2.019) $&
				\centering  $ (-6.008,\; 7.552) $&
				&
				\centering  $ -4.376 $&
				\centering  $ 170.9 $
				\cr
				$ 2000 $& \centering  $ y_0 $&
				\centering  $ -5 $ &
				\centering  $ -5.07$ \par   $(-5.13,\; -5.01) $&
				\centering  ($-1.2 \cdot 10^{4},\;  3.0 \cdot 10^{4}$)&
				\centering  $ 1051 $&
				\centering $ -71.8 $&
				\centering  $ 7112 $
				\cr
				&\centering  $ y_m $&
				\centering  $ 100 $&
				\centering  $ 100.1$ \par   $(99.70,\; 100.44) $&
				\centering  ($-1.2 \cdot 10^{8}, \; 9.9 \cdot 10^{6}$)&
				&
				\centering  $ -111 $&
				\centering  $ 12841 $
				\cr 
				&\centering  $K_m $&
				\centering  $ 0.15 $&
				\centering $  1.000$   \par $(0.099,\; 0.100) $&
				\centering  $ (-118,\; 30.8) $&
				&
				\centering  $ -71.8 $&
				\centering  $ 7112 $
				\cr
				&\centering  $ n $&
				\centering  2 &
				\centering  $ 1.999$ \par $(1.998,\; 2.000) $&
				\centering  $ (-115,\; 33) $&
				&
				\centering  $ -71.8 $&
				\centering  $ 7111 $
				\cr
				\hhline{*{8}=} 
			\end{supertabular} 
		\end{small}
	\end{center}
	\begin{footnotesize}
		Sim: indicates parameters used for the Monte Carlos simulations; Pred; indicates parameter values obtained from th simplex optimizations on the Monte Carlo simulated curves. The concentrations required by Eq. (\ref{E:HillMod}) were defined as: $ 0.001 $, $ 0.003 $, $ 0.01 $, $ 0.03 $, $ 0.1 $, $ 0.3 $ and $ 1 $; 	$ m $ is the number of $ d_{i,j} $ values used to calculate medians, 95\% confidence interval and ranges. Parameters and heading have same meaning as used in Table \ref{T:SkwKurt}; $ r $ indicates the number of random Cauchy values of type  $ f\left( U(0,1)_{i}\; \given \;\gamma,0 \right)\;+\; y \left( [D]_i\; \given \;\{y_0,y_m,K_m,n \} \right ) $ [Eq. (\ref{E:SimCauchyHill})] which were simulated for each concentration. Please notice that the units of $ K_m $ are irrelevant as long as they are equal to the units of [D]. See the text for further discussion.
	\end{footnotesize}
\end{table}

Table \ref{T:HillMonGaus1} presents Monte Carlo simulation of Gaussian random data distributed about Eq. (\ref{E:HillMod}). Data was calculated setting the concentration term at the following values (in arbitrary units): $ 0.001 $, $ 0.003 $, $ 0.01 $, $ 0.03 $, $ 0.1 $, $ 0.3 $ and $ 1 $. The number of replicates were simulated for each concentration were: 3, 10, 100 and 2000. 

\subsection{Monte Carlo simulation of data described by  Hill\textquoteright{}s equation modified as in Eq.  (\ref{E:HillMod}).}\label{S:MonCarSimHill}

\subsubsection{Fitting Cauchy-distributed data to the modified Hill equation.}\label{S:ModCauhData}

The analysis based on the data of Quintana \cite{Quintana2017} suggests that using the first derivatives of the objective function may produce confidence limits for stochastically independent parameters obtained from non-linear regressions. Yet, to simulate this kind of data with Monte Carlo methods faces an important problem. The quotient of two Gaussian variates having $ \mu=0 $ and variance $ \sigma^2=1$, the standard normal probability density function$ N(0,1) $, is distributed following the Cauchy distribution (also called Lorentz, Cauchy-Lorentz or Breit–Wigner distribution) \cite{Cramer1991, Pitman1993, Wolfram2003}  which has a pdf 
\begin{equation}\label{E:DensCauchy}
	\mf{c}( x\given \{\gamma,\widehat{\mu}\} ) = \frac{1}{\pi} \cdot \frac{\gamma}{\gamma ^2 + (x-\widehat{\mu}) ^2} 
\end{equation} 
where $(x ,  \widehat{\mu} , \gamma) \in \R \text{ and } \gamma > 0$. The probability distribution function (\textit{PDF}) is
\begin{equation}\label{E:DistribCauchy}
	\mf{C}(x \given \{\gamma,\widehat{\mu}\})=\frac{1}{\pi} \cdot \arctan\left( \frac{x-\widehat{\mu}}{\gamma}\right) +\frac{1}{2},
\end{equation} 
$ \widehat{\mu} $ is the \textit{median and mode} of the distribution, and the distribution is symmetric about $ \widehat{\mu} $. The maximum value or amplitude of the Cauchy pdf is $ \tfrac{1}{\pi \gamma} $, located at
$ x = \widehat{\mu} $, $ \gamma $ is called the \textit{scale factor}. Using Eq. (\ref{E:DistribCauchy}), it is easy to calculate the probability  $ C[ x \in (\widehat{\mu} \pm \text{\textbf{\underline{1}}}\gamma )\given \{\gamma,\widehat{\mu}\}]=0.5 $, thus $ \widehat{\mu} \pm \text{\textbf{\underline{1}}}\gamma$ is the 50\% CI of $ \widehat{\mu} $, the 69\% CI (like the CI $ \mu\pm \text{\textbf{\underline{1}}} \sigma $ in Gaussian statistics) is $ \widehat{\mu} \pm  \text{\textbf{\underline{1.89}}} \gamma$. The broadness the Cauchy distribution \textquotedblleft{}shoulders\textquotedblright{} becomes dramatically clear when a 95\% CI is desired, since this interval is approximately $ \widehat{\mu} \pm  \text{\textbf{\underline{12.7}}} \gamma $ for a Caucy variable, in contrast with $ \mu \pm \text{\textbf{\underline{1.96}}} \sigma $ required for Gaussian variables. 

\begin{lemma}
	\textbf{Wilks [\cite{Wilks1962}, pg. 156]}
	If $x$ is a variate having a PDF $F(x)$ then the variate $y = F(x)$ has the  rectangular distribution $R(\tfrac{1}{2}, 1)$.
	\begin{proof}\label{Lem:PDF_Unif}
		This follows at once from the fact that the PDF of $y$ is 
		\begin{equation}
			G(y)= P \left [ F(x) \leqq y \right ] =  \begin{cases} 1, &  y > 1   \\  y, & 0 < y \leqslant 1 \\  0, & y \leqslant 0 \end{cases}
		\end{equation}
		which is the pdf of the rectangular distribution $R(\tfrac{1}{2}, 1)$.
	\end{proof}
\end{lemma}

Wilks\textquoteright{ }\cite{Wilks1962} $ R(\tfrac{1}{2}, 1) $, is denoted $  U[0,1] $ in this paper. Lemma \ref{Lem:PDF_Unif} enables to simulate variates distributed as $ c(x \given \{\gamma,\widehat{\mu}\}) $ using uniform variates $ U(0,1) $ and the following expression
\begin{equation}\label{E:SimDistribCauchyGamma0}
	\varPsi_{\mf{C}} \left[U(0,1)_i \given \{\gamma,\widehat{\mu}\} \right]  =  \widehat{\mu}\;+\; \gamma \cdot \tan \left( \pi \cdot \left[U(0,1)_i-\tfrac{1}{2}\right]\right).
\end{equation}

\begin{figure}[h!]
	\begin{small}
		\centering
		\includegraphics[width=12cm]{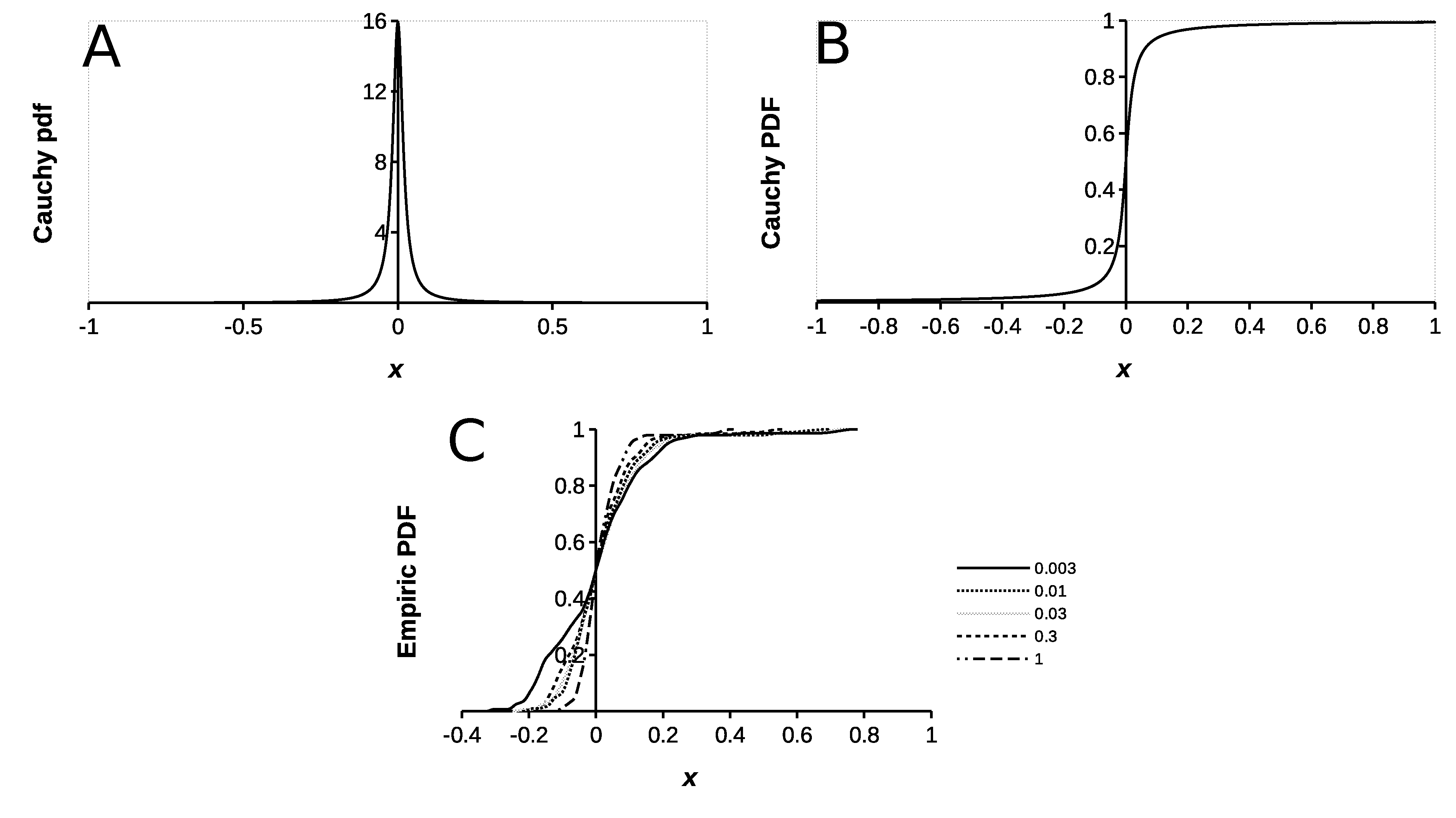}
		\caption{\textbf{\textbf{The Cauchy probability distribution an a set of 
					representative empirical distribution functions for FIII from 
					Quintana \cite{Quintana2017}} A- Cauchy probability density [Eq. 
				(\ref{E:DensCauchy})] function (pdf) calculated with $ \gamma = 1/50 $ and $ 
				\widehat{\mu} =0 $. B- Cauchy probability distribution function (PDF) [Eq. 
				(\ref{E:DistribCauchy})] calculated with $ \gamma = 1/50 $ and $ 
				\widehat{\mu} =0 $.  C- Several empirical probability distribution functions 
				\cite{Shorack1986,Pitman1993} estimated for FIII at concentrations of 
				0.003, 0.01, 0.03 and 1 mg/ml. This curves were selected because are 
				representative of the ones obtained with other fractions. The empirical curves 
				were plotted after subtracting the median of each killed cells fraction from the 
				remaining values in the set. Other details in the text of the communication.} }
		\label{F:CauchyDis}
	\end{small}
\end{figure}

Figure \ref{F:CauchyDis} presents a plot of a Cauchy probability density function (pdf) [Eq. (\ref{E:DensCauchy})] calculated with $ \gamma = 1/50 $ and $ \widehat{\mu} =0 $ (Panel \ref{F:CauchyDis}A), and a Cauchy probability distribution function (PDF) [Eq. (\ref{E:DistribCauchy})] also calculated with $ \gamma = 1/50 $ and $ \widehat{\mu} =0 $ (Panel \ref{F:CauchyDis}B). Also in Figure \ref{F:CauchyDis} (Panel \ref{F:CauchyDis}C) is a selection of empirical distribution functions \cite{Shorack1986,Pitman1993} determined for the sets of killed cells fractions observed with FIII \cite{Quintana2017}; this set was representative of other observed with the remaining fractions in Figures \ref{F:FracNoFit} and \ref{F:FracFit}. Sets $ \{L\} $ and $ \{F\} $ [Eqs. (\ref{E:LCorr}) and (\ref{E:FCorr})] were found not to be Gaussian using the Jarque-Bera \cite{Bera1981} and Shapiro-Wilks \cite{Shapiro1965} test, this only means that in addition to their most likely \textquotedblleft{}pathological\textquotedblright{ }distribution the precise nature of this distribution remains unknown. Yet, Figure \ref{F:CauchyDis}C suggests that the empirical PDF of data in Figures \ref{F:FracNoFit} and \ref{F:FracFit} resemble the Cauchy PDF in Figure \ref{F:CauchyDis}B.

To test the procedure discussed in this paper Cauchy-distributed data sets were generated using Monte Carlo simulation combining Eqs. (\ref{E:HillMod}) and (\ref{E:SimDistribCauchyGamma0}) as 
\begin{equation}\label{E:SimCauchyHill}
	\begin{split}
		\varPsi_{\mf{H}_o\mf{C}} \left\langle   [D]_i \; \given \{\gamma, \widehat{\mu}=y \left( [D]_i \given \{y_0, y_m,K_m,n\} \right)\} \right\rangle_{i=1, \, \dots \,, m}
		= \, \dots \, & \\ \, \dots \, = \left\lbrace y_0 + \dfrac{y_m}{1+\left(\tfrac{K_m}{[\mathrm{D}]_i}\right)^n}+   \gamma \cdot \tan \left( \pi \cdot \left[ U(0,1)_i-\tfrac{1}{2} \right] \right)\right\rbrace_{i=1, \, \dots \,, m} 
	\end{split}
\end{equation}
and used to calculate data sets, and processed as done with the data of Quintana \cite{Quintana2017}.

Some results of the fits of Eq. (\ref{E:HillMod}) to Cauchy data appear in Table \ref{T:HillMonCau1} together with their apparent sample skewedness \cite{Pearson1894, Pearson1963} and kurtosis calculated wuth Eqs. (\ref{E:Sk}) and (\ref{E:Kr}).

\begin{figure}[h!]
	\begin{small}
		\centering
		\includegraphics[width=12cm]{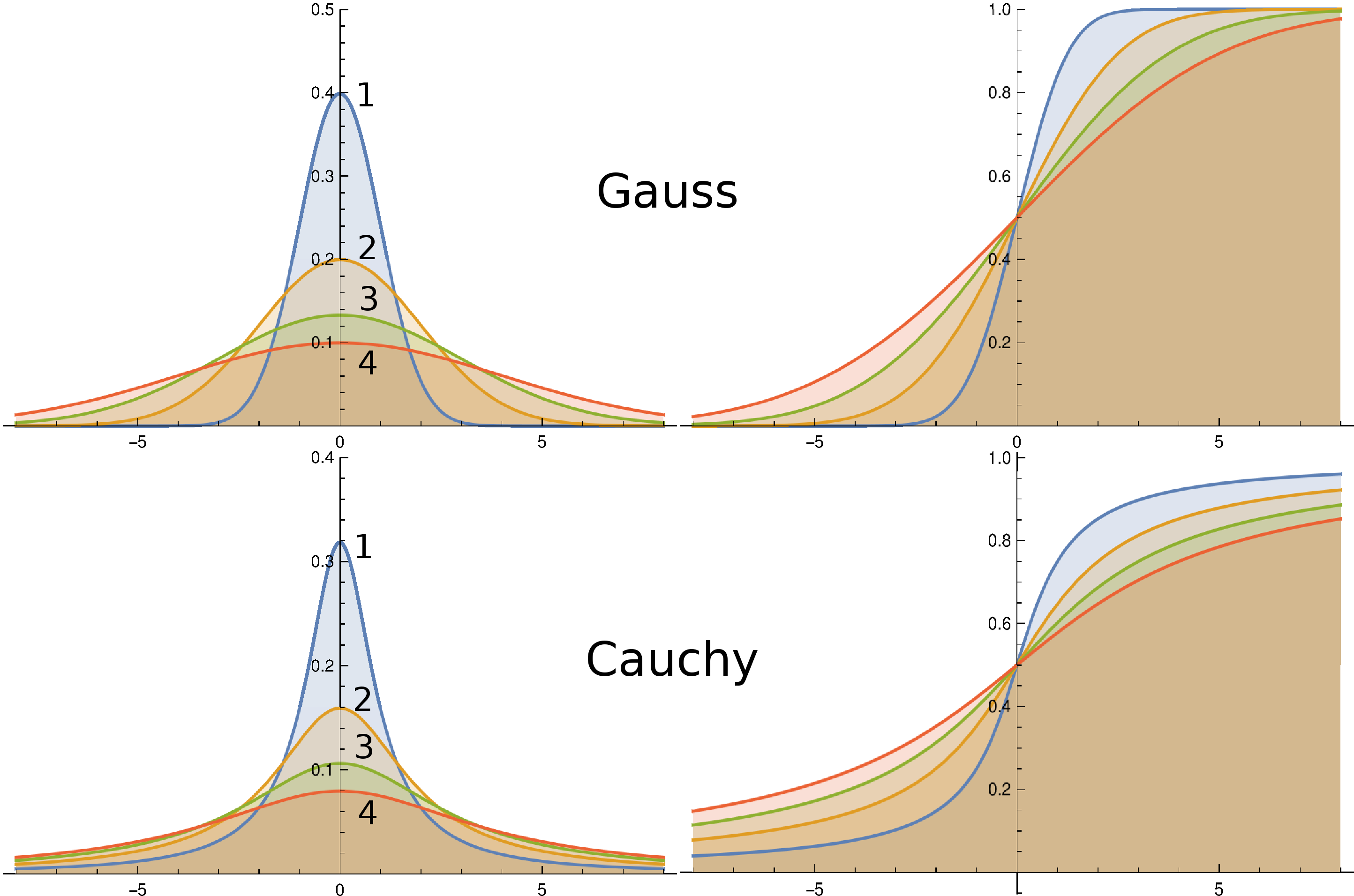}
		\caption{\textbf{Gauss and Cauchy probability density (pdf, left panels) and probability distribution (PDF, right panels) functions.} The functions are centered at the Gauss mean ($ \mu=0 $) and the Cauchy median ($ \widehat{\mu}=0 $). Upper left: Gauss pdf; upper right: Gauss PDF; lower left: Cauchy pdf; lower right: Cauchy PDF. Numbers near the Gauss curves indicate the value of the variance ($ \sigma^2 =$1, 2, 3 or 4) for each curve, numbers near the Cauchy curves indicate the value of the width factor ($ \gamma =$ 1, 2, 3 or 4). The figure enables a naked eye comparison of the two distributions, and shows that the Cauchy distribution has a sharper peak   (most evident when $ \gamma=1 $ in the figure) at the median ($ \widehat{\mu} $) of the distribution, and has broader shoulders (particularly evident in the Cauchy PDF plot) as $ \abs{x}\rightarrow \infty $), both factors explain apparently higher kurtosis ($ Kr $) sample estimates. Other details in the text of the communication.} 
		\label{F:Gauss_Cauchy}
	\end{small}
\end{figure}

In spite of their undefined central moments, both conditions occur in Cauchy-distributed variables if mean is replaced by median in the preceding Moors\textquoteright{} quote (see Figure \ref{F:Gauss_Cauchy}). Estimated $ Sk $ and  $ Kr $ were incompatible to the ones expected for  Gaussian variables  ($ Sk = 0,\; Kr =3 $), this is not surprising since the data subject of the simplex optimization were generated for Cauchy-distributed variates. The values of $ Sk > 0 $ indicate that the parameter estimates are asymmetrically distributed about the mean (median?), and $Sk$ increases with sample size. Thus the parameter estimates are not exactly Cauchy-distributed either \cite{Cramer1991, Pitman1993}. Data in Tables \ref{T:HillMonCau1} through \ref{T:HillMonCau3} are very leptokurtic, clustered about the median, but extreme values are observed as indicated by the parameters\textquoteright{} ranges  \cite{DeCarlo1997}. Yet, in spite of the wide ranges, 95\% confidence intervals are relatively narrow suggesting strong clustering of data around the median. It may seem startling that raising the initial $n$ did not improve consistently the final estimate of $n$, and worsened the estimations of the other parameters too. 
[h!]
Table \ref{T:HillMonCau1} shows that the parameters $ \{ \theta_i\}_{sim}= \{y_0=5\%,\; y_m=100\%,\; \text{\textit{\textbf{K\textsubscript{m}=0.15, n=2}}}  \} $ used to simulate data distributed as Cauchy are well predicted if a simplex optimization is used. At least if the simplex is initiated with a \textquotedblleft{}reasonable\textquotedblright{} set of parameters ($ \{  \theta_i \}_{init}  =  \{y_0=10\%,\; y_m=100\% \text{, \textbf{\textit{K\textsubscript{m}=0.1, n=2}}} \} $).

Table \ref{T:HillMonCau2} presents data generated with Eq. (\ref{E:SimCauchyHill}) using a  parameter set $ \{ \theta \}_{sim}= \{y_0=5\%,\; y_m=100\%,\;  \text{, \textbf{\textit{K\textsubscript{m}=0.01, n=15}}}  \} $, and the simplex optimization started from $ \{ \theta_i \}_{init}= \{y_0=10\%,\; y_m=100\% \text{, \textbf{\textit{K\textsubscript{m}=0.1, n=2}}} \} $ as in Table \ref{T:HillMonCau1}.  As seen in the Table \ref{T:HillMonCau2}, $ y_0, \;y_m,\text{ and } K_m $ are well estimted using the simplex minimization described. Yet, $ n \ne 15 $ in all instances presented in Table \ref{T:HillMonCau2}.

To check if starting the simplex from higher values of $ n $ improves $n$ estimate, in Table \ref{T:HillMonCau3} presents data calculated for \textit{\textbf{exactly the same Monte Carlo data used in Table \ref{T:HillMonCau2}}}, but starting the simplex optimization with $  \{\vect{\theta}_i\}_{init} =\{y_0=10\%,\; y_m=100\%  \text{, \textbf{\textit{K\textsubscript{m}=0.1, n=10}}}  \} $.  As indicated by the discussion in Section \ref{S:HillFstDer} and shown in Figure \ref{F:VariancesOptimum}, raising the initial value of $ n=10 $ did not improve the estimation of the parameter. More surprising was that \textbf{rising the initial value of  \textit{n}  worsened the estimates of all the other parameters characterizing Eq. (\ref{E:HillMod})}. Taken together the results in tables \ref{T:HillMonCau2} and \ref{T:HillMonCau3}, suggests that $ n $ is the most difficult parameter to estimate in Equation (\ref{E:HillMod}). The difficulty to estimate $ n $ correctly agrees with the discussion on $ \mf{H}'(n) $ or $ \$ \mf{H}_{opt}'(n) $ done in Section \ref{S:HillFstDer}.

It is necessary to point out several features, some very evident and others not so much, of data in Tables \ref{T:HillMonCau1} -- \ref{T:HillMonCau3}. In the three tables, specially in optimizations with larger $ r $, some sets of Monte Carlo simulated data did not reach Condition \ref{Cnd:StopOpt}, and the optimization stopped according to Condition \ref{Cnd:TooLong} (numbers with asterisks). One instance of this situation appears in each Table \ref{T:HillMonCau1} -- \ref{T:HillMonCau3}, but each of these data sets could have been replaced by results obtained for other sets of pseudorandom data generated under similar conditions where Condition \ref{Cnd:StopOpt} was indeed achieved, i.e., failure to comply with Condition \ref{Cnd:StopOpt} did not always occurred. The instances where Condition \ref{Cnd:TooLong} stopped calculations were included in the tables to show that they do occur. Another feature apparent in Tables \ref{T:HillMonCau1} -- \ref{T:HillMonCau3} is that although all set of parameters are leptokurtic and skewed, $ Kr $ increased with the number of points ($ r $) used for each concentration in the optimization, $ Sk $ did not increase as much as $Kr$ but was noticeably large wit $ r=2000 $. $ Sk $ and $ Kr $ depend on the 2\textsuperscript{nd}, 3\textsuperscript{rd} and 4\textsuperscript{th} central moments of a distribution which do not exist for the Cauchy distribution, thus sample values cannot converge towards any population value as required by sampling theory \cite{Wilks1962}, since population mean, variance, skewness and kurtosis do not exist for Cauchy distributed data. Sample variance, for example, grows with sample size. Since the probability density function of the Cauchy distribution has long tails, the odds for large values to occur are not negligible. This makes the mean jump considerably even when several hundred or thousand random numbers are averaged \cite{Rothenberg1964,Fama1968,Lohninger2012}. Yet, with all the uncertainties, the $ Sk $ and $ Kr $ estimates clearly indicate that data in tables  \ref{T:HillMonCau1} -- \ref{T:HillMonCau3} are non-Gaussian.

\begin{table}
	\begin{center}
		\begin{small}
			\caption{\textbf{Fitting Cauchy data generated with Eq.
					(\ref{E:SimCauchyHill}) setting $ \mathbf{\gamma = 1/50} $ and simplex optimization to the modified Hill equation. In all cases the optimization started with $ \mathbf{y_0 = 
						-10\%} $, $\mathbf{y_m = 100\%}$, $\mathbf{K_m=0.1}$ and 
					$\mathbf{n=2}$; $ \Delta_{init}$ was set as $ 0.1 $.}}\label{T:HillMonCau2}
			\begin{supertabular}{
					*{1}{m{0.8cm}}
					*{1}{m{0.8cm}}
					*{1}{m{1.5cm}}
					*{1}{m{3.cm}}
					*{1}{m{3.5cm}}
					*{1}{m{1.5cm}}
					*{2}{m{1.cm}}}
				\hhline{*{8}=}
				\centering$r$&
				\centering $ \theta_j $&
				\centering \textbf{Sim}&
				\centering \textbf{Pred}&
				\centering \textbf{Range} &
				\centering \textbf{Loops}&
				\centering \textbf{\textit{Sk}}&
				\centering \textbf{\textit{Kr}} \cr
				\hhline{*{8}-}		
				\centering $ 3 $&
				\centering $ y_0 \; \bullet $&
				\centering $  -5 $ &
				\centering  $ 8.5$  \par $ (-5.5,\; 41.0) $ &
				\centering  $ (-182.5,\; 95.9) $&
				\centering $ 41029 $&
				\centering $ 1.26 $&
				\centering $ 3.145 $
				\cr
				&\centering $ y_m $ &
				\centering  $  100 $ &
				\centering  $ 93.9$ \par $(-10^{21},\; 7 \cdot 10^{5})$& 
				\centering  $(-9 \cdot 10^{22},\; 7 \cdot 10^{11})$&
				&
				\centering  $ -4.092 $&
				\centering  $ 18.1 $
				\cr
				&\centering$ Km\: \bullet$&
				\centering  $ 0.01 $ &
				\centering  $ 0.20$ \par    $(0.06,\; 0.53) $&
				\centering $ (-0.07,\; 1.08) $&
				&
				\centering $ 1.260 $&
				\centering $ 3.145 $
				\cr
				&\centering$ n \; \bullet$&
				\centering  $ 15 $ &
				\centering  $ 11.8 $ \par $ (11.7,\; 12.2) $&
				\centering  $ (11.6,\; 12.7) $&
				&
				\centering $ 1.026 $&
				\centering $ 5.352 $
				\cr
				\centering $ 10 $ &
				\centering $ y_0 $&
				\centering $  -5 $ &
				\centering $  2.4$ \par $(1.2,\; 4.1) $& 
				\centering  $ (-358,\; 621) $&
				\centering  $ 2176 $&
				\centering $ -4.995 $&
				\centering $ 30.0 $
				\cr
				&\centering$ y_m $&
				\centering  $ 100 $ &
				\centering  $ 97.3 $ \par $(-34.6,\; 100.0) $& 
				\centering  $(-6 \cdot 10^{40},\; 8 \cdot 10^{40})$&
				&
				\centering $  2.065 $&
				\centering  $ 25.7 $
				\cr
				& \centering  $ K_m $&
				\centering  $ 0.01 $ &
				\centering  $ 0.05$ \par $(0.02,\; 0.05) $&
				\centering  $ (-3.57,\; 0.63) $&
				&
				\centering $ -4.995 $&
				\centering  $ 30.0 $
				\cr
				&\centering$ n $&
				\centering  $ 15 $ &
				\centering  $ 28.55 $ \par $ (28.54,\; 28.56) $&
				\centering  $ (24.9,\; 29.1) $&
				&
				\centering  $ $ 4.995 $ $&
				\centering  $ $ 30.0 $ $
				\cr 
				100 &$ y_0 $&
				\centering $  -5 $  &
				\centering $   -5.9 (-6.2,\; -5.6) $& 
				\centering $  (-19.3,\; 9.7) $&
				\centering  $ 26297 $&
				\centering  $ -12.2 $ &
				\centering  $ 255.4 $
				\cr
				&$ 	y_m $&
				\centering  $ 100 $&
				\centering  $ 102.6 (101.9,\; 103.7) $ &
				\centering  ($-10^{13},\; 10^{13}$)&
				\centering  $ 25.6 $&	
				\centering  $ 670.7 $
				\cr
				&$ K_m $&
				\centering  $ 0.01 $&
				\centering  $ 0.013$ \par $(0.010,\; 0.016) $&
				\centering  $ (-0.3,\; 0.2) $&
				\centering  $ -12.2 $&
				\centering  $ 255.4 $
				\cr
				&$ n $&
				\centering  $ 15 $ &
				\centering  $ 11.196$ \par $ (11.192,\; 11.199) $&
				\centering  $ (-8.0,\, 21.0) $&
				\centering  $ -12.2 $&
				\centering  $ 255.4 $
				\cr
				\centering $ 2000 $ & 
				\centering$ y_0 $&
				\centering  $ -5 $ &
				\centering  $ 26.5$ \par $(26.2,\; 26.8) $&
				\centering  $(-1.9 \cdot 10^{4},\; 10^{5})$&
				\centering  $ 932 $&
				\centering  $ 106.3 $&
				\centering  $ 12272 $
				\cr
				&\centering$ y_m $&
				\centering  $ 100 $&
				\centering  $ 59.3$ \par $(28.4,\; 63.6) $&
				\centering  $ (-2 \cdot 10^{61}, \; 2 \cdot 10^{60} )$&
				&
				\centering $  -113.6 $&
				\centering  $ 13261.4 $
				\cr 
				&\centering  $ K_m $&
				\centering  $ 0.01 $&
				\centering  $ 0.063 $ \par $ (0.060,\; 0.065) $&
				\centering  $ (-197.2,\; 1105.3) $&
				&
				\centering  $ 106.3 $&
				\centering  12272
				\cr
				&\centering $ n $&
				\centering  $ 15 $ &
				\centering  $ 31.9 $ \par $ (31.9,\; 31.9) $&
				\centering  $ (-165.4,\; 1137.1) $&
				&
				\centering  $ 106.3 $&
				\centering  $ 12272 $
				\cr
				\hhline{*{8}=} 
			\end{supertabular} 
		\end{small}
	\end{center}
	\begin{footnotesize}
		Parameters and heading have same meaning as used in Table \ref{T:SkwKurt}; $ r $ indicates the number of random Cauchy values fitted to Eq. (\ref{E:HillMod}) which were exactly the same used in Table \ref{T:HillMonCau1}. Please notice that the units of $ K_m $ are irrelevant as long as they are equal to the units of [D]. Parameter names ($ \vect{\theta} $) with a bullet ($\bullet$) indicated that, tested with the Jarque-Bera test, the probability that the parameter is not Gaussian by chance is not too low ($ P=0.03 $). Number of loops with an asterisk indicates the optimization stopped after fulfilling Condition \ref{Cnd:TooLong}. Other conditions as in Table \ref{T:HillMonCau1}. Please notice that $ n \vrdif 15 $ in all instances. See the text of the communication for further discussion. 
	\end{footnotesize}
\end{table}

\begin{table}
	\begin{center}
		\begin{small}
			\caption{\textbf{Fitting Cauchy data generated with Eq.
					(\ref{E:SimCauchyHill}) setting $ \mathbf{\gamma = 1/50} $ and simplex optimization to the modified Hill equation. In all cases the optimization started with $ \mathbf{y_0 = 
						-10\%} $, $\mathbf{y_m = 100\%}$, $\mathbf{K_m=0.1}$ and 
					$\mathbf{n=2}$; $ \Delta_{init}$ was set as $ 0.1 $.}}\label{T:HillMonCau2}
			\begin{supertabular}{
					*{1}{m{0.8cm}}
					*{1}{m{0.8cm}}
					*{1}{m{1.5cm}}
					*{1}{m{3.cm}}
					*{1}{m{3.5cm}}
					*{1}{m{1.5cm}}
					*{2}{m{1.cm}}}
				\hhline{*{8}=}
				\centering$r$&
				\centering $ \theta_j $&
				\centering \textbf{Sim}&
				\centering \textbf{Pred}&
				\centering \textbf{Range} &
				\centering \textbf{Loops}&
				\centering \textbf{\textit{Sk}}&
				\centering \textbf{\textit{Kr}} \cr
				\hhline{*{8}-}
				\centering $ 3 $ & \centering $ V_{\tfrac{1}{2}} $&
				\centering  $ -40 $ &
				\centering $ -40.395 $ \par $ (-40.399, -40.394) $&
				\centering  $ (-40.419, -40.394) $&
				\centering $ 210 $&
				\centering $ -1.866 $&
				\centering $ 5.381 $
				\cr
				& \centering  $\kappa$&
				\centering  $ 10 $ &
				\centering  $ 9.631 $ \par $ (9.630, 9.632) $& 
				\centering  $ (9.610, 9.643) $&
				&
				\centering  $ -0.181 $&
				\centering  $ 4.104 $
				\cr
				\centering  $ 10 $& \centering $ V_{\tfrac{1}{2}} $&
				\centering  $ -40 $ &
				\centering  $ -39.985 $ \par $ (-39.989, -39.984) $& 
				\centering  $ (-40.007, -39.983) $&
				\centering  $ 508 $&
				\centering  $ -1.676 $&
				\centering  $ 4.672 $
				\cr
				&$ \centering\kappa$&
				\centering  $ 10 $ &
				\centering  $ 10.5380 $ \par $ (10.5378, 10.5380) $& 
				\centering  $ (10.5176, 10.5584) $&
				&
				\centering  $ 0.011 $&
				\centering  $ 4.232 $
				\cr
				\centering $ 100 $&  \centering $ V_{\tfrac{1}{2}} $&
				\centering  $ -40 $ &
				\centering  $ -40.055 $ \par $ (-40.058, -40.054) $& 
				\centering  $ (-40.079, -40.053) $&
				\centering  $ 317 $&
				\centering  $ -1.777 $&
				\centering  $ 5.045$
				\cr
				& \centering  $ \kappa $&
				\centering  $ 10 $&
				\centering  $ 10.0451 $ \par $ ( 10.0451, 10.0452) $ &
				\centering  $ (10.024, 10.066) $&
				&
				\centering  $ 0.01 $7&	
				\centering  $ 4.437 $
				\cr
				\centering $ 2000 $ & $ \mathbf{V_{\text{\textonehalf}}}$&
				\centering  $ -40 $ &
				\centering  $ -39.978  $\par $ (-39.978, -39.978) $&
				\centering  $ (-40.00, -39.98) $&
				\centering  $ 635 $&
				\centering  $ -1.790 $&
				\centering  $ 5.094 $
				\cr
				&\centering $ \kappa $&
				\centering  $ 10 $&
				\centering  $ 9.9841  $\ \par $ ( 9.9841, 9.9841) $&
				\centering  $ (9.96, 10.01) $&
				&
				\centering  $ 0.005 $&
				\centering  $ 4.461 $
				\cr 
				\hhline{*{8}=} 
			\end{supertabular} 
		\end{small}
	\end{center}
	\begin{footnotesize}
		Table heading have same meaning as used in Table \ref{T:SkwKurt}; $ r $ indicates the number of random Gaussian values of type $N(\mu, \sigma^2=0.05 \cdot \mu)$ with $\mu=y(x_i \given \{V_{\text{\textonehalf}},\kappa\}) = \tfrac{1}{1+\e^{-(x_i-V_{\text{\textonehalf}})/\kappa}} $ which were simulated for each concentration. See the text for further discussion. Other details as in Table \ref{T:HillMonCau1}.
	\end{footnotesize}
\end{table}

\subsubsection{Fitting Gauss distributed data to the modified Hill equation.}\label{S:ModGausData}

To evaluate the behavior of Eq. (\ref{E:HillMod}) when Gaussian variables are adjusted to the equation, normal variates $ N(u_i ,s^2) $ generated as 
\begin{equation}\label{E:GaussHill}
	\begin{split}
		\varPsi_{\mf{H}G} \left\langle    [D]_{i} \given \left[  \sigma^2=\varsigma,\;\mu=y \left( [D]_{i} \given \{\vect{y_0,y_m,K_m,n }\} \right )\right]  \right\rangle_{i=1, \, \dots \,,m} = \cdots & \\ \cdots = \left\lbrace y_0+\frac{y_m}{1+(K_m/[D]_{i})^n} + \varsigma \cdot N(0,1)_i   \right\rbrace_{i=1, \, \dots \,,m},
	\end{split}
\end{equation}
was used to test the ability of the method proposed here to determine the parameters, with $ \varsigma = \tfrac{s^2[y(x_i)]}{\overline{y}(x_i)} $, in which $ s^2[y(x_i)] $ and $ \overline{y}(x_i) $ were the simulated effects variances and means, respectively.  Tables \ref{T:HillMonGaus1} to \ref{T:HillMonGaus3} follow the same sequence of changes of simulating sets $\{ \vect{\theta}_i\}_{sim} $ and initial values $ \{\vect{\theta}_i\}_{init} $ as Tables \ref{T:HillMonCau1} to \ref{T:HillMonCau3}, but data adjusted to Eq. (\ref{E:HillMod}) was Gaussian and generated as indicated in the prior paragraph. As in the case of Cauchy distributed data in Section \ref{S:ModCauhData}, $ y_0, \;y_m,\text{ and } K_m $ are well estimated using the simplex minimization described. Yet, $ n \vrdif 15 $ ($ \vrdif$ symbol is used to mean very significantly different) in all instances presented in Table \ref{T:HillMonCau2}, this suggests that $ n $ is the most difficult parameter to estimate if the initiating value of $ n $ in the simplex is very different from the value used in the Monte Carlo simulation. 

$ \Delta_{init}=0.1 $ (used as initiation value in Tables \ref{T:HillMonCau1} to \ref{T:HillMonGaus3}) means that the initial parameters begin increasing by 10\%. In calculations, not published here, the simulations in Tables \ref{T:HillMonCau1} to \ref{T:HillMonGaus3} were carried out setting $ \Delta_{init}=0.5$, and yet $ n \vrdif 15 $, and the estimates of the other parameters became worse. This agrees with the discussion about Eqs. (\ref{E:dy0}) to  (\ref{E:dn}) in Section \ref{S:HillFstDer} which holds for Gaussian variates, since the discussion in Section \ref{S:HillFstDer} is distribution independent.

Tabla aqui 

\subsection{The Boltzmann probability distribution function.}

A function commonly used to describe data from diverse empirical sources is the Boltzmann \textit{probability distribution function} (\textit{pdf}), one of its forms is Eq. (\ref{E:BoltzElect}), revised in \cite{Sevcik2017a}. Boltzmann function first partial derivatives, for a form commonly used in electrophysiology \cite{Hodgkin1952e, Sevcik2017a}, are Eqs. (\ref{E:BoltDerVmed}) and (\ref{E:BoltDerKappa}). In addition to the properties of the Boltzmann function considered in \cite{Sevcik2017a}, more details on this distribution are presented in Section \ref{S:BoltzEqu} and Figure \ref{F:BolVariancesOptimum}. 

\subsubsection{Using the Boltzmann distribution function to fit Gaussian data.}\label{S:BoltzMon}

To evaluate the behavior of Eq. (\ref{E:BoltzElect}) Gaussian variables are adjusted to the equation, normal variates $ N(u_i ,s^2) $ generated as 
\begin{equation}\label{E:BoltzGau}
	\varPsi_{\mf{B}G} \left\langle V_{i} \given \left\lbrace \varsigma=\sigma^2,\;y \left( V_{i}\; \given \{\;V_{\text{\textonehalf}},\kappa \}\right )\right\rbrace   \right\rangle_{i=1, \, \dots \,,m} = \left\lbrace \dfrac{1}{1+\exp{\left( -\tfrac{V_{i}-V_{\text{\textonehalf}}}{k}\right)}}+  \varsigma N(0,1)_i \right\rbrace _{i=1, \, \dots \,,m}
\end{equation}  
setting $ \varsigma =   \tfrac{s^2[y(x_i)]}{\overline{y(x_i)}}  $, where $ s^2[y(x_i)] $ and $ \overline{y(x_i)} $ are sample variance and mean, respectively, of the simulated effects. Resukts are shown in Tables \ref{T:BoltzMonGau1} and \ref{T:BoltzMonGau2}. 

Tables \ref{T:BoltzMonGau1} and \ref{T:BoltzMonGau2} shows considerable independence between initial and predicted $\{ \theta_j\} $ and $ r $ values, since in all cases the optimization produced remarkably similar estimates of $ V_{\text{\textonehalf}} $ and $ \kappa $ in agreement with the discussion of Eqs. (\ref{E:BoltDerVmed}) and (\ref{E:BoltDerKappa}). Optimizations presented in Tables \ref{T:BoltzMonGau1} and \ref{T:BoltzMonGau2} was started with $ \Delta_{init}= 0.5 $, and was even better when $ \Delta_{init}= 0.1 $. With $ \Delta_{init}= 0.1 $, it made no relevant difference for predicting the parameters used to simulate $ \{V_{\text{\textonehalf}}, \kappa \}=\{ -40, 10\} $ to start optimization with $ \{\vect{\theta}_i\}_{init} =  \{V_{\text{\textonehalf}}, \kappa \}_{init}$ set as :  $\{ -20,\; 1\} $, $\{ -20,\; 5\} $, $\{ -20,\; 20\} $ or $\{ -60,\; 20\} $. With $ \Delta_{init}= 0.1 $, it was to any practical purpose irrelevant, if the number of replicates ($ r $) was: 3, 10, 100 or 2000. Tables \ref{T:BoltzMonGau1} and \ref{T:BoltzMonGau2} show that parameters are also leptokurtic and not Gaussian (as shown by Jarque-Bera and Shapiro-Wilks tests), but with low $ Sk $, and that $ Sk $ and $ Kr $ seem largely independent of sample sizes ($ r $) and $\{  \vect{\theta}_j\}_{init} $. Ranges of predicted parameters in Tables \ref{T:BoltzMonGau1} and \ref{T:BoltzMonGau2} were quite narrow, the Boltzmann function [Eq. (\ref{E:BoltzElect})] seems easier to optimize to the correct parameters values that the Hill equation [Eq. (\ref{E:HillMod})]. 

\begin{table}
	\begin{center}
		\begin{small}
			\topcaption{\textbf{Parameters characterizing the \{\textit{d\textsubscript{j,i}}\}\textsubscript{\textit{i=1,2, \, \ldots \, ,m}} sets used to calculate parameter uncertainty in Table \ref{T:HillSimplexPar}}. $ \Delta_{init}$ was set as 0.1.}\label{T:SkwKurt}
			\begin{supertabular}{*{3}{m{1.5cm}}*{2}{m{1.5cm}}{m{3cm}}}
				\hhline{*{6}=}
				\textbf{Fraction} &
				\textbf{Loops} &
				\textbf{Parameter} &
				\centering \textbf{\textit{Sk}} &
				\centering \textbf{\textit{Ku}} &
				\centering \textbf{Range} \cr
				\hhline{*{6}-}\\
				\textbf{FI} &$ 2867 $&\centering  \textit{y\textsubscript{0}}&
				\centering  $ -3.95 $ &
				\centering  $ 11.5 $& \centering  $ -1.49,\; 0.21 $ 
				\cr
				&&\centering  \textit{y\textsubscript{m}}&
				\centering  $ -6.30 $ &
				\centering  $ 75.5 $& 
				\centering  $-1.9 \cdot 10^{3}$, $6 \cdot 10^{2}$
				\cr
				&&\centering  \textit{K\textsubscript{m}}&
				\centering  $ -2.95 $ &
				\centering  $ 11.5 $&
				\centering  $ -1.43,\; 0.28 $
				\cr
				&&\centering  \textit{n}&
				\centering  $ -2.95 $ &
				\centering  $ 11.5 $&
				\centering  $ 3.9,\; 5.6 $
				\cr
				\textbf{FII} &$ 1328 $&\centering  \textit{y\textsubscript{0}}&
				\centering  $ -2.90 $ &
				\centering  $ 49.1 $& 
				\centering  $ -5.7,\; 0.3 $ 
				\cr
				&&\centering  \textit{y\textsubscript{m}}&
				\centering  $ -5.66 $ &
				\centering  $ 66.8 $& 
				\centering  $ -492,\; 145 $
				\cr
				&&\centering  \textit{K\textsubscript{m}}&
				\centering  $ -2.90 $ &
				\centering  $ 49.1 $&
				\centering  $ -5.38,\; 0.62 $
				\cr
				&&\centering  \textit{n}&
				\centering  $ -2.57 $&
				\centering  $ 40.3 $&
				\centering  $ -3.9,\; 2.0 $
				\cr
				
				\textbf{FIII}  &$ 13277 $&\centering  \textit{y\textsubscript{0}}&
				\centering  $ -0.87 $ &
				\centering  $ 27.42 $ & 
				\centering  $ -695,\; 697 $
				\cr
				&&\centering  \textit{y\textsubscript{m}}&
				\centering  $ -4.35 $&
				\centering  $ 67.5 $&
				\centering  ($-10^{26},\; 5 \cdot 10^{25}$)
				\cr 
				&&\centering  \textit{K\textsubscript{m}}&
				\centering  $ -0.87 $&
				\centering  $ 27.4 $&
				\centering  $ -695,\; 697 $
				\cr
				&&\centering  \textit{n}&
				\centering  $ -0.70 $&
				\centering  $ 23.1 $&
				\centering  $ -681.6,\; 710.4 $
				\cr
				\textbf{FIV}  &$ 1024007 $*&\centering  \textit{y\textsubscript{0}}&
				\centering  $ -2.48 $ &
				\centering  $ 15.04 $ & 
				\cr
				&&\centering  \textit{y\textsubscript{m}}&
				\centering  $ -6.56 $&
				\centering  $ 73.6 $&
				\centering  ($ -2 \cdot 10^{6},\; 5 \cdot 10^{5} $)
				\cr 
				&&\centering  \textit{K\textsubscript{m}}&
				\centering  $ -2.48 $&
				\centering  $ 15.0 $&
				\centering  $ -0.68,\; 0.47 $
				\cr
				&&\centering  \textit{n}&
				\centering  $ -2.20 $ &
				\centering  $ 12.4 $ & 
				\centering $  2.6,\; 3.8 $
				\cr
				\textbf{FV} &$ 21432 $&\centering  \textit{y\textsubscript{0}}&
				\centering  $ -3.85 $ &
				\centering  $ 18.1 $ & 
				\centering  $ -1.6,\; 0.24 $
				\cr
				&&\centering  \textit{y\textsubscript{m}}&
				\centering  $ -2.79 $&
				\centering  $ 60.0 $&
				\centering  ($ -7\cdot 10^{14},\; 3 \cdot 10^{14} $)
				\cr 
				&&\centering  \textit{K\textsubscript{m}}&
				\centering  $ -3.85 $&
				\centering  $ 18.1 $&
				\centering  $ -1.6,\; 0.3 $
				\cr
				&&\centering  \textit{n}&
				\centering  $ -3.46 $ &
				\centering  $ 14.9 $&
				\centering  $ 14.6,\;16.4 $
				\cr
				\hhline{*{6}=} 
			\end{supertabular} 
		\end{small}
	\end{center}
	\begin{footnotesize}
		Parameters of the modified Hill Equation (\ref{E:HillMod}): $ y_0 $, offset parameter; $ y_m $, maximum effect; $ K_m $, concentration producing half maximum effect, $ n $, is called Hill constant; \textit{Ku}, kurtosis and \textit{Sk}, skewedness. \textit{Range}, \textit{newness} and \textit{kurtosis} have the usual statistical meanings. \textit{Loop}, indicates the number of times a parameter was changed during the simplex optimization \cite{Nelder1965}. For fractions I, II, III and V optimization stopped when when Condition \ref{Cnd:StopOpt} was fulfilled. In case of FIV the optimization was topped fulfilling Condition \ref{Cnd:TooLong} after 3 h attempting unsuccessfully to fulfill Condition \ref{Cnd:StopOpt}. See the text for further discussion.
	\end{footnotesize}
\end{table}

\begin{table}
	\begin{center}
		\begin{small}
			\topcaption{\textbf{Fitting Gaussian data generated Eq. (\ref{E:BoltzGau}) to the Boltzmann equation with the simplex optimization. In all cases the optimization started with $ \mathbf{V_{\text{\textonehalf}}=-20}, $ $\mathbf{\kappa=1}$ and $ \mathbf{\Delta_{init}}=0.5. $. }}\label{T:BoltzMonGau1}
			\begin{supertabular}{
					*{1}{m{0.8cm}}
					*{1}{m{0.8cm}}
					*{1}{m{1.5cm}}
					*{1}{m{3.cm}}
					*{1}{m{3.5cm}}
					*{1}{m{1.5cm}}
					*{2}{m{1.cm}}}
				\hhline{*{8}=}
				\centering$r$&
				\centering $ \theta_j $&
				\centering \textbf{Sim}&
				\centering \textbf{Pred}&
				\centering \textbf{Range} &
				\centering \textbf{Loops}&
				\centering \textbf{\textit{Sk}}&
				\centering \textbf{\textit{Kr}} \cr
				\hhline{*{8}-}			
				\centering$ 3 $ &
				\centering $ V_{\tfrac{1}{2}} $&
				\centering  $ -40 $ &
				\centering  $ -41.531 $ \par $ ( -41.536, -41.530) $&
				\centering  $ (-41.56, -41.53) $&
				\centering $ 1766 $&
				\centering $ -1.9 $&
				\centering $ 5.48 $
				\cr
				&\centering $\kappa$&
				\centering  $ 10 $ &
				\centering  $ 9.403 $ \par $ (9.401, 9.404) $& 
				\centering  $ (9.38, 9.43) $&
				&
				\centering  $ 0.331 $&
				\centering  $ 4.555 $
				\cr
				\centering$ 10 $ &
				\centering $ V_{\tfrac{1}{2}} $&
				\centering  $ -40 $ &
				\centering  $ -40.0585 $ \par $ (-40.062, -40.058) $& 
				\centering $  (-40.08, -40.06) $&
				\centering  $ 478 $&
				\centering $  -1.8 $&
				\centering  $ 5.128 $
				\cr
				& \centering $ \kappa$&
				\centering  $ 10 $ &
				\centering  $ 9.9416 $ \par $ (9.9415, 9.9417) $& 
				\centering  $ (9.92, 9.96) $&
				\centering &
				\centering  $ 0.02 $&
				\centering  $ 4.478 $
				\cr
				\centering  $ 100 $ &
				 \centering $ V_{\tfrac{1}{2}} $&
				\centering  $ -40 $ &
				\centering  $ -39.922 $ \par $ (-39.925, -39.921) $& 
				\centering  $ (-39.95, -39.93) $&
				\centering  $ 381 $&
				\centering  $ -1.779 $&
				\centering  $ 5.055 $
				\cr
				&
				\centering $ \kappa$ &
				\centering  $ 10 $ &
				\centering  $ 10.0322 $ \par $ (10.0322, 10.0323) $ &
				\centering  $ (10.01, 10.05) $&
				&
				\centering  $ -0.004 $&	
				\centering  $ 4.441 $
				\cr
				\centering  $ 2000 $ & \
				\centering $ V_{\tfrac{1}{2}} $&
				\centering  $ -40 $ &
				\centering  $ -39.991 $ \par $ (-39.991, -39.990) $&
				\centering  $ (-40.014, -39.989) $&
				\centering  $ 242 $&
				\centering  $ -1.785 $&
				\centering  $ 5.078 $
				\cr
				&
				\centering  $ \kappa$&
				\centering  $ 10 $&
				\centering  $ 10.005 $ \par $ (10.005, 10.005) $&
				\centering  $ (9.98, 10.03) $&
				&
				\centering  $ 0.007 $&
				\centering  $ 4.453 $
				\cr 
				\hhline{*{8}=} 
			\end{supertabular} 
		\end{small}
	\end{center}
	\begin{footnotesize}
		The $V$ values required by Eq. (\ref{E:BoltzElect}) were defined as: -100, -80, -60, -40, -20, 0, 20, 40, 50, 80 and 100; 	$ m $ is the number of $ d_{j,i} $ values used to calculate medians, 95\% confidence interval and ranges. Table heading have same meaning as used in Table \ref{T:SkwKurt}; $ r $ indicates the number of random Gaussian values of type $N(\mu, \sigma^2=0.05 \cdot \mu)$ with $\mu=y(x_i \given \{V_{\text{\textonehalf}},\kappa\}) = \tfrac{1}{1+\e^{-(x_i-V_{\text{\textonehalf}})/\kappa}} $ which were simulated for each concentration, other definitions as in Table \ref{T:HillMonCau1}. See the text for further discussion.
	\end{footnotesize}
\end{table}

\begin{table}
	\begin{center}
			\begin{small}
				\topcaption{\textbf{Fitting Gaussian data generated Eq. (\ref{E:BoltzGau}) to the Boltzmann equation with the simplex optimization. In all cases the optimization started with $ \mathbf{V_{\text{\textonehalf}}=-60}, $ $\mathbf{\kappa=20}$ and $ \mathbf{\Delta_{init}}=0.5. $. }}\label{T:BoltzMonGau2}
				\begin{supertabular}{
					*{1}{m{0.8cm}}
					*{1}{m{0.8cm}}
					*{1}{m{1.5cm}}
					*{1}{m{3.cm}}
					*{1}{m{3.5cm}}
					*{1}{m{1.5cm}}
					*{2}{m{1.cm}}}
					\hhline{*{8}=}
					\centering$r$&
					\centering $ \theta_j $&
					\centering \textbf{Sim}&
					\centering \textbf{Pred}&
					\centering \textbf{Range} &
					\centering \textbf{Loops}&
					\centering \textbf{\textit{Sk}}&
					\centering \textbf{\textit{Kr}} \cr
					\hhline{*{8}-}	
					\centering $ 3  $& \centering $ V_{\tfrac{1}{2}} $&
					\centering $  -4 $0 &
					\centering  $ -38.493 $ \par $ (-38.498, -38.492) $&
					\centering  $ (-38.52, -38.49) $&
					\centering $ 1024003 $*&
					\centering $ -1.654 $&
					\centering $ 4.567 $
					\cr
					& \centering  $\kappa$&
					\centering $ 10 $ &
					\centering  $ 10.591 $ \par $ ( 10.590, 10.593) $& 
					\centering  $ (10.57, 10.61) $&
					&
					\centering  $ -0.181 $&
					\centering  $ 4.104 $
					\cr
					\centering  $ 10 $ & \centering $ V_{\tfrac{1}{2}} $&
					\centering  $ -40 $ &
					\centering  $ -39.927 $ \par $ (-39.931, -39.927) $& 
					\centering  $ (-39.950, -39.926) $&
					\centering  $ 1768 $&
					\centering  $ -1.768 $&
					\centering  $ 5.014 $
					\cr
					& \centering  $ \kappa$&
					\centering  $ 10 $ &
					\centering  $ 9.9416 $ \par $ (9.9415, 9.9417) $& 
					\centering  $ (9.92, 9.96) $&
					&
					\centering $  -0.004 $&
					\centering  $ 4.420 $
					\cr
					 \centering $ 100 $ & \centering $ V_{\tfrac{1}{2}} $&
					\centering  $ -40 $ &
					\centering  $ -40.16 $3 \par $ (-40.166, -40.162) $& 
					\centering  $ (-40.19, -40.16) $&
					\centering  $ 340 $&
					\centering  -$ 1.798 $&
					\centering  $ 5.127 $
					\cr
					&\centering $ \kappa$&
					\centering  $ 10 $&
					\centering  $ 9.942 $ \par $ (9.942, 9.942)  $&
					\centering $  (9.92, 9.96) $&
					&
					\centering  $ 0.038 $&	
					\centering  $ 4.477 $
					\cr
					\centering $ 2000 $ & \centering $ V_{\tfrac{1}{2}} $ &
					\centering  $ -40  $&
					\centering $  -40.005 $ \par $ (-40.005, -40.004) $&
					\centering  $ (-40.028, -40.003) $&
					\centering  $ 327 $&
					\centering  $ -1.786 $&
					\centering $  5.080 $
					\cr
					&\centering $\kappa$&
					\centering  $ 10 $&
					\centering  $ 10.002 $ \par $ (10.003, 10.002) $&
					\centering  $ (9.98, 10.02) $&
					&
					\centering  $ 0.010 $&
					\centering  $ 4.454 $
					\cr 
					\hhline{*{8}=} 
			\end{supertabular} 
		\end{small}
	\end{center}
	\begin{footnotesize}
			Table heading have same meaning as used in Table \ref{T:SkwKurt}; $ r $ indicates the number of random Gaussian values of type $N(\mu, \sigma^2=0.05 \cdot \mu)$ with $\mu=y(x_i \given \{V_{\text{\textonehalf}},\kappa\}) = \tfrac{1}{1+e^{-(x_i-V_{\text{\textonehalf}})/\kappa}} $ which were simulated for each concentration. See the text for further discussion. Number of loops with an asterisk indicates the the optimization stopped after fulfilling Condition \ref{Cnd:TooLong}. Other details as in Table \ref{T:HillMonCau1}
	\end{footnotesize}
\end{table}

\begin{table}
	\begin{center}
		\begin{small}
			\topcaption{\textbf{Fitting Cauchy data generated with Eq. 
					(\ref{E:SimCauchyHill}) setting $ \mathbf{\gamma = 1/50} $ and simplex optimization to the modified Hill equation.  In all cases the optimization started with $ \mathbf{y_0 = 
						-10\%} $, $\mathbf{y_m = 100\%}$, $\mathbf{K_m=0.1}$ and 
					$\mathbf{n=10}$; $ \Delta_{init}$ was set as 0.1. }}\label{T:HillMonCau3}
			\begin{supertabular}{
					*{1}{m{0.8cm}}
					*{1}{m{0.8cm}}
					*{1}{m{1.5cm}}
					*{1}{m{3.cm}}
					*{1}{m{3.5cm}}
					*{1}{m{1.5cm}}
					*{2}{m{1.cm}}}
				\hhline{*{8}=}
				\centering$r$&
				\centering $ \theta_j $&
				\centering \textbf{Sim}&
				\centering \textbf{Pred}&
				\centering \textbf{Range} &
				\centering \textbf{Loops}&
				\centering \textbf{\textit{Sk}}&
				\centering \textbf{\textit{Kr}} \cr
				\hhline{*{8}-}	
				\centering$ 3 $ & \centering$ y_0 $&
				\centering  $ -5 $ &
				\centering  $ 8.64 $ \par$  (-5.45, 41.02) $&
				\centering  $ (-18.1, 95.9) $&
				\centering $ 31829 $&
				\centering $ 1.260 $&
				\centering $ 3.147 $
				\cr
				&\centering $ y_m $&
				\centering  $ 100 $ &
				\centering  $ 93.7 $ \par $ (-1.6 \cdot 10^{21}, 1.8 \cdot 10^{6}) $& 
				\centering  $ (-10^{22}, 2.2 \cdot 10^{11}) $&
				&
				\centering  $ -4.072 $&
				\centering  $  18.0  $
				\cr
				&\centering  $ K_m $&
				\centering  $ 0.01 $ &
				\centering  $ 0.20 $ \par $ (0.06, 0.52) $&
				\centering  $ (-0.07, 1.07) $&
				&
				\centering $ 1.260 $&
				\centering $ 3.147 $
				\cr
				&\centering   $ n $&
				\centering  $ 15 $ &
				\centering  $ 11.84 $ \par $ (11.70, 12.17) $&
				\centering  $ (11.6, 12.72) $&
				&
				\centering $ 1.260 $&
				\centering $ 3.147 $
				\cr
				\centering $ 10 $& \centering $ y_0 $&
				\centering  $ -5 $ &
				\centering  $ 2.38 $ \par $ (1.22, 4.05) $& 
				\centering  $ (-357.8, 62.1) $&
				\centering  $ 2176 $&
				\centering  $ -4.995 $&
				\centering  $ 30.0 $
				\cr
				&\centering   $ y_m $&
				\centering  $ 100 $ &
				\centering  $ 84.6 $ \par $ (-3.0 \cdot 10^{5}, 92.0) $& 
				\centering  $ (-10^{11}, 4 \cdot 10^{7}) $&
				&
				\centering  $ -4.89 $&
				\centering  $ 30 $.79
				\cr
				&\centering  $ K_m $&
				\centering  $ 0.01 $ &
				\centering  $ -0.007 $ \par $ (-0.023, 0.007) $&
				\centering  $ (-0.334, 0.363) $&
				&
				\centering  $ -0.352 $&
				\centering  $ 8.142 $
				\cr
				&\centering  $ n $&
				\centering  $ 15 $ &
				\centering  $ 9.46 $ \par $ (9.45, 9.48) $&
				\centering  $ (9.3, 9.8) $&
				&
				\centering  $ -0.352 $&
				\centering  $ 8.142 $
				\cr
				\centering  $ 100 $& \centering $ y_0 $&
				\centering  $ -5  $&
				\centering  $ -4.19 $ \par $ (-4.50, -3.88) $& 
				\centering  $ (-1925.1, 972.1) $&
				\centering  $ 1060 $&
				\centering  $ -12.2 $&
				\centering  $ 255.4 $
				\cr
				& \centering $ y_m $&
				\centering  $ 100 $&
				\centering  $ 97.61 $ \par $ (96.66, 98.39) $&
				\centering  $ (-9 \cdot 10^{16},\; 10^{18}) $&
				&
				\centering  $ 25.6 $&	
				\centering  $ 669.7 $
				\cr
				& \centering  $ K_m $&
				\centering  $ 0.01 $&
				\centering  $ 0.006 $ \par $ (0.003, 0.009) $&
				\centering  $ (-19.2, 9.8) $&
				&
				\centering  $ -12.2 $&
				\centering $  255.4 $
				\cr
				&\centering $ n $&
				\centering $  15 $ &
				\centering  $ 14.99 $ \par $ (14.99, 15.00) $&
				\centering  $ (-4.2, 24.8) $&
				&
				\centering  $ 106.3 $&
				\centering  $ 255.4 $
				\cr
				\centering  $ 2000 $ &\centering  $ y_0 $&
				\centering  $ -5 $ &
				\centering  $ 26.5 $ \par $ (26.2, 26.8) $&
				\centering  $ (-1.9 \cdot 10^{4},\; 10^{5}) $&
				\centering  $ 2010 $&
				\centering  $ 32.9 $&
				\centering  $ 12272.1 $
				\cr
				&\centering  $ y_m $&
				\centering $  100 $&
				\centering  $ 59.3 $ \par $ (28.4, 63.6) $&
				\centering  $ ( -3 \cdot 10^{61}, \; 3 \cdot 10^{61}) $&
				&
				\centering $  -113.6 $&
				\centering  $ 13261.5 $
				\cr 
				&\centering $ K_m $&
				\centering  $ 0.01 $&
				\centering  $ 0.07 $ \par $ (0.06, 0.07) $&
				\centering  $ (-197.2, 1105.3) $&
				&
				\centering  $ 106.3 $&
				\centering  $ 12272.1 $
				\cr
				&\centering$ n $&
				\centering  $ 15 $ &
				\centering  $ 31.98 $ \par $ (31.97, 31.98) $&
				\centering  $ (-165.2, 1137.2) $&
				&
				\centering  $ 106.3 $&
				\centering  $ 12272.1 $\cr
				\hhline{*{8}=} 
			\end{supertabular} 
		\end{small}
	\end{center}
	\begin{footnotesize}
			Table \ref{T:HillMonCau3} presents results calculated for \textit{\textbf{exactly the same }}Monte Carlo data used in Table \ref{T:HillMonCau2}, but starting the simplex optimization with $ \{y_0=10\%,\; y_m=100\%,\; K_m=0.1,\; \mathbf{ \mathit{n}=10}  \} $. Parameters and heading have same meaning as used in Table \ref{T:SkwKurt}; $ r $ indicates the number of random Cauchy values calculated with Eq. (\ref{E:SimCauchyHill}) were simulated for each concentration. Please notice that the units of $ K_m $ are irrelevant as long as they are equal to the units of [D]. Other conditions as in Table \ref{T:HillMonCau1}. Please notice that $ n \vrdif 15 $ in all instances. See the text of the communication for further discussion.
	\end{footnotesize}
\end{table}

\begin{table}  
	\begin{center}
		\begin{small}
		\topcaption{\textbf{Fitting Gauss data ($ \varPsi_{HG} $) generated with Eq. 
				(\ref{E:GaussHill}) setting $ \mathbf{\varsigma = 0.05} $ and simplex optimization to the modified Hill equation.  In all cases the optimization started with $ \mathbf{y_0 = -10\%} $, $\mathbf{y_m = 100\%}$, $\mathbf{K_m=0.1}$ and $\mathbf{n=2}$; $ \Delta_{init}$ was set as 0.1.}}\label{T:HillMonGaus1}
			\begin{supertabular}{
					*{1}{m{0.8cm}}
					*{1}{m{0.8cm}}
					*{1}{m{1.5cm}}
					*{1}{m{3.cm}}
					*{1}{m{3.5cm}}
					*{1}{m{1.5cm}}
					*{2}{m{1.cm}}}
				\hhline{*{8}=}
				\centering$r$&
				\centering $ \theta_j $&
				\centering \textbf{Sim}&
				\centering \textbf{Pred}&
				\centering \textbf{Range} &
				\centering \textbf{Loops}&
				\centering \textbf{\textit{Sk}}&
				\centering \textbf{\textit{Kr}} \cr
				\hhline{*{8}-}	
				\centering $ 3 $& $ y_0 $ *&
				\centering  $ -5  $&
				\centering  $ -3.68 $ \par $ (-6.44, -10.16) $&
				\centering  $ (-14.03, 9.22) $&
				\centering $ 230969 $&
				\centering $ 0.314 $&
				\centering $ 2.944 $
				\cr
				&\centering $ y_m $&
				\centering  $ 100  $&
				\centering $ 82.8 $ \par $ (-937.4, 71.8) $& 
				\centering  $ (-4 \cdot 10^7,\; 2 \cdot 10^{7}) $&
				&
				\centering  $ -1.864 $&
				\centering  $ 11.1 $
				\cr
				&\centering  $ K_m $*&
				\centering $ 0.15 $&
				\centering  $ 0.12 $ \par $ (0.09, 0.15) $&
				\centering  $ (0.02, 0.25) $&
				&
				\centering $ 0.314 $&
				\centering $ 2.944 $
				\cr
				&\centering $ n $*&
				\centering  $ 2 $ &
				\centering $ 3.20 $ \par $ (3.17, 3.22) $&
				\centering  $ (3.1, 3.3) $&
				&
				\centering $ 0.314 $&
				\centering $ 2.944 $
				\cr
				\centering $ 10 $& \centering $ y_0 $*&
				\centering  $ -5 $ &
				\centering  $ -4.09 $ \par $ (-5.48, -2.68) $& 
				\centering  $ (-152.2, 7.5) $&
				\centering  $ 6253 $&
				\centering  $ -0.167 $&
				\centering  $ 2.277 $
				\cr
				&\centering  $ y_m $&
				\centering  $ 100 $ &
				\centering  $ 101.7 $ \par $ (65.2, 141.1) $& 
				\centering  $ ( -10^5, 1.7 \cdot 10^{5}) $&
				&
				\centering $ 0.946 $&
				\centering  $ 11.7 $
				\cr
				&\centering $ K_m $*&
				\centering  $ 0.15 $ &
				\centering  $ 0.15 $ \par $ (0.14, 0.16) $&
				\centering  $ (0.04, 0.27) $&
				&
				\centering  $ -0.167 $&
				\centering  $ 2.277 $
				\cr
				&	\centering $ n $*&
				\centering  $ 2 $ &
				\centering  $ 1.91 $ \par $ (1.90, 1.93) $&
				\centering  $ (1.8, 2.0) $&
				&
				\centering  $ -0.167 $&
				\centering  $ 2.277 $
				\cr
				\centering  $ 100 $ &\centering  $ y_0 $*&
				\centering $  -5 $ &
				\centering $ -4.96 $ \par $( -5.33, -4.58) $ & 
				\centering  $ (-20.3, 9.7) $&
				\centering  $ 745 $&
				\centering  $ -0.069 $&
				\centering  $ 2.912 $
				\cr
				& \centering $ y_m $&
				\centering  $ 100 $&
				\centering  $ 97.5 $ \par $ (94.0, 100.9) $ &
				\centering  $ ( -6 \cdot 10^5, 5 \cdot 10^5) $&
				&
				\centering $ -0.081 $&	
				\centering  $ 21.6 $
				\cr
				&\centering  $ K_m $*&
				\centering  $ 0.15 $&
				\centering  $ 0.143 $ \par $ (0.139, 0.146) $&
				\centering $  (-0.01, 0.29) $&
				&
				\centering $  -0.069 $&
				\centering  $ 2.912 $
				\cr
				&\centering  $ n $*&
				\centering  $ 2 $ &
				\centering  $ 2.121 $ \par $ (2.117, 2.124) $&
				\centering  $ (2.0,4 2.3) $&
				&
				\centering $  -0.069 $&
				\centering  $ 2.912 $
				\cr
				\centering $ 2000 $&\centering $ y_0 $*&
				\centering  $ -5  $&
				\centering  $ -5.18 $ \par $ (-5.27, -5.09) $&
				\centering  $ (-5.3 -5.1) $&
				\centering  $ 1024002 $*&
				\centering  $ 0.032 $&
				\centering  $ 2.951 $
				\cr
				&\centering  $ y_m $&
				\centering  $ 100 $&
				\centering  $ 101.9 $ \par $ (101.2, 102.7) $&
				\centering  $ (-3 \cdot 10^5, 4 \cdot 10^5) $&
				&
				\centering  $ 0.250 $&
				\centering  $ 20.1 $
				\cr 
				&\centering  $ K_m $*&
				\centering  $ 0.15 $&
				\centering  $ 0.152 $ \par $ (0.151, 0.152) $&
				\centering  $ (-0.05, 0.34) $&
				&
				\centering  $ 0.032 $&
				\centering  $ 2.951 $
				\cr
				& \centering  $ n $*&
				\centering  $ 2 $ &
				\centering  $ 2.013 $ \par $  (2.012, 2.014) $&
				\centering  $ (1.8, 2.2) $&
				&
				\centering  $ 0.032 $&
				\centering  2$ .951 $
				\cr
				\hhline{*{8}=} 
			\end{supertabular} 
		\end{small}
	\end{center}
	\begin{footnotesize}
			Parameters and heading have same meaning as used in Table \ref{T:SkwKurt}; $ r $ indicates the number of random Gaussian values of type $N(\mu, 0.05 \cdot \mu)$ with $\mu=y \left( [D] \given \{y_0,y_m,K_m,n\}\right)$ which were simulated for each concentration. Parameter names ($ \theta_j $) with an asterisk indicated that Gaussianity cannot be ruled out when Jarque-Bera test is used ($ 0.95 > P > 0.05 $), for all $y_m$\textquoteright{s} $ P<10^{-6} $ using the same test. Number of loops with an asterisk indicates the optimization stopped after fulfilling Condition \ref{Cnd:TooLong}.  Other conditions as in Table \ref{T:HillMonCau1}, see the text of the communication for further discussion.
	\end{footnotesize}
\end{table}

\begin{table}
	\begin{center}
		\begin{small}
			\topcaption{\textbf{Fitting Gauss data ($ \varPsi_{HG} $) generated with Eq. 
					(\ref{E:GaussHill}) setting $ \mathbf{\varsigma = 0.05} $ and simplex optimization to the modified Hill equation.  In all cases the optimization started with $ \mathbf{y_0 = -10\%} $, $\mathbf{y_m = 100\%}$, $\mathbf{K_m=0.1}$ and $\mathbf{n=2}$; $ \Delta_{init}$ was set as 0.1.}}\label{T:HillMonGaus2}
				\begin{supertabular}{
					*{1}{m{0.8cm}}
					*{1}{m{0.8cm}}
					*{1}{m{1.5cm}}
					*{1}{m{3.cm}}
					*{1}{m{3.5cm}}
					*{1}{m{1.5cm}}
					*{2}{m{1.cm}}}
				\hhline{*{8}=}
				\centering$r$&
				\centering $ \theta_j $&
				\centering \textbf{Sim}&
				\centering \textbf{Pred}&
				\centering \textbf{Range} &
				\centering \textbf{Loops}&
				\centering \textbf{\textit{Sk}}&
				\centering \textbf{\textit{Kr}} \cr
				\hhline{*{8}-}				
				\centering  $ 3 $&\centering  $ y_0 $*&
				\centering  $ -5 $ &
				\centering  $ -9.3 $ \par $ (-10.9, -5.6) $&
				\centering  $ (-19.9, 3.6) $&
				\centering $ 70873 $&
				\centering $ 0.420 $&
				\centering $ 2.941 $
				\cr
				&\centering$  y_m $&
				\centering  $ 100 $ &
				\centering  $ 109.2 $ \par $ (104.1, 115.9) $& 
				\centering  $ (-902.9, 4401.8) $&
				&
				\centering  $ 3.693 $&
				\centering  $ 16.4 $
				\cr
				&\centering$ K\_m $*&
				\centering  $ 0.01 $ &
				\centering  $ 0.02 $ \par $ (0.00, 0.05) $&
				\centering  $ (-0.09, 0.14) $&
				&
				\centering $ 0.420 $&
				\centering $ 2.941 $
				\cr
				&\centering$ n $*&
				\centering  $ 15 $ &
				\centering  $ 2.55 $ \par $ (2.54, 2.59) $&
				\centering  $ (2.5 2.7) $&
				&
				\centering $ 0.420 $&
				\centering $ 2.941 $
				\cr
				\centering $ 10 $&\centering $ y_0 $*&
				\centering  $ -5 $ &
				\centering  $ -6.40 $ \par $ (-7.72, -5.16) $& 
				\centering  $ (-14.2, -0.11) $&
				\centering  $ 17486 $&
				\centering  $ -0.381 $&
				\centering $  2.843 $
				\cr
				&\centering$ y_m $&
				\centering  $ 100 $ &
				\centering  $ 101.6 $ \par $ (98.7, 104.2) $& 
				\centering  $ (5 \cdot 10^4, 5 \cdot 10^4) $&
				&
				\centering  $ 1.089 $&
				\centering  $ 14.2 $
				\cr
				&\centering$ K_m $*&
				\centering  $ 0.01 $ &
				\centering  $ 0.005 $ \par $ (-0.008, 0.01747) $&
				\centering  $ (-0.13, 0.11) $&
				&
				\centering  $ -0.381 $&
				\centering  $ 2.843 $
				\cr
				&	\centering  $ n $*&
				\centering  $ 15 $ &
				\centering  $ 3.71 $ \par $ (3.69, 3.72) $&
				\centering  $ (3.57, 3.81) $&
				&
				\centering  $ -0.381 $&
				\centering $  2.843 $
				\cr
				\centering$ 100 $&\centering $ y_0 $*&
				\centering  $ -5 $ &
				\centering  $ -6.32 $ \par $ (-6.71, -5.95) $& 
				\centering  $ (-20.9, 10.3) $&
				\centering  $ 31287 $&
				\centering  $ 0.037 $&
				\centering  $ 2.950 $
				\cr
				&\centering$ y_m $&
				\centering  $ 100 $&
				\centering  $ 101.4 $ \par $ (100.5, 102.3) $ &
				\centering  $ (-4 \cdot 10^5, 4 \cdot 10^5) $&
				&
				\centering  $ 0.997 $&	
				\centering  $ 18.0 $
				\cr
				&\centering  $ K_m $*&
				\centering  $ 0.01 $&
				\centering $  0.010  $ \par $ (0.006, 0.013) $&
				\centering  $ (-0.14, 0.18) $&
				&
				\centering  $ 0.037 $&
				\centering  $ 2.950 $
				\cr
				&\centering  $ n $*&
				\centering  $ 15 $ &
				\centering  $ 4.604 $ \par $ (4.601, 4.608) $&
				\centering  $ (4.5, 4.8) $&
				&
				\centering $ 0.037 $&
				\centering  $ 2.950 $
				\cr
				\centering$ 2000 $&\centering$ y_0\bullet $&
				\centering  $ -5 $ &
				\centering  $ 2.97 $ \par $ (-5.46, 11.43) $&
				\centering  $ (-2 \cdot 10^3, 2 \cdot 10^3) $&
				\centering  $ 1106 $1&
				\centering  $ 0.046 $&
				\centering  $ 3.019 $
				\cr
				&\centering  $ y_m $&
				\centering  $ 100 $&
				\centering  $ 92.3 $ \par $ (71.6, 113.0 $)&
				\centering  $ (-2 \cdot 10^6, 2 \cdot 10^6) $&
				&
				\centering $ 0.028 $&
				\centering  $ 20.8 $
				\cr 
				&\centering  $K_m \bullet$&
				\centering  $ 0.01 $&
				\centering  $ -0.013 $ \par $ ( -9.78, 7.11) $&
				\centering  $ (-19.5, 19.8) $&
				&
				\centering  $ 0.046 $&
				\centering  $ 3.019 $
				\cr
				&\centering  $n\bullet$&
				\centering  $ 15  $&
				\centering  $ 2.8 $ \par $ (2.7, 2.9) $&
				\centering  $ (-16.8, 22.5) $&
				&
				\centering  $ 0.046 $&
				\centering  $ 3.019 $
				\cr
				\hhline{*{8}=} 
			\end{supertabular} 
		\end{small}
	\end{center}
	\begin{footnotesize}
			Parameter names ($ \theta_i $) with an asterisk indicated that Gaussianity cannot be ruled out when Jarque-Bera test is used ($ 0.95 > P > 0.05 $), for all $y_m$\textquoteright{s} $ P\ll10^{-6} $ using the same test. Parameter names with a bullet ($\bullet$) indicate weak not Gaussianity ($ P=0.037 $, Jarque-Bera test). Number of loops with an asterisk indicates the optimization stopped after fulfilling Condition \ref{Cnd:TooLong}.  Other conditions as in Table \ref{T:HillMonCau1}. Please notice that $ n \vrdif 15 $ in all instances. See the text of the communication for further discussion.
	\end{footnotesize}
\end{table}

\begin{table}
	\begin{center}
		\begin{small}
			\topcaption{\textbf{Fitting Gauss data ($ \varPsi_{HG} $) generated with Eq. (\ref{E:GaussHill}) setting $ \mathbf{\varsigma = 0.05} $ and simplex optimization to the modified Hill equation.  In all cases the optimization started with $ \mathbf{y_0 = -10\%} $, $\mathbf{y_m = 100\%}$, $\mathbf{K_m=0.1}$ and $\mathbf{n=10}$. }}\label{T:HillMonGaus3}
				\begin{supertabular}{
					*{1}{m{0.8cm}}
					*{1}{m{0.8cm}}
					*{1}{m{1.5cm}}
					*{1}{m{3.cm}}
					*{1}{m{3.5cm}}
					*{1}{m{1.5cm}}
					*{2}{m{1.cm}}}
				\hhline{*{8}=}
				\centering$r$&
				\centering $ \theta_j $&
				\centering \textbf{Sim}&
				\centering \textbf{Pred}&
				\centering \textbf{Range} &
				\centering \textbf{Loops}&
				\centering \textbf{\textit{Sk}}&
				\centering \textbf{\textit{Kr}} \cr
				\hhline{*{8}-}\cr			
				\centering $ 3 $&\centering $y_0 \bullet$&
				\centering  $ -5 $ &
				\centering  $ 1.50 $ \par $ (-1.06, 1.67) $&
				\centering  $ (-11.0, 50.8) $&
				\centering $ 38091 $&
				\centering $ 1.716 $&
				\centering $ 4.595 $
				\cr
				&\centering $ y_m $&
				\centering $  100 $ &
				\centering  $ 93.74 $ \par $ (-1.7 \cdot 10^{19}, 1.8 \cdot 10^{6} $)& 
				\centering  $ (10^{23}, 2 \cdot 10^{11} $)&
				&
				\centering  $ -4.072 $&
				\centering  $ 18.0 $
				\cr
				&\centering  $K_m \bullet$&
				\centering  $ 0.01 $ &
				\centering  $ 0.20  $\par$  (0.06, 0.52) $&
				\centering  $ (-0.06, 1.07) $&
				&
				\centering $ 1.260 $&
				\centering $ 3.147 $
				\cr
				&\centering $n \bullet$&
				\centering  $ 15 $ &
				\centering  $ 1.42 $ \par $ (1.40, 1.48) $&
				\centering  $ (1.4, 1.6) $&
				&
				\centering $ 0.704 $&
				\centering $ 2.905 $
				\cr
				\centering $ 10 $&\centering $ y_0 $&
				\centering  $ -5 $ &
				\centering  $ 0.46 $ \par $ (-1.40, 3.04) $& 
				\centering  $ (-9.8, 53.0) $&
				\centering  $ 8577 $&
				\centering  $ 1.751 $&
				\centering  $ 4.582 $
				\cr
				&\centering$ y_m $&
				\centering  $ 100 $ &
				\centering  $ 94.5 $ \par $ (-6 \cdot 10^{12}, 98.8) $& 
				\centering  $ (-4\cdot 10^{23},\; 5 \cdot 10^{15} $)&
				&
				\centering  $ -2.854 $&
				\centering  $ 9.716 $
				\cr
				&\centering  $ K_m $&
				\centering  $ 0.01 $ &
				\centering  $ 0.03 $ \par $ (0.01, 0.06) $&
				\centering  $ (-0.07, 0.56) $&
				&
				\centering  $ 1.751 $&
				\centering  $ 4.582 $
				\cr
				&\centering  $ n $&
				\centering  $ 15 $ &
				\centering  $ 16.6 $ \par $ (16.5, 16.6) $&
				\centering  $ (16.5, 17.1) $&
				&
				\centering  $ 1.751 $&
				\centering  $ 4.582 $
				\cr
				\centering  $ 100 $&\centering  $ y_0 $&
				\centering  $ -5 $ &
				\centering  $ 15.0 $ \par $ (12.4, 18.9) $& 
				\centering  $ (-19.7, 108.4) $&
				\centering  $ 1262 $&
				\centering  $ 1.264 $&
				\centering  $ 3.355 $
				\cr
				&\centering  $ y_m $&
				\centering  $ 100 $&
				\centering  $ 78.9 $ \par $ (-3 \cdot 10^{78}, 82.4)  $&
				\centering  $ (-3 \cdot 10^{107},\; 10^{48}) $&
				&
				\centering  $ -2.558 $&	
				\centering  $ 8.519 $
				\cr
				&	\centering$ K_m $&
				\centering  $ 0.01 $&
				\centering $ 0.10 $ \par $ (0.08, 0.14) $&
				\centering  $ (-0.2, 1) $&
				&
				\centering  $ 1.264 $&
				\centering  $ 3.355 $
				\cr
				&\centering  $ n $&
				\centering  $ 15 $ &
				\centering  $ 60.1 $ \par $ (60.1, 60.2 $&
				\centering  $ (60, 61) $&
				&
				\centering  $ 1.264 $&
				\centering  $ 3.355 $
				\cr
				\centering  $ 2000 $&\centering  $ y_0 $&
				\centering  $ -5 $ &
				\centering  $ 19.2$  \par  $(18.8, 19.6) $&
				\centering  $ (-22.5, 112.7) $&
				\centering  $ 3226 $&
				\centering  $ 1.140 $&
				\centering  $ 3.261 $
				\cr
				&\centering  $ y_m $&
				\centering  $ 100 $&
				\centering  $ 72.7 $ \par $ (68.9, 74.0) $&
				\centering  $ ( -4 \cdot 10^{84},\; 5 \cdot 10^{39} ) $&
				&
				\centering  $ -2.280 $&
				\centering  $ 6.609 $
				\cr 
				&\centering  $ K_m $&
				\centering  $ 0.01 $&
				\centering  $ 0.094  $\par $ (0.091, 0.098) $&
				\centering  $ (-0.32, 1.03) $&
				&
				\centering  $ 1.140 $&
				\centering  $ 3.261 $
				\cr
				&\centering  $ n $&
				\centering  $ 15 $ &
				\centering  $ 45.1 $ \par $ (45.0, 45.1) $&
				\centering  $ (44.6, 46.0) $&
				&
				\centering  $ 1.140 $&
				\centering  $ 3.261 $
				\cr
				\hhline{*{8}=} 
			\end{supertabular} 
		\end{small}
	\end{center}
		\begin{footnotesize}
			Table \ref{T:HillMonGaus3} presents data calculated for \textit{exactly the same }Monte Carlo data used in Table \ref{T:HillMonGaus2}, but starting the simplex optimization with $ \{y_0=10\%,\; y_m=100\%,\; K_m=0.1,\; n =10  \} $.  and heading have same meaning as used in Table \ref{T:SkwKurt}; $ r $ indicates the number of random Gaussian values of type $N(\mu, \sigma^2=0.05 \cdot \mu)$ with $\mu=y(x_i \given \{V_{\text{\textonehalf}},\kappa\}) = \tfrac{1}{1+e^{-(x_i-V_{\text{\textonehalf}})/\kappa}} $ which were simulated for each concentration. Parameter names ($ \theta_j $) with an bullet ($\bullet$) indicated that the probability of Gaussianity when Jarque-Bera test is used  is $ P \approxeq 9 \cdot 10 ^{-4}$, for all parameters without a bullet $ P\ll10^{-6} $ using the same test. Please notice that $ n \vrdif 15 $ in all instances. See the text for further discussion.
	\end{footnotesize}
\end{table}

\subsubsection{Using the Boltzmann distribution function to fit Cauchyan data.}\label{S:BoltzCau}

The most common use of the Boltzmann function in electrophysiology is to fit normalized ionic currents \cite{Hodgkin1952e,Peigneur2012,Sevcik2017a}. To do this, ionic currents are measured at a broad range of membrane potentials, and the currents recorded are divided by the maximum value observed at the most negative potentials tested, in the case of sodium current in excitable cells. Normalized currents (or any other parameter) which results from variate quotients are likely to obey a Cauchy resembling distribution.

Table \ref{T:BoltzMonCau} present data calculated as in Table \ref{T:BoltzMonGau2} except for the data fit to the Boltzmann equation wich was distributed a Cauchy, generated as
\begin{equation}
	\begin{split}
		\varPsi_{\mf{BC}} \left\langle V_{i} \given \left[ \gamma,y \left( V_{i} \given \{ V_{\text{\textonehalf}},\kappa \}\right ) \right] \right\rangle_{i=1, \, \dots \,,m}   & = \cdots \\ 
		\cdots = \left\lbrace  \frac{1}{1+\exp{\left( -\frac{V_{i}-V_{\text{\textonehalf}}}{\kappa}\right)}}+ \gamma \cdot\tan \left( \pi \cdot \left[U(0,1)_i-\tfrac{1}{2}\right] +1\right)   \right \rbrace_{i=1, \, \dots \,,m},
	\end{split}
\end{equation} 
$ \gamma = \tfrac{2}{50}$  for all data in Table  \ref{T:BoltzMonCau}. Comparing Tables \ref{T:BoltzMonGau2} and \ref{T:BoltzMonCau} may be appreciated that data predicted did not differ much whether the input is Gaussian or Cauchyan when fitted to the Boltzmann equation. As it was the case with Gaussian data, the fit with Cauchyan data was faster and better using $ \Delta_{init}=0.5 $. In all cases sample $ Kr > 3 $ and sample $ Sk < 0 $ for predicted $ V_{\text{\textonehalf}} $. Jarque-Bera and Shapiro-Wilks tests indicated that predicted parameters were not Gaussian  ($ P < 10^{-6} $).

\begin{table}
	\begin{center}
		\begin{small}
			\topcaption{\textbf{Fitting Gaussian data generated Eq. (\ref{E:BoltzGau}) to the Boltzmann equation with the simplex optimization. In all cases the optimization started with $ \mathbf{V_{\text{\textonehalf}}=-60}, $ $\mathbf{\kappa=20}$ and $ \mathbf{\Delta_{init}}=0.5. $. }}\label{T:BoltzMonCau}
					\begin{supertabular}{
						*{1}{m{0.8cm}}
						*{1}{m{0.8cm}}
						*{1}{m{1.5cm}}
						*{1}{m{3.cm}}
						*{1}{m{3.5cm}}
						*{1}{m{1.5cm}}
						*{2}{m{1.cm}}}
					\hhline{*{8}=}
					\centering$r$&
					\centering $ \theta_j $&
					\centering \textbf{Sim}&
					\centering \textbf{Pred}&
					\centering \textbf{Range} &
					\centering \textbf{Loops}&
					\centering \textbf{\textit{Sk}}&
					\centering \textbf{\textit{Kr}} \cr
					\hhline{*{8}-}\cr			
					\centering $ 3 $&\centering $V_{\tfrac{1}{2}}$&
					\centering  $ -40 $ &
					\centering $ -40.395  $ \par $ (-40.399, -40.394) $&
					\centering  $ (-40.419, -40.394) $&
					\centering $ 210 $&
					\centering $ -1.866 $&
					\centering $ 5.381 $
					\cr
					&\centering $\kappa$ &
					\centering  $ 10 $ &
					\centering  $ 9.631 $ \par $ (9.630, 9.632) $& 
					\centering  $ (9.610, 9.643) $&
					&
					\centering  $ -0.181 $&
					\centering  $ 4.104 $
					\cr
					\centering  $ 10 $& \centering $V_{\tfrac{1}{2}}$&
					\centering  $ -40 $ &
					\centering  $ -39.985 $ \par $ (-39.989, -39.984) $& 
					\centering  $ (-40.007, -39.983) $&
					\centering  $ 508 $&
					\centering  $ -1.676 $&
					\centering  $ 4.672 $
					\cr
					&\centering $\kappa$&
					\centering  $ 10 $ &
					\centering  $ 10.5380 $ \par $ (10.5378, 10.5380) $& 
					\centering  $ (10.5176, 10.5584) $&
					&
					\centering  $ 0.011 $&
					\centering  $ 4.232 $
					\cr
					\centering$ 100 $ & \centering $V_{\tfrac{1}{2}}$&
					\centering  $ -40 $ &
					\centering  $ -40.055 $ \par $ (-40.058, -40.054) $& 
					\centering  $ (-40.079, -40.053) $&
					\centering  $ 317 $&
					\centering $  -1.777 $&
					\centering  $ 5.045 $
					\cr
					&\centering  $\kappa$&
					\centering  $ 10 $&
					\centering  $ 10.0451 $ \par $ ( 10.0451, 10.0452) $ &
					\centering  $ (10.024, 10.066) $&
					&
					\centering  $ 0.017 $&	
					\centering  $ 4.437 $
					\cr
					\centering $ 	2000 $ & \centering $V_{\tfrac{1}{2}}$&
					\centering  $ -40 $ &
					\centering  $ -39.978 $ \par $ (-39.978, -39.978) $&
					\centering  $ (-40.00, -39.98) $&
					\centering  $ 635 $&
					\centering  $ -1.790 $&
					\centering  $ 5.094 $
					\cr
					&\centering  $\kappa$&
					\centering  $ 10 $&
					\centering  $ 9.9841 $ \par $ ( 9.9841, 9.9841) $&
					\centering  $ (9.96, 10.01) $&
					&
					\centering  $ 0.005 $&
					\centering  $ 4.461 $
					\cr 
					\hhline{*{8}=} 
				\end{supertabular} 
		\end{small}	
	\end{center}
	\begin{footnotesize}
			Table heading have same meaning as used in Table \ref{T:SkwKurt}; $ r $ indicates the number of random Gaussian values of type $N(\mu, \sigma^2=0.05 \cdot \mu)$ with $\mu=y(x_i \given \{V_{\text{\textonehalf}},\kappa\}) = \tfrac{1}{1+e^{-(x_i-V_{\text{\textonehalf}})/\kappa}} $ which were simulated for each concentration. See the text for further discussion. Other details as in Table \ref{T:BoltzMonGau1}.
\end{footnotesize}
\end{table}

\section{Concluding remarks.}\label{S:ConcRem}

Estimating the uncertainty of objective function parameters which are not linearly independent is a challenging problem of regression analysis \cite{Seber1989, Schittkowski2002}.  Iterative processes used in nonlinear optimization need a starting set of parameters $ \{ \theta\}_{init} $ which, if close enough to the global maximum or minimum, enables the algorithm to render $  \vect{\{\theta\}}_{opt} $, the best possible set of parameters and to minimize residual differences between empirical points and the objective function. Part of the difficulty is the existence of local minima or maxima towards which the iterative optimization processes (such as the simplex algorithm \cite{Nelder1965}) may converge, failing to reach the global minimum or maximum. At any of these local minima or maxima the objective function gradient respect to the independent variable(s) becomes null.

Estimating parameter uncertainties in linear regression analysis may be simpler, and is usually done by least squares analysis (also called $ \ell_2$-\textit{norm}) which minimizes the sum of residuals squares and produces a set of simultaneous linearly independent equations, which may be solved to determine regression parameters. The least squares procedure has the pitfall of giving undue weight to outliers. The undue weight of outliers may be prevented by minimizing the sum of absolute values of residuals (also called $ \ell_1$\textit{-norm}) but it has no analytical solution for neither parameter  nor parameter uncertainty determination \cite{Cadzow2002, Donoho2006}.

In many real word situations the fluctuating nature of the \textit{obf} makes lots of sense. Hill and Boltzmann equations are both used to describe interactions between particles or molecules, the structures of those molecules and their interactions fluctuate at any temperature distinct from 0\textdegree{}K \cite{Hill1910b, Hill1913, Sevcik2017a}. At the subatomic level, quantum physics is totally based on random processes \cite{Ballentine1970}. Neurotransmitter release is a Poisson process \cite{Augustine2007} and cell physiology is critically dependent on random cell membrane ionic permeability changes \cite{Neher1976}. As indicated by this small and arbitrary selection of physical realities indicates, demanding that the \textit{obf} is static and reality fluctuates randomly about it, is only an arbitrary choice.

Even at global optimum of a regression, residual differences between the objective function and empirical data remain. Here it is proposed that residuals may be seen as a measure of uncertainty of an objective function to describe a set of empirical data. That is, empirical data are taken as variables produced by the \textit{obf} which fluctuates randomly describing fuzzily the relation between dependent and independent variables. Fluctuations remain no matter if we know the objective function\textquoteright{s} parameters at the global optimum. 

Sets of  empirical variables \cite{Quintana2017} and two functions widely used to  describe data in science, the Boltzmann \cite{Boltzmann1896,Boltzmann1964}  and the Hill functions \cite{Hill1910b, Hill1913} are used here to evaluate the \textit{first derivative at the optimum} analysis (\textit{fdao}) usefulness. The Boltzmann function was  used in a form common in electrophysiology [Eq. (\ref{E:BoltzElect})] \cite{Hodgkin1952e, Cronin1987, Sevcik2017a} and Hill function [Eq. (\ref{E:HillMod})] modified to include shift in the baseline often occurring in  experimental situations. The first derivatives of those functions and the analysis at the optimum properties (Sections \ref{S:HillFstDer} and \ref{S:BoltzFstDer}.

In Section \ref{S:CitotDat} empirical data shown in Figures \ref{F:FracNoFit}  and \ref{F:FracFit}, as well as Tables \ref{T:HillSimplexPar} and   \ref{T:SkwKurt} present median values ($\bullet$) and 95\% confidence intervals (CI, bars) of anti-neoplastic effects produced by compounds isolated from \textit{P.  constellatum} ({S}avigny, 1816), a marine  animal \cite{Quintana2017}. As  seen in Figure \ref{F:FracNoFit}, 95\% CI are very asymmetric and  broad suggesting negative outliers.  

When data in Fig. \ref{F:FracNoFit} was plotted as in Fig. \ref{F:FracFit}, clipping the lower axis at -20\%, a sigmoid resemblance of the median data at the different concentrations became evident. Lines in Fig. \ref{F:FracFit} were drawn using Eq. (\ref{E:HillMod}) fitted using simplex optimization (described in Sections \ref{S:CurveFit} and \ref{S:CitotDat}), are close to the median determined at each concentration. The parameters used are in Table \ref{T:HillSimplexPar}, and some additional sample statistical properties are in Table \ref{T:SkwKurt}. As it would be expected if the data would be Cauchy- or Cauchy-like-distributed, parameter ranges fluctuate between wide and huge, data appear to be strongly skewed and very leptokurtic. In case of fraction FIV the simplex stopped on Condition \ref{Cnd:TooLong} since Condition \ref{Cnd:StopOpt} could not be achieved in $ \approxeq 3 $ h, but in the other cases Condition \ref{Cnd:StopOpt} was reached with $ \leqq 21432 $ algorithm iterations in few minutes. Perhaps the most interesting feature of the parameters describing the curves in Table \ref{T:HillSimplexPar} is that the 95\% CIs are narrow in spite of the parameter ranges, this can indeed be expected if the parameters are strongly leptokurtic and mostly packed around the medians as it is the case for the Cauchy distribution.

Figure \ref{F:CauchyDis} presents Cauchy probability density [Eq.  (\ref{E:DensCauchy})] function (pdf) and the Cauchy PDF [Eq. (\ref{E:DistribCauchy})] calculated with $ \gamma = 1/50 $  and $ \widehat{\mu} =0 $.  The figure also depicts several empirical probability  distribution \cite{Shorack1986,Pitman1993} curves estimated for FIII at diverse  concentrations. This curves were selected because are representative of the ones  obtained with other fractions. Ccomparing panels  \ref{F:CauchyDis}B and \ref{F:CauchyDis}C in the figure, it is apparent that there is a good  agreement between the empirical PDF and the Cauchy PDF. To check how does the ve analysis performs when applied to Cauchy data, Cauchyan [D] values were generated as explained in Section \ref{S:ModCauhData} and used as input to Eq. (\ref{E:HillMod}) using variable $ \theta_{j} $ and $ \Delta_{init} $ starting increments for the simplex algorithm, the results are summarized in Tables \ref{T:HillMonCau1} to \ref{T:HillMonCau3}. The empirical data \cite{Quintana2017} and parameters determined imputing Cauchy variables to the Hill equation [Eq. (\ref{E:HillMod})] have several characteristics in common. Parameter ranges are broad, and extremely so in some cases, the parameters are apparently skewed and leptokurtic but the median parameter 95\% CI are relatively narrow. In some cases the optimization stopped on Condition \ref{Cnd:TooLong}, the parameters calculated when this happened seemed \textquotedblleft{}reasonable\textquotedblright{} since they did not look too different from the parameters obtained when the optimization ended on Condition \ref{Cnd:StopOpt}. Although it was not extensively studied here, no $ \{\vect{\theta}\} $ or $ \Delta_{init} $ values prone to produce endings with Condition \ref{Cnd:TooLong} were identified. It is the author\textquoteright{s} impression that ending on Condition \ref{Cnd:TooLong} was more likely with larger sample sizes, and when samples were per chance more disperse and thus harder to optimize.

Optimizations using Cauchy data summarized in Tables \ref{T:HillMonCau1} to \ref{T:HillMonCau3} were all initiated with the same $ \Delta_{init}=0.1 $ but different $ \vect{\{\theta\}}_{init} $. It was a surprising finding that $ y_0, \, y_m \text{ and } K_m $ could be easily determined in the optimizations, but that high $ n $ values were very difficult, if possible, to determine accurately. Still, the fdao analysis provides and explanation to this as discussed in details in Section \ref{S:HillFstDer} and is presented graphically in Figure \ref{F:VariancesOptimum} which shows that the contribution $ n $ to the residuals decreases as $ n $ grows, determining a minor role of $ n $ uncertainty in the optimization process (Figure \ref{F:VariancesOptimum}D). 

Sample theory states that estimates of parameters such as mean, variance, skewedness and kurtosis, become less variable and converge towards population values as sample size grows \cite{Wilks1962}.  Data in Tables \ref{T:HillMonCau1} to \ref{T:HillMonCau3} show that estimated $ Sk $ and $ Kr $ do not converge but grow with sample size. Lack of convergence of sample mean, variance, skewedness and kurtosis for Cauchyan variates reflects that there are no population parameters to converge towards.

The Hill equation is not always used to fit Cauchyan data. Many, if not most, of the situations where the Hill equation is fitted to data, are direct measurements of a drug effect, an enzyme catalytic rate, or gas-metal surface interactions . Hence, no quotients are calculated, and there is no reason to deal with Cauchyan variates. Results of fitting Gaussian data to the modified Hill equation are shown in Tables \ref{T:HillMonGaus1} to \ref{T:HillMonGaus3}. As seen in Tables \ref{T:HillMonGaus1} to \ref{T:HillMonGaus3} even though the data submitted to the simplex optimization were Gaussian, all parameter estimates in the tables are leptokurtic and somewhat skewed, to a degree that all of them tested non-Gaussian with the Jarque-Bera and Shapiro-Wilk tests \cite{Shapiro1965,Jarque1980,  Bera1981, Giles2014}. Some of the parameter ranges were quite wide. Still, in contrast with data in Tables \ref{T:HillMonCau1} to \ref{T:HillMonCau3}, neither $ Sk $ nor $ Kr $ seem to depend on sample sizes in Tables \ref{T:HillMonGaus1} to \ref{T:HillMonGaus3}, which suggests that, whichever their distribution, their central moments are defined, and their sample estimates converge towards population values as sample size grows.

A function also subjected to fdao analysis (Section \ref{S:BoltzMon}) was the Boltzmann distribution function [Eq. (\ref{E:BoltzElect}), revised in \cite{Sevcik2017a}]. Data in Tables \ref{T:BoltzMonGau1} and \ref{T:BoltzMonGau2} shows considerable independence between initial and predicted ($\{ \theta\} $, Section \ref{S:Intro}), and sample size  ($ r $) values. in agreement with the discussion of Eqs. (\ref{E:BoltDerVmed}) and (\ref{E:BoltDerKappa}) and Figure \ref{F:BolVariancesOptimum}. Predicted parameters were also leptokurtic. their distribution had low $ Sk $, and $ Sk $ and $ Kr $ values seemed independent from sample sizes ($ r $) and $ \vect{\{\theta\}}_{init}$.  Ranges of predicted parameters were quite narrow. Thus, the Boltzmann function seems easier to optimize to the correct parameter values that the Hill equation [Eq. (\ref{E:HillMod})]. This agrees with the fdao analysis done in Sections \ref{S:HillFstDer} and \ref{S:BoltzFstDer}. 

\subsection{Caveats on \EqF.}\label{S:CaveatsEqF}

\EqF have several properties we must be aware of:

\begin{enumerate}
	\item \textit{\textbf{\EqF are equal approximations to calculate $ \Delta \theta_i $. It is always reassuring when you reach the same conclusion starting from two separate different premises}}.
	\item The coincide between \EqF suggests that their validity is probably very general, at least if $ \delta_{j,i} $ [Eq. (\ref{E:delta_ji})] is not exceedingly large. 
	\item \EqF are an operations on three random variables of some kind: $ \theta_{i_{opt}}\left(x_i \right) $, $ \Delta \mf{f}(x_i) $, and $ \mf{f}_{\theta_i}' (x_i) $, with, generally, unknown pdfs. 
	\item Even in the (Unlikely?) case that the three variables in \EqF are Gaussian, the ratio
	\begin{equation*}
		\frac{\Delta \mf{f}(x_i)}{\mf{f}_{\theta_i}' (x_i)}
	\end{equation*}
	will most likely have and unknown, probably pathological pdf, with undefined statistical moments and wide outliers such as the Cauchy pdf  \cite{Cramer1991, Pitman1993, Wolfram2003} for which concepts like mean, variance, skewedness and kurtosis are undefined and meaningless (See also section \ref{S:ModCauhData}). 
	\item The use of nonparametric statistics \cite{Holcomb2001} becomes mandatory, since Cauchy-like distributions are symmetric about their median and have nonparametricaly estimable width functions to estimate dispersion [See Eqs (\ref{E:DensCauchy}) and (\ref{E:DistribCauchy})]. The Cauchy pdf has a spiky central part and very broad shoulders. 
\end{enumerate} 

\subsection{A summary of the procedure.}\label{S:Procedure}

Taken together Eqs. (\ref{E:ScaCartGrad}) --- (\ref{E:kappa_ji}) provide a pathway to estimate parameter uncertainty. It will be summarized as follows:
\begin{enumerate}
	\item Fit the  objective function  $\mf{f}\left( x \given \vect{\{\theta_j\}}_{j=1,\cdots,k}\right) $  [$ \mf{f}(x) $, in brief] to $ m $ pairs $ \{x_i, y_i\}_{i=1, \ldots, m} $ pairs of data using an efficient optimization procedure.
	\item  Estimate $  $ optimized model variables $ \theta_j  $ using \EqF  equations.
	\item \textit{\textbf{Now you will have a  supersets, of $ \boldsymbol{l} $ sets, $  \boldsymbol{\left\lbrace  \{\theta_{i,1}\}, \{\theta_{i,2}\}, \ldots, \{\theta_{i,l}\} \right\rbrace}  $, for the set $  \boldsymbol{\{x_i, y_d\}_{d=1,2,\ldots,l}} $ of $ \boldsymbol{l} $ values of $ \boldsymbol{y_d} $ measured at $ \boldsymbol{x_i} $ pairs,  to gauge uncertainty, probably non parametric statistical tools. $ \lqqd $}}
\end{enumerate}

\section{Appendices.}

\begin{appendices}
	\section{Sampling theory and \comillas{pathological }distributions.}\label{E:PholDist}
	
It is generally impossible to study any property of a complete population, if for no other reason, because populations are commonly very large ant the study would be costly, or because the study may de destructive, ie., requiring that the subject is destroyed or killed. This determinates that a subset of the population, as small as possible preserving accuracy, is randomly selected and studied. This subset is a \textit{sample}.Each time a sample is drawn from s population the question remains, as to how well the population is represented by the sample. Most basic statistics textbooks \cite{Lipschutz1998} pent arithmetic expressions for sample mean an variance, mean standard deviation (usually called \textit{s}tandard \textit{e}rror of the \textit{m}ean or \textit{sem}), skewedness and kurtosis [See Eqs. (\ref{E:Sk}) and (\ref{E:Kr})], with little more than a definition for the student to memorize; the vacuum is usually left unfilled ba a large number of teachers. 

Yet, the definition mentioned in the preceding paragraph have meaning because sample parameters converge towards population parameters as simple size grows towards population\textquoteright{s} size. But even if it is possible to apply the recipes to a sample drawn from a population with undefined statistical moments, such as the Cauchy pdf,  the estimates are meaningless because there are no population parameters to converge towards. 

All statistical central moments of the Cauchy distribution are undefined, mean, variance, kurtosis and skewedness of the Cauchy distribution are undefined. The Cauchy distribution is considered an example of a \textquotedblleft{}pathological\textquotedblright{} distribution function. Thus even when populations $ \{ L \} $ and $ \{ F \} $ are Gaussian, population $ \{ p \} $ [see Eqs. (\ref{E:LCorr} -- \ref{E:LivFract})] should be \textquotedblleft{}pathologically\textquotedblright{} distributed and its mean and variance should be undefined. Sample values will be concentrated about $\widehat{\mu}$, but the sample mean ($ \overline{x} $) will be increasingly variable as the number observations increases, due to the increased probability of encountering sample points with a large absolute value (\textquotedblleft{outliers}\textquotedblright). The distribution of the sample mean will be equal to the distribution of the outlying observations; i.e., the sample mean is just an estimator of any single outlying observation from the sample. Similarly, calculating the sample variance will result in values which grow larger as more observations are considered \cite{Rothenberg1964,Fama1968,Lohninger2012}.

Skewedness and kurtosis are essential for the Jarque-Bera Gaussianity test \cite{Jarque1980, Bera1981, Bera1981a, Giles2014}, and even in case of data following pathological distributions where their interpretation may be controvertible, if widely different from their values in pdfs where their value is known, they may be clear indicators showing that those distributions do not describe some data of interest.

In tables of this paper skewedness and kurtosis	and their apparent sample kurtosis calculated with the following functions is presented,  skewedness sample estimate is
\begin{equation}\label{E:Sk}
	Sk=\frac{\tfrac{1}{m}\sum_{i=1}^{m}(x_i-\overline{x})^3}{\left[\tfrac{1}{m}\sum_{i=1}^{m}(x_i-\overline{x})^2 \right] ^{3/2}}
\end{equation}
and kurtosis sample estimate is
	\begin{equation}\label{E:Kr}
		Kr=\frac{\tfrac{1}{m}\sum_{i=1}^{m}(x_i-\overline{x})^4}{\left[\tfrac{1}{m}\sum_{i=1}^{m}(x_i-\overline{x})^2 \right] ^{2}}.
	\end{equation}
	In both expressions $ m $ is sample size, $ x_i $ are data, and $ \overline{x}$ is the sample mean estimate. The definition represented by Eq. (\ref{E:Kr}) is presented here since there are controversies and discrepancies in the definition and interpretation of \textquotedblleft{}kurtosis\textquotedblright{} and \textquotedblleft{}excess kurtosis\textquotedblright{} or \textquotedblleft{}Pearson\textquoteright{}s kurtosis\textquotedblright{}, in the literature \cite{Pearson1905, Faleschini1948, Pearson1963, Proschan1965, Darlington1970, Hildebrand1971, Ali1974, Johnson1980, Moors1986, Ruppert1987, Westfall2014}, kurtosis is used here in sense of Moors \cite{Moors1986}:
	\begin{quote}
		\textquotedblleft{}High kurtosis, therefore, may arise in two situations: (a) concentration of probability mass near $ \mu $ (corresponding to a peaked unimodal distribution) and (b) concentration of probability mass in the tails of the distribution.\textquotedblright
	\end{quote}
	
	\section{Brief description of some biochemical aspects and procedures.}\label{S:Biochem}
	
	\subsection{Brief description of a method to determine cell apoptosis under the action of antineoplastic drugs.}\label{S:Apoptosis}
	
	\subsubsection{Brief description of the colorimetric procedure to detect cell mortality.}\label{S:ColorProc}
	
	As an example of determining the modified Hill Eq. (\ref{E:HillMod}) parameters data from a study on potentially anti-neoplastic compounds by fractions isolated from the marine organism \textit{Polyclinum constellatum} \cite{Quintana2017}. The procedure is a colorimetric assay \cite{Slater1963, Mosmann1983, Denizot1986} with a compound that has a pale yellow color, but if it penetrates into living cells it is turned into dark purple--blue crystals by an enzymatic mechanism. Dark purple-blue color is indicative of cell life. Cell death is determined measuring light absorbance \cite{Swinehart1962} of one cell thick layers (called \textit{monolayes}) in wells where the cells are seeded. A set of wells (called blank here $\{ B_h \}_{h=1, 2, \, \dots \, ,nb}$, \textit{nb} = 10 replicates) was pretreated with a detergent which kills and removes the cells from the wells prior to dye addition. A second set ($ \{L_h\}_{h=1,2,\cdots, nd} $, \textit{nd} = 48 replicates) of wells contained cells, exposed only to the dye used to identify living cells, without any putative cell killing fraction; the purple-blue product of the reaction in these wells is taken to represent 100\% living cells. Finally, there is a number of absorbancy sets ($ \{F_{h} \}_{h=1,2.\cdots, nf} $) measured in wells with dye and various concentrations ([D\textsubscript{i}]  = 0.01, 0.03, 0.1, 0.3 and 1 mg/mL) of fractions under study (named FI -- FV), again, the purple-blue color is proportional to the fraction of cells not killed at the concentration tested.
	
	\subsubsection{Brief description of absorbance corrections and their use for dose--response curves.}\label{S:AbsAn}
	Absorbances were corrected for blank absorbance by subtraction as
	\begin{eqnarray} 
		\{ L_h^{*} \}_{h=1,\, \dots \,,nd \cdot nb}= \{ L_h \}_{h=1,\, \dots \,,nd}-\{ B_h \}_{h=1,\, \dots \,,nb} \label{E:LCorr} & \\
		\{ F_h^{*} \}_{h=1,\, \dots \,,nf \cdot nb}= \{ F_h \}_{h=1,\, \dots \,,nf}-\{ B_h \}_{h=1,\, \dots \,,nb}.  \label{E:FCorr} 
	\end{eqnarray}
	Equations (\ref{E:LCorr}) and (\ref{E:FCorr}) indicate that each element of $ \{ B \} $ (background absorbance) was subtracted from each measurements in the other two sets to produce two sets corrected for cell layer background absorbance (labeled with an asterisk). The colorimetric procedure establishes \cite{Slater1963, Mosmann1983, Denizot1986} that the fraction of living cells in presence of drug is linearly proportional to the ratio
	\begin{equation}\label{E:LivFract}
		\{p_h\}_{h=1,\, \dots \,,nf \cdot nd \cdot nb^2} =\frac{\{F_h^* \}_{h=1,\, \dots \,,nf \cdot nb}}{\{L_h^*\}_{h=1,\, \dots \,,nd \cdot nb}}
	\end{equation}
	thus $ nf \cdot nd \cdot nb^2 = 24000 $ estimates of the fraction of living cells were obtained in Quintana \cite{Quintana2017} and used for statistical processing at each fraction concentration. Drug effect, expressed as percentage of cell death was calculated as
	\begin{equation}\label{E:PercEf}
		\{y_i\}_{i=1,\, \dots \,,nf \cdot nd \cdot nb^2}=100 \cdot \left (1-\{p_h\}_{h=1,\, \dots \,,nf \cdot nd \cdot nb^2} \right ).
	\end{equation}
	
	The main difference between the analyses described in Eqs, (\ref{E:LCorr} -- \ref{E:LivFract}) and analyses in the literature, is that here the data sets were processed nonparametricaly with Moses statistics \cite{Hollander1973}, while most authors use a parametric approach without considering the non-Gaussianity of data involving ratios such as Eq. (\ref{E:LivFract}).
	
	\section{The Hessisan matrix.}\label{S:Hessian}
	
	The \textit{Hessian} matrix of a function $y(x\given \{\vect{\theta}\})$  is a matrix of second partial derivatives of the form
	\begin{equation}\label{E:GenHessian}
		\boldsymbol{\mbb{H}}[ y(x\given \{\vect{\theta}\})] =\left[ 
		\begin{matrix}
			\tfrac{\partial^2 y(x\given \{\vect{\theta}\})}{\partial \theta_1^2}
			&\tfrac{\partial^2 y(x\given \{\vect{\theta}\})}{\partial \theta_1\partial \theta_2}
			&\tfrac{\partial^2 y(x\given \{\vect{\theta}\})}{\partial \theta_1\partial \theta_3}
			&\tfrac{\partial^2 y(x\given \{\vect{\theta}\})}{\partial \theta_1 \partial \theta_4}\\
			
			\tfrac{\partial^2 f(y(x\given \{\vect{\theta}\})}{\partial \theta_2 \partial \theta_1}
			&\tfrac{\partial^2 y(x\given \{\vect{\theta}\})}{\partial \theta_2^2}
			&\tfrac{\partial^2 f(y(x\given \{\vect{\theta}\})}{\partial \theta_2 \partial \theta_3}
			&\tfrac{\partial^2 y(x\given \{\vect{\theta}\})}{\partial \theta_2 \partial \theta_4}\\
			
			\tfrac{\partial^2 f(y(x\given \{\vect{\theta}\})}{\partial \theta_3 \partial \theta_1}
			&\tfrac{\partial^2 f(y(x\given \{\vect{\theta}\})}{\partial \theta_3 \partial \theta_2}
			&\tfrac{\partial^2 y(x\given \{\vect{\theta}\})}{\partial \theta_3^2}
			&\tfrac{\partial^2 f(y(x\given \{\vect{\theta}\})}{\partial \theta_3 \partial \theta_4}\\
			
			\tfrac{\partial^2 y(x\given \{\vect{\theta}\})}{\partial \theta_4\partial \theta_1}
			&\tfrac{\partial^2 y(x\given \{\vect{\theta}\})}{\partial \theta_4\partial \theta_1}
			&\tfrac{\partial^2 y(x\given \{\vect{\theta}\})}{\partial \theta_4\partial \theta_3}
			& \tfrac{\partial^2 y(x\given \{\vect{\theta}\})}{\partial \theta_4^2}
		\end{matrix} \right]
	\end{equation}
	where $ x_j = [\text{D}_j] $ and $\{\vect{\theta}\}=\{\vect{y_0,y_m,K_m,n}\}$ for Equation (\ref{E:HillMod}). If $ \{\vect{\theta}\} $ are all linearly 
	independent, then $Hf \left(y(x\given \{\vect{\theta}\}) \right)$ is the \textit{diagona}l matrix:
	\begin{equation}\label{E:HessianInd}
		\boldsymbol{\mbb{H}}^L f[ y(x\given \{\vect{\theta}\}) ]=
		\left[ 
		\begin{matrix} 
			\tfrac{\partial^2 y(x\given \{\vect{\theta}\})}{\partial\theta_1^2}
			&0
			&\, \dots \,
			&0\\
			0
			&\tfrac{\partial^2 y(x\given \{\vect{\theta}\})}{\partial\theta_2^2}
			&\, \dots \,
			&0\\
			\vdots
			&\vdots
			&\ddots
			&\vdots\\
			0
			&0
			&\, \dots \,
			& \tfrac{\partial^2 y(x\given \{\vect{\theta}\})}{\partial\theta_n^2}
		\end{matrix}  \right]
	\end{equation}

	\section{Case study one: A modified Hill equation, or the Hill equation with offset.}\label{S:Hill}
	
	\subsection{Modified Hill equation first derivatives  at an optimum in  presence of uncertainty.}\label{S:HillFstDer}
	
	Equation (\ref{E:HillMod}) has the following first derivatives:
	
	\begin{eqnarray}
		\mf{H}_{opt}'(y_{0})=\frac{\partial \mf{H}_o([\mathrm{D}] \given \{\vect{y_0,y_m,K_m,n }\})}{\partial  y_0}=1 &\label{E:dy0}\\
		\mf{H}_{opt}'(y_m)=\frac{\partial \mf{H}_o([\mathrm{D}] \given  \{ \{\vect{y_0,y_m,K_m,n }\}) }{\partial y_m} = \frac{1}{1+\zeta}& \label{E:dym}\\
		\mf{H}_{opt}'(K_m)=\frac{\partial \mf{H}_o([\mathrm{D}] \given \{ \{\vect{y_0,y_m,K_m,n }\}) }{\partial K_m}= -\frac{\zeta \cdot n \cdot y_m}{K_m \cdot  \left(1+\zeta\right)^2} &\label{E:dkm}\\
		\mf{H}_{opt}'(n)=\tfrac{\partial \mf{H}_0([\mathrm{D}] \given \{\vect{y_0,y_m,K_m,n }\}) }{\partial  n}=-\frac{\zeta \cdot y_m \cdot \log (\zeta)}{n \cdot  \left(1+\zeta\right)^2},&\label{E:dn}
	\end{eqnarray}
	where $ \zeta = \left(\tfrac{K_m}{[\mathrm{D}]}\right)^n = \mf{D}^n$ as used in Eq. (\ref{E:GradVect}). In the work of Quintana \cite{Quintana2017} used here as a practical example of curve fitting, $ y_0 $ is probably due to uncertainties in background subtraction.
	
\begin{figure}[h!]
		\centering
		\includegraphics[width=12cm]{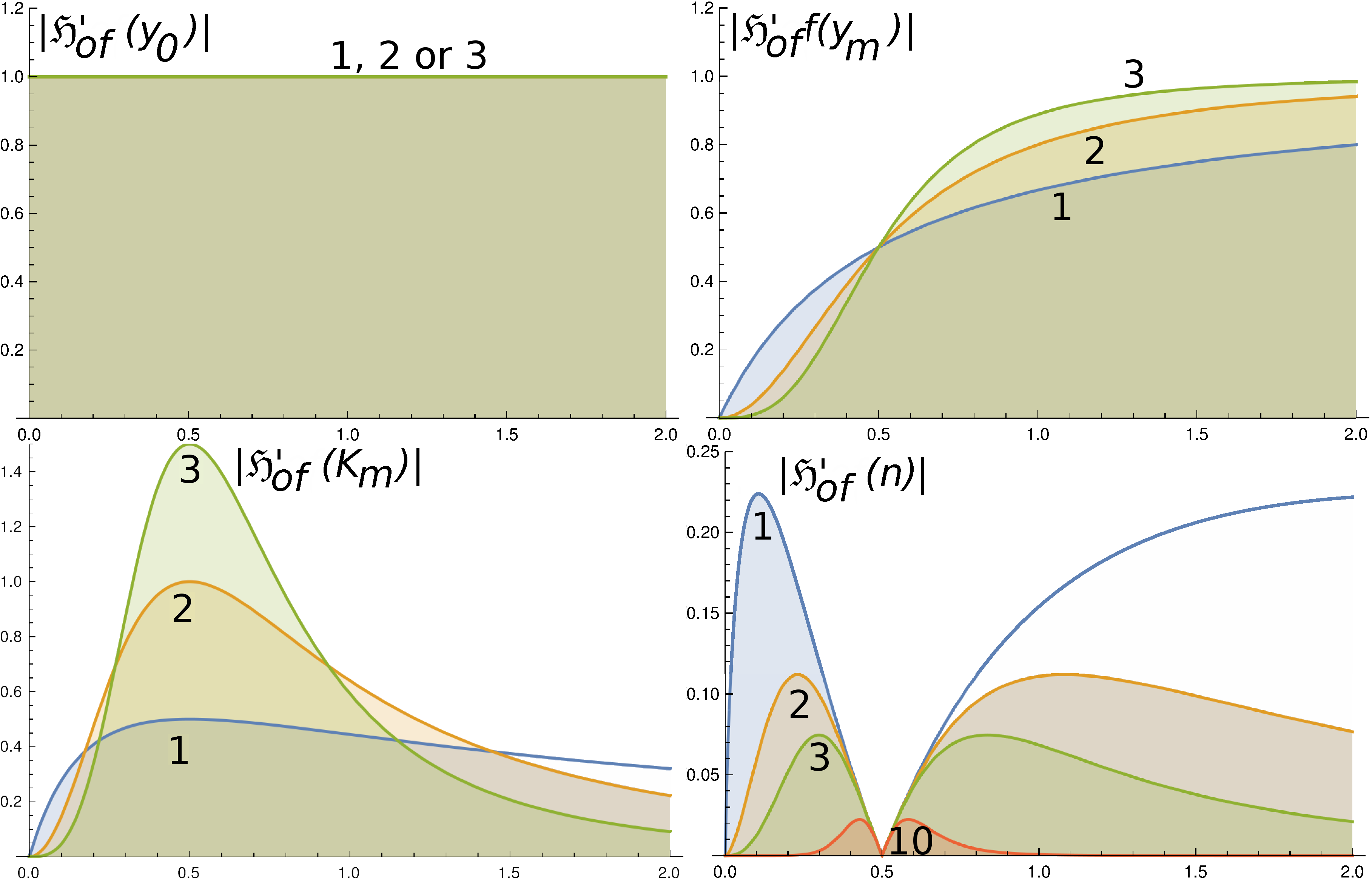}
		\caption{ \textbf{Rate of change absolute value as function of  $ \boldsymbol{y_0\text{,} y_m\text{,}  K_m \text{,} n    } $ .} Parameters used to calculate the derivatives were  $ y_0=-0.1$,   $y_m=1$, $K_m=0.5$, $n=\text{1, 2,3} $ except for $ \$ \mf{H}_{opt}' (n)$ where $n=\text{1, 2, 3, 10}$. Letterings indicate the absolute partial derivatives of $ \mf{H}_o(\theta_i) $ respect to $\theta_i$. The figure present absolute values of derivatives calculated with Eqs. (\ref{E:dym}) to (\ref{E:dn}). Please notice that the abscissa has same value in all panels, top panels have the same ordinate scale, but lower panels have different ordinate scales scales between them, and also different from the top panels. Other  details in the  text of the  communication.}
		\label{F:VariancesOptimum}
	\end{figure}

	Figure \ref{F:VariancesOptimum} presents absolute values plots  of Eqs. (\ref{E:dy0}) through  (\ref{E:dn}) as functions of [D]. To facilitate comprehension, absolute value of  the derivatives are plotted, please  notice that Eqs. (\ref{E:dkm}) and (\ref{E:dn}) have negative signs. Parameters used to calulate the derivatives in Figure \ref{F:VariancesOptimum} were  $ \{y_0=-0.1,\;  y_m=1,\; K_m=0.5,\; n=\text{1, 2, 3 or 10}  \} $. Numbers  near the  curves indicate the value of $ n $ used to calculate each curve. Eq. (\ref{E:dy0}) tells that $ \mf{H}_{opt}'(y_0) $ is constant for any [D], but the other derivatives are more sophisticated functions of [D]. Next, $ \mf{H}_{opt}'(y_m) =0$ when [D]=0, grows with [D] and becomes increasingly sigmoid as $ n $ increases, all curves describing $ \mf{H}_{opt}'(y_m) $ intercept at $ \text{[D]}=K_m $ (or $ \mf{D}=1 $), at this point the curves increase in slope as $ n $ increases, and

	There is another difference between $ \$ \mf{H}_{opt}'(n) $ and the other three derivatives:  $  \displaystyle\lim_{n \to +\infty} \abs{\$ \mf{H}_{opt}'(n)}=0  $, the bigger  $n$ gets, the least it contributes to $\mf{H}_{opt}([\mathrm{D}] \given \{\vect{y_0,y_m,K_m,n }\})$ uncertainty, making it harder to guess in any optimization procedure when $n \gg 1$ since there is less and less to minimize in regard to $n$ as it increases. With the same reasoning since
	\begin{equation*}
		\left\lbrace \abs{\$ \mf{H}_{opt}'(y_0)}>\abs{\$ \mf{H}_{opt}'(y_m)}\right\rbrace  \; \forall \; \text{[D]}
	\end{equation*}	  
	it would be easier to determine $ y_0 $ with more accuracy than $ y_m $; in colloquial terms $ y_0 $ inreoduces the same uncertainty over all the [D] range, while $ y_m $ contribution increases with [D]. Data in Figure \ref{F:VariancesOptimum} suggest also that accuracy of $ K_m $ is higher if  $ n $ is higher and when enough data is collected around $ K_m $. 
	
	The most uncertain parameter to estimate seems to be $ n $, data in Figure \ref{F:VariancesOptimum} suggests, however that accuracy of $ n $ estimates improve when data between $ \text{\textonehalf}K_m $ and $ 2 K_m $ is more available, but that even then a good estimate of a high $ n $ would be difficult (if possible) to get, as it is suggested by the curve calculated setting $ n=10 $.

	\subsection{The modified Hill equation Hessian matrix.}
	
	The Hessian matrix, of Eq. (\ref{E:HillMod}) is the \textit{non diagonal} matrix:
	
	\begin{equation}\label{E:HillHessian}
		\boldsymbol{\mbb{H}} \mf{H}_o([\mathrm{D}] \given \{\vect{y_0,y_m,K_m,n }\})=\Xi\cdot 
		\left[
		\begin{matrix}
			\begin{array}{cccc}
				0 & 0 & 0 & 0 \\
				0 & 0 & -\tfrac{n}{K_m} & -\tfrac{\log \left(\zeta\right)}{n} \\
				0 & 0 & \tfrac{n \left[\zeta+\left(\zeta-1\right) n+1\right] y_m}{K_m^2 \left(\zeta+1\right)} & -\tfrac{y_m \left[\zeta-\left(\zeta-1\right)  \log \left(\zeta\right)-1\right]}{K_m \left(\zeta+1\right)} \\
				0 & -\tfrac{\log \left(\zeta\right)}{n} & -\tfrac{y_m \left[\zeta-\left(\zeta-1\right)  \log \left(\zeta\right)-1\right]}{K_m \left(\zeta+1\right)} & \tfrac{\left(\zeta-1\right) y_m \log ^2\left(\zeta\right)}{n^2 (\zeta+1)} \\
			\end{array}
		\end{matrix}
		\right]
	\end{equation}
	where
	\begin{equation}
		\Xi =\frac{\zeta}{\left(1+\zeta\right)^2}. 
	\end{equation}
	Equation(\ref{E:HillHessian}) shows the Hill equation non linearity, Also if $[\mathrm{D}]=K_m$ then $ \Xi =\tfrac{1}{4} $ and 
	
	\begin{equation}
		\boldsymbol{\mbb{H}}\mf{H}_o([\mathrm{D}] \given \{\vect{y_0,y_m,K_m,n }\})=\frac{y_m}{4K_m} \cdot
		\left[
		\begin{matrix}
			0& 0& 0& 0\\
			0& 0& -\tfrac{n}{y_m}& 0\\
			0& 0& \tfrac{n}{K_m}& -1\\
			0& 0& -1& 0\\
		\end{matrix}
		\right]
	\end{equation}
	where $ \left \{0,0,   \tfrac{\left( n - \sqrt{4 K_m^2+n^2} \right) y_m}{8 K_m^2} , \frac{\left(n + \sqrt{4 K_m^2+n^2}\right) y_m}{8 K_m^2}  \right \} $ are the eigenvalues of $ \boldsymbol{\mbb{H}}\mf{H}_o([\mathrm{D}] \given \{\vect{y_0,y_m,K_m,n }\})$, and
	\begin{equation}
		\textrm{Disc} \left \{ \boldsymbol{\mbb{H}}\mf{H}_o([\mathrm{D}] \given \{\vect{y_0,y_m,K_m,n }\}) \right \}=\frac{y_m}{4K_m}
		\left (
		\begin{matrix}
			0& 0& 0& 0\\
			0& 0& -\tfrac{n}{y_m}& 0\\
			0& 0& \tfrac{n}{K_m}& -1\\
			0& 0& -1& 0\\
		\end{matrix}
		\right)=0
	\end{equation}
	where $ \textrm{Disc} $ stands for \textit{Discriminant}, a determinant form of the Hessian matrix. Since $ \textrm{Disc} =0 $ when $ [\mathrm{D}] = K_m $ the function reaches a \textit{degenerate critical point} \cite{Bolis1980} where $ \mf{H}_o([\mathrm{D}] \given \{\vect{y_0,y_m,K_m,n }\}) $ has an inflection.
	
	\section{Case study two: The Boltzmann equation.}\label{S:BoltzEqu}
	
	A common form of the Boltzmann equation used in biology is \cite{Sevcik2017a}:
	\begin{equation}\label{E:BoltzElect}
		\mf{B} \left( V \given \{\vect{ V_{ \text{\textonehalf{}} } , \, \kappa } \} \right)= \frac{1}{1+\e^{-(V-V_{\text{\textonehalf{}}})/ \kappa}}\;. 
	\end{equation}
	When Eq. (\ref{E:BoltzElect}) is used in the original fashion of Hodfkin and Huxley \cite[pg 501, Eq. 1]{Hodgkin1952e}, to represent trans membrane distribution of some charged particle, $B$ is expressed in respect to the potential at which 50\% of the particles are in one side of the membrane, and 50\% is at the other side. Eq. (\ref{E:BoltzElect}) is thus reduced to a situation where a dependent variable $B$ may be fitted by some nonlinear optimization procedure to an independent variable $V$ (usually expressed in mV) using Eq. (\ref{E:BoltzElect}). The optimization procedure enables to estimate the parameters $V_{\text{\textonehalf}}$ and $\kappa$.
	
	\subsection{Boltzmann equation first derivatives  at an optimum in  presence of uncertainty.}\label{S:BoltzFstDer}
	
	Mathematical properties of the Boltzmann equation are discussed elsewhere 
	\cite{Sevcik2017a}. But since the first derivatives of Eq. (\ref{E:BoltzElect}) 
	respect to $ \left\lbrace \theta_{i}\right\rbrace  $ are crucial for this paper, they 
	are presented here:
	\begin{eqnarray}
		\mf{B}'(V_{\text{\textonehalf}})=\frac{\partial \mf{B} \left( V \given \{\vect{ V_{ \text{\textonehalf{}}}, \kappa }   \} \right)}{\partial V_{\text{\textonehalf{}}}} &=-\dfrac{1}{2\kappa \cosh \left(\frac{\upsilon}{k}\right)+2 \kappa \label{E:BoltDerVmed} }\\
		\mf{B}'(\mf{\kappa})=\frac{\partial \mf{B} \left( V \given \{\vect{ V_{ \text{\textonehalf{}}}, \kappa }   \} \right)}{\partial  \kappa}&=-\dfrac{\upsilon \sech^2 \left(\tfrac{\upsilon}{2  \kappa } \right)}{4 \kappa^2} \label{E:BoltDerKappa} 
	\end{eqnarray}
	where $ \upsilon = V- V_{\text{\textonehalf}} $. Eqs. (\ref{E:BoltDerVmed}) and 
	(\ref{E:BoltDerKappa}) contrast sharply with the situation discussed in 
	connection with  Eqs. (\ref{E:dy0}) through (\ref{E:dn}), Even though Eq. 
	(\ref{E:BoltDerVmed}) reaches a maximum while Eq. (\ref{E:BoltDerKappa}) 
	reaches a minimum at $ K_m $, in the vicinity of this value the two derivatives 
	have similar high values which grow as $ \mf{v} $ gets higher and $ 
	\abs{V}\rightarrow +\infty $, this suggests that both parameters contribute 
	similar uncertainties to variates distributed around $ \mf {B} \left( V \given \left\lbrace  \vect{
	V_{\text{\textonehalf{}}} ,\kappa}\right\rbrace \right) $ and that that both $ V_{\text{\textonehalf}} $ and $ \kappa $ will 
	be determined with similar ac curacies after optimization.
	
	\begin{figure}[h!]
		\centering
		\includegraphics[width=12cm]{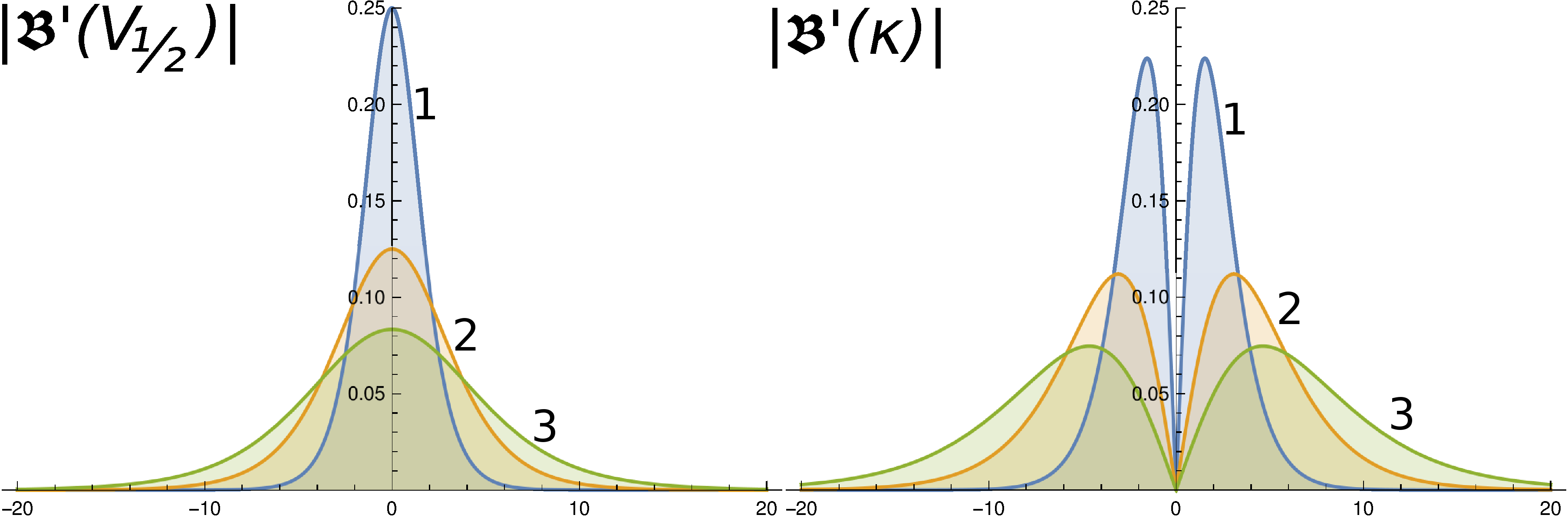}
		\caption{$  \boldsymbol{\mf{B} ( V \given V_{1/2}, \kappa )} $ \textbf{  rate of change absolute value as function of} $ \boldsymbol{V_{\tfrac{1}{2}}}$, and $\boldsymbol{\kappa} $.  Parameters used to calculate the derivatives were  $ V_{1/2}=0 $ and ${v}=\text{1, 2 and 3} $. Lettering in the panels indicate absolutes values of: $\mf{B}' ( V_{1/2})= \tfrac{\partial \mf{B} ( V \given V_{1/2}, \kappa)}{\partial V_{1/2} }$ and  $ \mf{B}' ( \kappa)= \tfrac{\partial  \mf{B} ( V \given V_{1/2}, \kappa)}{\partial \kappa }$. Numbers near the curves indicate the value of  v  used to calculate each curve. Other  details in the  text of the communication.} 
		\label{F:BolVariancesOptimum}
	\end{figure}
	
	\subsection{Boltzmann equation gradient.}
	
	\begin{equation}\label{E:BoltzGrad}
		\vect{\nabla \mf{B} \left( V \given \{\vect{ V_{ \text{\textonehalf{}}, \mf{v} } }  \} \right) }=-\frac{(\mf{v} - \upsilon) }{4 \mf{v}^2}  \sech^2\left( \frac{\upsilon}{2 \mf{v}}\right)
	\end{equation}
	
	if $  \upsilon=0 $ then
	\begin{equation}
		\vect{\nabla  \mf{B} \left( V=V_{\text{\textonehalf}} \given \lbrace \vect{V_{\text{\textonehalf}},\mf{v}}\rbrace \right) }= -\frac{1}{4 \mf{v}}
	\end{equation}
	which shows that the gradient at $ \upsilon $ only delends on $ \mf{v} $. 
	
\end{appendices}

\section*{Conflicts of interests.}

There are no conflicts of interest. 

\section*{Acknowledgments.}

Free and open-source software was used extensively. This manuscript was written in \LaTeX{ }using \TXs  for Ubuntu Linux{ }(Also , \url{http://www.texstudio.org}), an open source free  \TeX{ }editor under \TeX Live 21 (\url{https://www.tug.org/texlive/}). Most images were calculate using \LOv{ }Calc and assembled ad edited using GIMP 2.10.24 for Ubuntu Linux. the GNU Image Prosecution Manager, a free and open source program also available for Apple{} OS X and MicroSoft Windows{}.
	
%\bibliography{mybib}

\end{document}